\documentclass[twocolumn,showpacs,nofootinbib,preprintnumbers,amsmath,amssymb,showkeys,superscriptaddress]{revtex4}

\usepackage{epsfig}
\usepackage{color}
\usepackage{graphicx}
\usepackage{dcolumn}
\usepackage{bm}
\usepackage{dsfont}

\newcommand{\beq}{\begin{equation}}
\newcommand{\eeq}{\end{equation}}
\newcommand{\bea}{\begin{eqnarray}}
\newcommand{\eea}{\end{eqnarray}}
\newcommand{\nn}{\nonumber \\}

\newcommand\eqn[1]{(\ref{#1})}      
\newcommand\Eqn[1]{Eq.~(\ref{#1})}  
\newcommand\Fig[1]{Fig.~\ref{#1}}  

\newcommand{\tr}{\hbox{tr}}

\newcommand{\cb}{{\bar c}}

\begin{document}


\title{Two-loop study of the deconfinement transition\\ in Yang-Mills theories: SU($3$) and beyond
}

\author{U. Reinosa}%
\affiliation{%
Centre de Physique Th\'eorique, Ecole Polytechnique, CNRS, Universit\'e Paris-Saclay, F-91128 Palaiseau, France.
\vspace{.1cm}}%
\author{J. Serreau}%
\affiliation{APC, AstroParticule et Cosmologie, Universit\'e Paris Diderot, CNRS/IN2P3, CEA/Irfu, Observatoire de Paris, Sorbonne Paris Cit\'e,\\ 10, rue Alice Domon et L\'eonie Duquet, 75205 Paris Cedex 13, France.\vspace{.1cm}}
\author{M. Tissier}
\affiliation{Laboratoire de Physique Th\'eorique de la Mati\`ere Condens\'ee, UPMC, CNRS UMR 7600, Sorbonne
Universit\'es, 4 Place Jussieu,75252 Paris Cedex 05, France.
\vspace{.1cm}}
\affiliation{%
 Instituto de F\'{\i}sica, Facultad de Ingenier\'{\i}a, Universidad de la Rep\'ublica, \\J.H.y Reissig 565, 11000 Montevideo, Uruguay.
\vspace{.1cm}}%
\author{N. Wschebor}%
\affiliation{%
 Instituto de F\'{\i}sica, Facultad de Ingenier\'{\i}a, Universidad de la Rep\'ublica, \\J.H.y Reissig 565, 11000 Montevideo, Uruguay.
\vspace{.1cm}}%

\date{\today}

\begin{abstract}
We study the confinement-deconfinement phase transition of pure Yang-Mills theories at finite temperature using a simple massive extension of standard background field methods. We generalize our recent next-to-leading-order perturbative calculation of the Polyakov loop and of the related background field effective potential for the SU($2$) theory to any compact and connex Lie group with a simple Lie algebra. We discuss in detail the SU($3$) theory, where the two-loop corrections yield improved values for the first-order transition temperature as compared to the one-loop result. We also show that certain one-loop artifacts of thermodynamical observables disappear at two-loop order, as was already the case for the SU($2$) theory. In particular, the entropy and the pressure are positive for all temperatures. Finally, we discuss the groups SU($4$) and Sp($2$) which shed interesting light, respectively, on the relation between the (de)confinement of static matter sources in the various representations of the gauge group and on the use of the background field itself as an order parameter for confinement. In both cases, we obtain first-order transitions, in agreement with lattice simulations and other continuum approaches.
\end{abstract}

\maketitle



\section{Introduction}
\label{sec:intro}

Understanding the phase structure of non-Abelian gauge theories at finite temperature is of interest not only for phenomenological applications, e.g., quark-gluon plasma physics in heavy-ion collisions or early-Universe cosmology, but also for unraveling the essential aspects of confinement. On the one hand, direct numerical simulations of pure Yang-Mills (YM) theories with various gauge groups allow one to unambiguously assess the existence and the properties of a confinement-deconfinement phase transition (e.g., the order of the transition, the behavior of thermodynamical observables, etc.) \cite{Engels:1981qx,Boyd:1995zg,Bazavov:2014pvz,Borsanyi:2013bia}. 
On the other hand, continuum approaches necessarily involve approximations and the comparison with lattice results provides a valuable test for identifying the relevant features of the dynamics at work. 

Standard perturbation theory, based on the Faddeev-Popov (FP) quantization, generally fails near the transition temperature because the latter is typically of the order of the intrinsic scale of the theory, where the running coupling is large. Various nonperturbative approaches have been devised to tackle this problem, such as the nonperturbative/functional renormalization group (NPRG/FRG) \cite{Braun:2007bx,Marhauser:2008fz,Braun:2010cy,Fister:2013bh}, truncations of the Dyson-Schwinger equations \cite{Epple:2006hv,Alkofer:2006gz,Fischer:2009wc}, the Hamiltonian approach of Ref.~\cite{Reinhardt:2012qe}, or two-particle-irreducible techniques \cite{Fister:2013bh,Fukushima:2012qa}. In their pioneering work \cite{Braun:2007bx}, Braun {\it et al.} have pointed out the usefulness of background field techniques \cite{Abbott:1980hw,Abbott:1981ke} in keeping track of the relevant symmetries and have obtained a criterion relating the existence of a confined phase at low temperature to the low-momentum behavior of the gluon and ghost two-point functions; see also Ref.~\cite{Fister:2013bh}. Similar studies have been performed in the presence of dynamical quarks \cite{Fischer:2009gk,Fischer:2011mz,Fischer:2013eca,Fischer:2014vxa}.

Another possible approach consists in building effective models for the relevant order parameters of the transition, namely the Polyakov loops in the various representations of the gauge group \cite{Dumitru:2003hp,Dumitru:2012fw,Sasaki:2012bi,Dumitru:2013xna}. In such descriptions, mainly guided by symmetries, the nonperturbative aspects of the underlying dynamics is captured by some phenomenological parameters to be fitted against lattice data.

 A different line of development is based on modified quantization procedures \cite{Zwanziger89,Vandersickel:2012tz,Dudal08,Serreau:2012cg,Serreau:2013ila}. They are motivated by the fact that  the FP quantization is plagued by the issue of Gribov ambiguities, which are believed to be important for temperatures close to and below the transition. A prominent example is the Gribov-Zwanziger quantization \cite{Zwanziger89}, where the functional integral is restricted to the so-called first Gribov region---the field configurations for which the FP determinant is positive---possibly including some phenomenological condensates \cite{Dudal08}. This has been used to study the confinement-deconfinement transition of SU($N$) theories in Refs.~\cite{Zwanziger:2004np,Fukushima:2013xsa,Canfora:2015yia}.

Another approach, developed in Refs. \cite{Tissier_10,Tissier_11}, involves a simple massive extension of the FP action. On the practical side, this is motivated by the observation that lattice calculations of the vacuum propagators in the Landau gauge \cite{Cucchieri_08b,Cucchieri_08,Cucchieri09,Bogolubsky09,Dudal10} show that the gluon behaves as a massive field at low momentum. In fact, perturbative calculations of YM correlators in a massive extension of the Landau gauge action give a fairly accurate description of lattice results in the vacuum \cite{Tissier_10,Tissier_11,Pelaez:2013cpa,Pelaez:2014mxa,Pelaez:2015tba} (see also Ref.~\cite{Reinosa:2013twa} for a comparison at finite temperature). On a more fundamental level such massive extensions can be motivated by the quantization method put forward in Ref.~\cite{Serreau:2012cg}, where one averages over Gribov copies with a nontrivial weight involving a new dimensional (gauge-fixing) parameter, eventually related to the gluon mass. Some variants of this approach have been studied in Refs.~ \cite{Weber:2011nw,Siringo:2015aka}.

This massive extension has been used in the context of background field methods to study the phase diagram of SU($N$) YM theories at finite temperature in Refs.~\cite{Reinosa:2014ooa,Reinosa:2014zta} and of QCD with heavy quarks at finite temperature and chemical potential in Ref.~\cite{Reinosa:2015oua}, with remarkable results. A simple leading-order (LO) perturbative calculation accurately reproduces the main qualitative and even some quantitative aspects of the known phase diagrams of these theories. This follows from an incomplete cancellation between massive and massless degrees of freedom, as pointed out in Refs.~\cite{Braun:2007bx,Fister:2013bh} (a different massive model has been studied in Ref.~\cite{Meisinger:2001cq} which, however, does not lead to a phase transition due to the absence of massless modes). In Ref.~\cite{Reinosa:2014zta}, we have computed the next-to-leading-order (NLO) corrections to both the background effective potential and the Polyakov loop in the pure gauge SU($2$) theory, which undergoes a second-order phase transition. This gives improved values of the critical temperature and cures some unphysical features of the LO expressions of thermodynamical observables near the transition.

The main aim of the present work is to extend this NLO analysis to the SU($3$) YM theory, where the transition is of first order in four spacetime dimensions \cite{Yaffe:1982qf}. In fact, we generalize the perturbative approach and the NLO calculations of Ref.~\cite{Reinosa:2014zta} to any compact and connex Lie group with a simple Lie algebra. We explicitly discuss the NLO results for the background field potential and the Polyakov loop for the SU($3$) theory, where, as in the SU($2$) case, we obtain an improved value for the transition temperature as compared to the LO result and we find that NLO corrections lead to better-behaved thermodynamical observables. In particular, both the pressure and the entropy density are positive at all temperatures. We also discuss some limitations of the Landau gauge and of the present approach for what concern these observables. 
Finally, we discuss the phase structure of the Sp($2$) and SU($4$) theories at LO, which also undergo a first-order transition in four dimensions \cite{Moriarty:1981wc,Green:1984ks,Batrouni:1984vd,Wheater:1984wd,Holland:2003kg}. These are interesting because of the particular way center symmetry is realized/broken as compared to the SU($2$) and SU($3$) cases.

The paper is organized as follows. In Sec.~\ref{sec:gen}, we set the notations used throughout this work. In particular we introduce an intrinsic bilinear invariant form on the color algebra that allows for the construction of gauge-invariant quantities and facilitates the change from standard cartesian bases to so-called canonical or Cartan-Weyl bases. The latter allow for a simple generalization of usual Feynman rules in the presence of the nontrivial background field to be considered here. We also recall some generalities concerning the relation between the confinement of static sources, as measured by the appropriate Polyakov loops, and center symmetry.  We present the massive extension of the Landau-DeWitt (LDW) gauge in Sec.~\ref{sec:LdW}. There, we discuss some issues of background field methods in general and of the present massive extension in particular, and we derive the Feynman rules of the massive theory using the framework of canonical bases. We present a detailed discussion of the relevant symmetries of the problem in Sec.~\ref{sec:symmetries}. Our NLO calculations of the background field potential and of the Polyakov loop in any unitary representation for arbitrary compact and connex Lie group with a simple Lie algebra are presented in Secs.~\ref{sec:pot} and \ref{sec:pol}, respectively. The explicit NLO results for the SU($3$) theory are discussed in detail in Sec.~\ref{sec:su3}, while our LO analysis of the phase structure of the Sp($2$) and SU($4$) theories is presented in Sec.~\ref{sec:su4}. We summarize and conclude in Sec.~\ref{sec:conc}. Some technical discussions and material are gathered  in Appendices~\ref{appsec:mass}--\ref{appsec:ploopcounting}. Some parts of the article are technical and can be skipped in a first reading. The reader mainly interested in our physical results can omit the material of Secs.~\ref{sec:LdW} and \ref{sec:symmetries}, except for Sec.~\ref{sec:SSB}.

\section{Generalities}\label{sec:gen}

In this work, we consider a compact and connex Lie group $G$ with a simple Lie algebra ${\cal G}$. Without loss of generality, we can assume that $G$ is a subgroup of a matrix group U($n$) for some $n$ \cite{Knapp}, and thus that the elements of ${\cal G}$ are anti-Hermitian matrices.

\subsection{Notations}\label{sec:notations}
An infinitesimal group transformation corresponds to an element of the group close to the identity, $1+\theta$, where $\theta$ is an element of the algebra ${\cal G}$. Such a transformation acts on any other element $X\in{\cal G}$ as $X\to X+\delta X$, with
\beq
\delta X=[\theta,X]\,.
\eeq
The commutator $[X,Y]$ of two elements $X,Y\in{\cal G}$ transforms covariantly, that is $\delta [X,Y]=[[\theta,X],Y]+[X,[\theta,Y]]=[\theta,[X,Y]]$. This is so even when $X$ and $Y$ are ${\cal G}$-valued fields and the infinitesimal group transformation is gauged. In contrast, the derivative $\partial_\mu X$ of a ${\cal G}$-valued field $X$ is not covariant under gauged transformations. The covariant derivative is defined as $D_\mu X\equiv\partial_\mu X-[A_\mu,X]$, where the connection $A_\mu$ transforms as
\beq
\delta A_\mu=[\theta,A_\mu]+\partial_\mu\theta\,.
\eeq
From the connection $A_\mu$, one can define the field-strength tensor $F_{\mu\nu}\equiv\partial_\mu A_\nu-\partial_\nu A_\mu-[A_\mu,A_\nu]$, which is also covariant. 

One builds gauge-invariant combinations from the various covariant fields defined above by means of the (negative-definite) bilinear form\footnote{As the elements of ${\cal G}$ are anti-Hermitian matrices, we have $(X;X)=-2\,{\rm tr}\,X^\dagger X=-\sum_{ab} |X_{ab}|^2\leq 0$.}
\beq
\label{eq:invform}
(X;Y)=2\,{\rm tr}\,X Y\,.
\eeq
Indeed, it is easily checked that
\beq\label{eq:inv}
\delta (X;Y)=([\theta,X];Y)+(X;[\theta,Y])=0\,,
\eeq 
a property which we shall also use in the form of the cyclic identities
\beq\label{eq:cyclic}
(X_1;[X_2,X_3])=(X_2;[X_3,X_1])=(X_3;[X_1,X_2])\,,
\eeq
for any $X_1,X_2,X_3\in {\cal G}$. 

In terms of the invariant bilinear form, the pure YM Euclidean action associated to the simple Lie algebra ${\cal G}$ reads\footnote{All definite invariant forms on a simple Lie algebra are proportional to each other; see for instance Ref.~\cite{WeinbergBook}. The choice of negative-definite invariant form in Eq.~(\ref{eq:YM}) is then arbitrary and amounts to a redefinition of the bare coupling.}
\beq\label{eq:YM}
S_{\rm YM}=-\frac{1}{4g^2_0}\int_x (F_{\mu\nu};F_{\mu\nu})\,,
\eeq
where $g_0$ is the bare coupling, $\int_x$ denotes $\int_0^\beta d\tau\int d^{d-1}x$, where $\beta=1/T$ is the inverse temperature, and $d=4-2\varepsilon$, as required by the dimensional regularization that we consider in this work. In a basis $\{\tau_a\}$ of the algebra such that $(\tau_a;\tau_b)=-\delta_{ab}$ one recovers the standard form
\beq
S_{\rm YM}=\frac{1}{4g_0^2}\int_x F^a_{\mu\nu}F^a_{\mu\nu}\,.
\eeq

Although the use of the negative-definite form in (\ref{eq:YM}) might seem unnecessary, it is appropriate when the elements of a basis $\{\tau_a\}$ of the algebra are written $\smash{\tau_a=it_a}$, as is the common practice in the physics literature, in which case $(t_a;t_b)=\delta_{ab}$. Strictly speaking, the $t_a$'s need to be seen as elements of the complexified algebra ${\cal G}\times{\cal G}\equiv {\cal G}\oplus i{\cal G}$, over which the commutator $[\,\,,\,]$ and the form $(\,\,;\,)$ are extended by bilinearity; see Appendix A. The notion of a complexified algebra is useful for our later purposes in that it allows one to decompose the elements of the (real) algebra ${\cal G}$ on a larger class of bases, such as the Cartan-Weyl bases $\{it_\kappa\}$, that we discuss below. The use of an intrinsic (i.e. coordinate independent) structure such as the invariant bilinear form in (\ref{eq:YM}) eases the switch to these more general bases.

Finally, because the Lie group $G$ is assumed to be compact and connex, its Lie algebra ${\cal G}$ is mapped onto $G$ via the exponential map $\theta\mapsto e^\theta$. After exponentiation, the YM action (\ref{eq:YM}) is invariant under any finite gauged group transformation whose elements belong to $G$. These finite transformations take the form
\beq
X^U = UXU^{-1}\,,\quad A^U_\mu = UA_\mu U^{-1}-U\partial_\mu U^{-1}\,,\label{eq:two}
\eeq
and one has $(X^U;Y^U)=(X;Y)$.

\subsection{Center symmetry, Polyakov loops and confinement}

 Let us briefly recall some basic facts about the phase transition of pure gauge theories at nonzero temperature. The latter is governed by the spontaneous breaking of the global symmetries associated with the center of the gauge group, which can be interpreted in terms of a confinement-deconfinement transition of infinitely heavy matter fields (static sources) in group representations of nontrivial $N$-alities \cite{Svetitsky:1985ye,Pisarski:2002ji,Greensite:2003bk}. Indeed, the free energy $F_R$ of a static matter source in a given (complex and finite) irreducible representation $it^a\mapsto it^a_R$ of dimension $d_R$ is directly related to the (averaged) Polyakov loop 
\beq
  \label{eq_pol_trace}
\ell_R\equiv\frac{1}{d_R}\langle{\rm tr}L_R({\bf x})\rangle\propto e^{-\beta F_R}\,,
\eeq
where $L_R$ is the path-ordered exponential
\begin{equation}
  \label{eq_pol_untrace}
L_R({\bf x})=  P\exp\int_0^\beta \!\!d\tau\,iA_0^a(\tau,{\bf x})t^a_R.
\end{equation}
A confined state for a given source corresponds to an infinite free energy, that is, a vanishing Polyakov loop in the corresponding representation. 

The relation with the center symmetry stems from the fact that, at finite temperature, the measure (including the Yang-Mills action) of the functional integral is invariant under generalized gauge transformations $U(\tau,{\bf x})$, such that 
\beq\label{eq:per}
U(\tau+\beta,{\bf x})=U(\tau,{\bf x})Z\,,
\eeq
where $Z$ is an element of the center of $G$ \cite{Svetitsky:1985ye}. In the case of compact Lie groups with a simple Lie algebra, the latter is always a subgroup of the form $\{e^{i2\pi k/N}\mathds{1},k=0,\dots,N-1\}\simeq Z_N$, for some $N$. Under the above generalized gauge transformation with a center element $Z=z\mathds{1}$, the Polyakov loop \eqn{eq_pol_trace} transforms as $ \ell_R\to z^p\ell_R$, where the integer $0\le p\le N-1$ is the $N$-ality of the representation $R$. It follows that $\ell_R=0$---and, hence, that the corresponding matter sources are confined---for any representation $R$ with nonzero $N$-ality if the center symmetry is not broken. 

We mention that, if a manifest center symmetry clearly implies the confinement of all sources in representations with nonzero $N$-alities, the reverse is not completely obvious. If the nonvanishing of at least one Polyakov loop with nontrivial $N$-ality suffices to conclude that the center symmetry is spontaneously broken, it is not clear that all of them have to vanish simultaneously. Indeed, one could imagine situations where only some of these Polyakov loops vanish, e.g., in case of a partial breaking of the center, signaling a phase of partial deconfinement \cite{Cohen:2014swa}. We shall discuss this issue in more detail when analyzing the SU($4$) theory in Sec.~\ref{sec:su4}.

Clearly, it is of key importance to work in a framework where the center symmetry is as explicit as possible. In particular, since continuum approaches require one to fix a gauge (see, however, Ref.~\cite{Arnone:2005fb}) it is important to do so in a way that does not the break this symmetry explicitly. This is the main motivation behind the use of background field methods in the present context \cite{Braun:2007bx}. Here, we follow Ref.~\cite{Braun:2007bx} and consider the LDW gauge, the background field generalization of the Landau gauge. We wish to evaluate the Polyakov loops in the various representations of the gauge group using some sensible approximation scheme. In this work, we perform a NLO perturbative calculation in a simple massive extension of the LDW gauge. The rationale for this massive extension has been discussed at length elsewhere \cite{Tissier_10,Tissier_11,Serreau:2012cg,Reinosa:2014ooa} and we shall not repeat it here. In short, the mass term for the gluonic fluctuations about the background field is motivated by the issue of Gribov ambiguities (which presumably play a nontrivial role at low energies) and the associated (soft) breaking of the Becchi-Rouet-Stora-Tyutin (BRST) symmetry of the FP action.

\section{The massive LDW action}\label{sec:LdW}

In this section, we first introduce the massive LDW gauge for any compact and connex Lie group with a simple Lie algebra and mention some open issues concerning background field methods in general and the present massive extension in particular. We then specify to the case of a homogeneous background field in the temporal direction (which respects the space-time symmetries of the problem at hand). The latter introduces a preferred direction in color space and it proves convenient to work in a color basis where the corresponding covariant derivative is diagonal. This is the case of the so-called canonical or Cartan-Weyl bases. Because the latter play such a central role in the present framework, we discuss them in some detail; see also Refs.~\cite{Dumitru:2012fw,Dumitru:2013xna}. Finally, we derive the Feynman rules of the YM theory in the massive LDW gauge in these particular bases.

\subsection{The (massive) LDW gauge}\label{subsec:LdW}

The LDW gauge is defined as 
\beq
\label{eq:LDW}
 \bar D_\mu a_\mu=0,
\eeq
with $a_\mu=A_\mu-\bar A_\mu$ and $\bar A_\mu$ is a given (background) gauge field configuration. In terms of the invariant form \eqn{eq:invform}, the (massive) gauge-fixed action reads
\begin{align}\label{eq:intrinsic}
S = -\frac{1}{g^2_0}\int_x&\Big\{\frac{1}{4}(F_{\mu\nu};F_{\mu\nu})+\frac{m^2_0}{2}(a_\mu;a_\mu)\nonumber\\
&+(ih;\bar D_\mu a_\mu)+(\bar D_\mu\bar c;D_\mu c)\Big\}\,,
\end{align}
with $\bar D_\mu X\equiv \partial_\mu X-[\bar A_\mu,X]$. The mass term for the fluctuation field $a_\mu$ is a model input related  to the Gribov ambiguities of the LDW gauge. From the general discussion in Sec.~\ref{sec:notations}, it is clear that the gauge-fixed action, including the gluonic mass term, is invariant under the transformation
\bea
\label{eq:gen1}
\bar A^U_\mu & = & U\bar A_\mu U^{-1}-U\partial_\mu U^{-1}\,,\\
\label{eq:gen2}
X^U & = & UX\,U^{-1}\,,
\eea
where $X$ denotes one of the fields $a_\mu$, $c$, $\cb$, or $h$. Here, the local transformation matrices $U$ include generalized gauge transformations, periodic up to an element of the center of $G$; see the discussion around Eq.~(\ref{eq:per}).

Introducing a source $J$ for the gauge field $A$ and after Legendre transformation of the generating functional for connected Green's functions $W[J,\bar A]$, one obtains the effective action $\Gamma[A,\bar A]$. The limit of vanishing source corresponds to the minimization of $\Gamma[A,\bar A]$ with respect to $A$, that is finding $A_{\rm min}(\bar A)$ such that $\Gamma[A_{\rm min}(\bar A),\bar A]\leq\Gamma[A,\bar A]$, $\forall A$. Note that there might exist various degenerate absolute minima for a given $\bar A$, in which case the limit $J\to0$ may be nontrivial, possibly signaling the breaking of some symmetry. We shall come back to this point in Sec.~\ref{sec:SSB}.

The background $\bar A$ merely represents a choice of gauge within the class of LDW gauges and, in principle, one can choose it arbitrarily for the purpose of evaluating gauge-invariant observables. In a given approximation scheme however, a good choice of background might be relevant to unveil certain physical properties. Conversely, a bad choice of background might render certain properties hardly accessible with typical approximation schemes. It is convenient\footnote{The reason why self-consistent backgrounds are useful in the present context is discussed in Sec.~\ref{sec:SSB}.} to work with a self-consistent background $\bar A_s$, defined by $A_{\rm min}(\bar A_s)=\bar A_s$ (for a certain $A_{\rm min}(\bar A)$). Introducing $\tilde\Gamma[\bar A]\equiv\Gamma[\bar A,\bar A]$, we have
\beq
\tilde\Gamma[\bar A_s]=\Gamma[\bar A_s,\bar A_s]=\Gamma[A_{\rm min}(\bar A_s),\bar A_s]\,.
\eeq
The quantity $\Gamma[A_{\rm min}(\bar A),\bar A]$ is proportional to the free energy of the system and thus does not depend on the gauge-fixing ``parameter'' $\bar A$. We have then
\beq
\tilde\Gamma[\bar A_s]=\Gamma[A_{\rm min}(\bar A),\bar A]\leq\Gamma[\bar A,\bar A]=\tilde\Gamma[\bar A]\quad\forall\bar A\,.
\eeq
This shows that a self-consistent background is necessarily an absolute minimum of $\tilde\Gamma[\bar A]$. We can next show that, either the absolute minima of $\tilde\Gamma[\bar A]$ correspond exactly to the self-consistent backgrounds or there is no self-consistent background at all. Indeed, if there exists at least one self-consistent background $\bar A_s$  and $\bar A_m$ denotes an absolute minimum of $\tilde\Gamma[\bar A]$, we have $\Gamma[\bar A_m,\bar A_m]=\tilde\Gamma[\bar A_m]\leq\tilde\Gamma[\bar A_s]=\Gamma[A_{\rm min}(\bar A_s),\bar A_s]$. Using the fact that $\Gamma[A_{\rm min}(\bar A),\bar A]$ does not depend on $\bar A$, we end up with $\Gamma[\bar A_m,\bar A_m]\leq \Gamma[A_{\rm min}(\bar A_m),\bar A_m]\leq\Gamma[\bar A,\bar A_m]\,\,\,\forall\bar A$. This implies that $A_{\rm min}(\bar A_m)=\bar A_m$ for one of the possibly degenerate minima for a given $\bar A$ and, in turn, that $\bar A_m$ is self-consistent.\footnote{We have checked in a specific example, namely a one-loop calculation of $\langle A\rangle$ in the SU($2$) theory, that the absolute minimum of the potential corresponds to a self-consistent background.}

It follows from the above discussion that finding the self-consistent backgrounds (when they exist) boils down to finding the absolute minima of $\tilde\Gamma[\bar A]$. Once such a background is found, one can set $\smash{\langle A\rangle=\bar A_s}$ or, equivalently, $\smash{\langle a\rangle=0}$, in the calculation of any quantity, as follows from the identity $\smash{\langle A\rangle=A_{\rm min}(\bar A_s)}$ at vanishing source and the definition of a self-consistent background. We note also that the free-energy density can be obtained as $\tilde\Gamma[\bar A_s]$, from which one obtains all other thermodynamical quantities. We stress that in the case where certain symmetries are broken, a generic observable may depend on the minimum at which it is evaluated, (see Sec.~\ref{sec:SSB}), but this is not the case for the free-energy density.

It is worth mentioning possible loopholes in the previous arguments:
\begin{enumerate}

\item We have assumed that $\Gamma[A_{\rm min}(\bar A),\bar A]$ does not depend on $\bar A$. An all-order proof can be given in the case $m_0=0$; see Appendix \ref{appsec:mass}. However, things are less clear in the massive extension studied here. We prove in Appendix \ref{appsec:mass} that self-consistent backgrounds are necessarily extrema of $\tilde\Gamma[\bar A]$ (even when $m_0\neq 0$) but we were not able to show that they are the absolute minima without resorting to the independence of $\Gamma[A_{\rm min}(\bar A),\bar A]$ with respect to $\bar A$ in the first place.

\item We have assumed that we always have to minimize $\Gamma[A,\bar A]$ with respect to $A$ at a fixed $\bar A$. Standard justifications of this fact involve the positivity of the integration measure. This implies the convexity of the Euclidean generating functional for connected Green functions $W[J,\bar A]$, which in turn implies that the potential is minimal at the point corresponding to a vanishing source.\footnote{For instance, for a scalar theory, the convexity of $W[J]$ follows from the positivity of $e^{-S}$ and H\"older's inequality. The convexity implies $$W[J]\geq W[J_0]+\int_x (J-J_0)\delta W/\delta J|_{J_0},$$ from which it follows that
\begin{align}
\Gamma[\delta W/\delta J]&=-W[J]+\int_x J\delta W/\delta J\nn
&\leq -W[J_0]-\int_x(J-J_0)\delta W/\delta J|_{J_0}+\int_x J\delta W/\delta J\nn
&=\Gamma\left[\delta W/\delta J|_{J_0}\right]+\int_x J(\delta W/\delta J|_{J_0}-\delta W/\delta J).\nonumber
\end{align} 
Taking the limit $J\to 0$ in some way and irrespective of whether the symmetry breaks or not, we obtain $\Gamma\left[\delta W/\delta J|_{J\to 0}\right]\leq \Gamma\left[\delta W/\delta J|_{J_0}\right]$ for any $J_0$, which means that the effective action is minimal at any value of $\phi$ corresponding to a vanishing source.} In the present case, after integration over the ghost and the Nakanishi-Lautrup fields, one obtains a measure made of two factors, one positive (a delta function) and one equal to the Faddeev-Popov determinant which changes sign for large enough field configurations. We note that, if the (renormalized) gluon mass is large enough, those field configurations are suppressed \cite{Serreau:2012cg}. This makes it plausible that the generating functional for connected Green functions is convex but we do not have a proof for this. We mention that in approaches based on the standard (i.e., $m_0=0$) Faddeev-Popov action, the problem is even more severe because there is {\it a priori} no suppression factor.\footnote{This problem is absent from the Gribov-Zwanziger approach, in which the functional integration is restricted to the first Gribov region, where the ghost determinant at vanishing source is positive. Lattice simulations also have a positive measure and therefore avoid this ambiguity.}

\end{enumerate}

The above issues require further investigation. We shall leave this for future work and simply assume here that the correct recipe to obtain the self-consistent backgrounds is to minimize $\tilde \Gamma[\bar A]$, as usually done in the literature (see, however, Ref.~\cite{Reinosa:2015oua} for an interesting counterexample). We shall restrict to backgrounds which respect explicitly the spatial and temporal (in imaginary time) symmetries of the finite-temperature problem. These are homogeneous and in the temporal direction:
\beq\label{eq:restrict}
\bar A_\mu(x)=\bar A\,\delta_{\mu0}\,.
\eeq 
We have then
\beq
F_{\mu\nu}=\bar D_\mu a_\nu-\bar D_\nu a_\mu-[a_\mu,a_\nu]
\eeq
and, since
\beq
D_\mu c=\bar D_\mu c-[a_\mu,c]\,,
\eeq
the gauge-fixed action reads
\begin{align}\label{eq:S2}
S =  &-\frac{1}{g^2_0}\int_x\Big\{\frac{1}{2}\left(\bar D_\mu a_\nu;\bar D_\mu a_\nu\right)+\frac{m^2_0}{2}\left(a_\mu;a_\mu\right)\nn
&-\left(\bar D_\mu a_\nu;[a_\mu,a_\nu]\right)+\frac{1}{4}\left([a_\mu,a_\nu];[a_\mu,a_\nu]\right)\nn
&+\left(ih;\bar D_\mu a_\mu\right)+\left(\bar D_\mu\bar c;\bar D_\mu c\right)-\left(\bar D_\mu\bar c;[a_\mu,c]\right)\!\Big\},
\end{align}
where we have used the gauge condition \eqn{eq:LDW}. The effective action $\tilde\Gamma[\bar A]$ depends only on the variable $\bar A$ defined in Eq.~(\ref{eq:restrict}) and is proportional to $\beta\Omega$, where $\Omega$ is the spatial volume. It is then enough to consider the effective potential
\beq
\label{eq:effpot}
V(T,r)=\frac{\tilde\Gamma[\bar A]}{\beta\Omega}-V_{\rm vac}\,,
\eeq
where $r\equiv \beta g_0\bar A$ and $V_{\rm vac}$ is the value of the potential at zero temperature and fixed $\bar A$. This value is independent of the background as we show in Sec.~\ref{sec:symmetries}, where we also recall that $g_0\bar A$ is ultraviolet finite. Finally, without loss of generality, we can rotate $\bar A$ into a Cartan subalgebra of ${\cal G}$ by means of a global group transformation. In what follows, we then write $\bar A=i\bar A_j H_j$ where $\{iH_j\}$ is a basis of the Cartan subalgebra with
\beq
[H_j,H_k]=0\,.
\eeq
The labels $j$ and $k$ run from $1$ to $d_C$, the dimension of the Cartan subalgebra.

\subsection{Canonical bases}

We wish to derive the Feynman rules for the (massive) LDW gauge---that is, in the presence of a nontrivial background field---in their simplest form. To this aim, we note that any group transformation that leaves the background invariant is a symmetry of the gauge-fixed action to which Noether's theorem associates conserved charges. It is thus useful to work in a basis where such symmetries are explicit at the level of the Lagrangian density, which amounts to diagonalizing the action of the corresponding generators. The latter belong to the Cartan subalgebra and can be decomposed on the basis $\{iH_j\}$. We thus have to solve the eigenvalue problem $[H_j,X]=\lambda_{j,X}X$, simultaneously for all $j$. The solution, which we briefly review here, is well known~\cite{Zuber}. 

\begin{figure}[h]  
\vglue-20mm
\epsfig{file=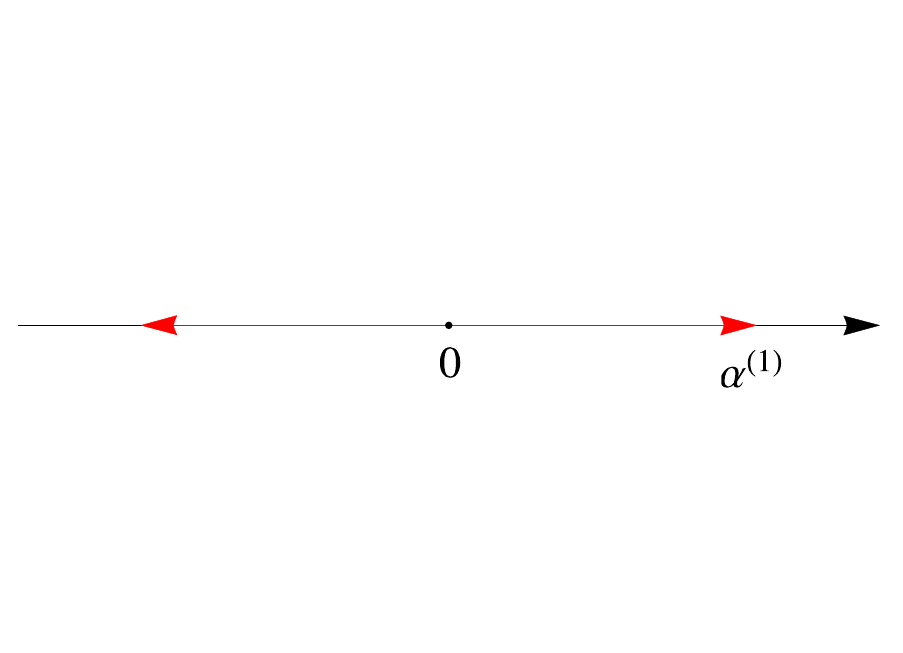,width=7cm}
\vglue-10mm
\epsfig{file=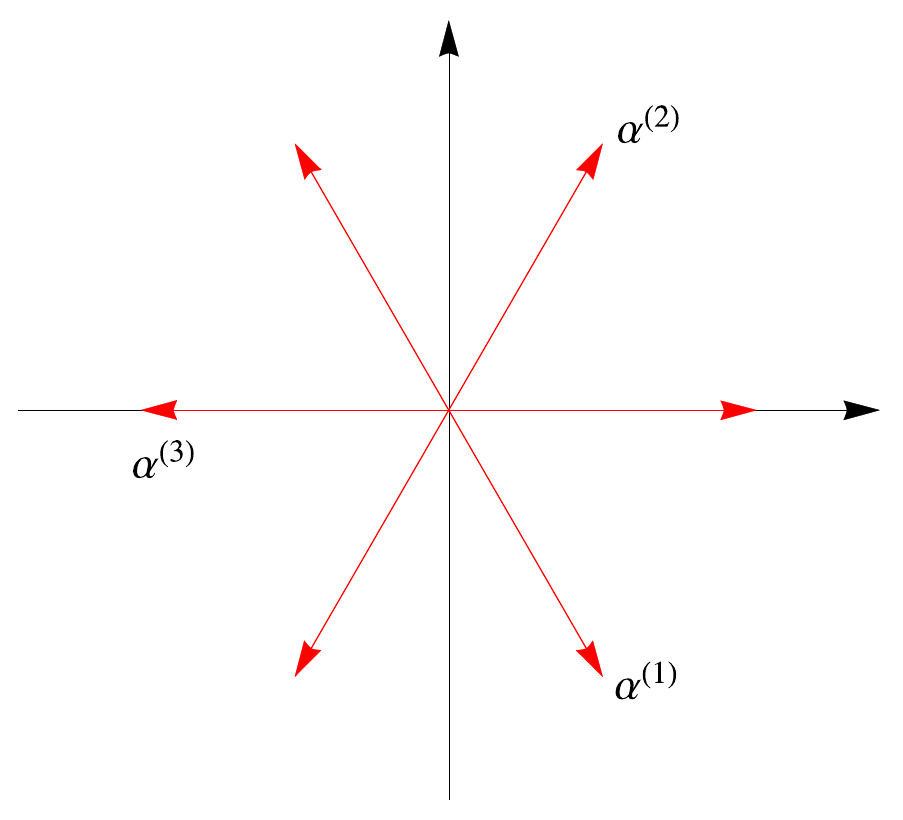,width=8cm}
 \caption{The root systems for the SU($2$) [upper graph] and SU($3$) [lower graph] algebras. The axes represent independent directions in the Cartan space which is one-dimensional for SU($2$) and two-dimensional for SU($3$). The roots always come in pairs $\pm\alpha$. For SU($2$), we have $\pm\alpha^{(1)}=\pm1$. For SU($3$), we have $\pm\alpha^{(1)}=\pm(1/2,-\sqrt{3}/2)$, $\pm\alpha^{(2)}=\pm(1/2,\sqrt{3}/2)$, and $\pm\alpha^{(3)}=\pm(-1,0)$. The normalization of the roots depends on that of the elements $H_j$.}\label{fig:roots}
\end{figure}

First, the elements of the Cartan subalgebra trivially solve the above eigenvalue equations with $\lambda_{j,X}=0$. Second, one can extend the basis $\{iH_j\}$ of the Cartan subalgebra into a basis $\{iH_j,iE_\alpha\}$ of the complexified algebra\footnote{A general element $Z$ of the complexified algebra is $Z=i(Z_jH_j+Z_\alpha E_\alpha)$, where $Z_j$ and $Z_\alpha$ are complex. For an element $X=i(X_jH_j+X_\alpha E_\alpha)$ of the original algebra ${\cal G}$, one shows that the $X_j$ are real while the $X_\alpha$ can be complex in general. In fact, for each $E_\alpha$ there exists also a $E_{-\alpha}$ and one can choose the normalizations such that $X_\alpha^*=X_{-\alpha}$. This means that when choosing to decompose an element $X\in {\cal G}$ in a canonical basis $\{iH_j,iE_\alpha\}$ rather than in a standard basis $\{it_a\}$, one switches from a description in terms of real fields $X_a$ to a description in terms of a set of real fields $X_j$ and pairs of complex-conjugated fields $(X_\alpha,X_{-\alpha})$, corresponding to pairs of opposite color charges. Of course the number of independent components is the same in both cases.} ${\cal G}\oplus i{\cal G}$ in such a way that
\beq\label{eq:can}
[H_j,E_\alpha]=\alpha_j E_\alpha\quad\forall j\,.
\eeq
The various $\alpha$, called roots, can be represented as nonzero real-valued vectors with components $\alpha_j$ in a space with the same number of dimensions as the Cartan subalgebra.  We shall refer to such a space as the Cartan space. For the sake of illustration and for later use, we represent in Fig.~\ref{fig:roots} the roots of the SU($2$) and SU($3$) algebras; see Sec.~\ref{sec:su3} for more details. 

It can be useful to give the following quantum-mechanical interpretation of the Cartan-Weyl bases: the complexified algebra ${\cal G}\oplus i{\cal G}$, equipped with a certain sesquilinear extension $\langle\,;\,\rangle$ of $(\,;\,)$ (see Appendix~\ref{appsec:canonical}) is a Hilbert space describing the color states of the system. Each element $iY\in i{\cal G}$ can also be associated to an ``observable'' ${\rm ad}_{iY}\,\cdot\equiv [iY,\,\cdot\,]$, which is a Hermitian operator with respect to the form $\langle\,;\,\rangle$. In particular, the operators ${\rm ad}_{H_j}$ form a set of commuting observables which can then be diagonalized in a common basis $\{iH_j,iE_\alpha\}$. The color states of the system can then be classified according to a set of quantum numbers associated to the observables ${\rm ad}_{H_j}$. The state $iE_\alpha$ has the quantum numbers $\alpha_j$ which we can conveniently gather into a vector $\alpha$ that can be directly represented in the Cartan  space. As the notation indicates, each state $E_\alpha$ is nondegenerate with respect to the set of observables $H_j$, in the sense that there are no two $E_\alpha$ having the same quantum numbers. Moreover, at least one of the $\alpha_j$ is nonzero.  In contrast, the states $iH_j$ are degenerate, having all their quantum numbers equal to zero and are thus to be represented by the zero vector in the Cartan subalgebra. For later convenience, we shall make this degeneracy explicit by denoting the zero vector associated to $iH_j$ as $0^{(j)}$.\footnote{The vectors $0^{(j)}$, with $j=1,\ldots,d_C$, are null vectors in the Cartan space. Here, the subscript ${(j)}$ does not denote the components in Cartan space but rather labels the $d_C$ null vectors: by construction, there are as many such null vectors as there are directions in the Cartan space.} We shall also refer to the $H_j$'s as the neutral modes and to the $E_{\pm\alpha}$'s as the charged modes. For instance, in the SU($2$) case, we have one neutral and two charged modes.

As we recall in Appendix~\ref{appsec:canonical}, the elements $E_\alpha$ obey the following commutation relations:
\beq
\label{eq:sadeft}
[E_\alpha,E_\beta]=\left\{
\begin{array}{ll}
N_{\alpha\beta} E_{\alpha+\beta} & \mbox{if $\alpha+\beta$ is a root}\\
u_j H_j & \mbox{if $\alpha+\beta=0$}\\
0 & \mbox{otherwise}
\end{array}\right.\,,
\eeq
where $N_{\alpha\beta}$ and $u_j$ are some coefficients. The $H_j$ and the $E_\alpha$ can be chosen such that $(H_j;H_k)=\delta_{jk}$, $(H_j;E_\alpha)=0$ and $(E_\alpha;E_\beta)=\delta_{\alpha+\beta,0}$, in which case $u_j=\alpha_j$. The $N_{\alpha\beta}$ can also be determined. In the following, we shall only need to know that $|N_{\alpha\beta}|^2=\frac{1}{2}(1-p)q\alpha^2$, where $q$ and $-p$ are the largest positive integers such that both $\beta+q\alpha$ and $\beta+p\alpha$ are roots \cite{Zuber}.

\subsection{Feynman rules}
\label{sec:Frules}

From here on, we rescale all fields by a factor $g_0$, as appropriate for a perturbative expansion. To obtain a compact form of the Feynman rules, it is convenient to introduce a generalized index $\kappa$ which can be either a root, $\kappa=\alpha$, or a zero, $\kappa=0^{(j)}$. Accordingly, denoting $X_{0^{(j)}}\equiv X_j$, $t_{0^{(j)}}\equiv H_j$ and $t_\alpha\equiv E_\alpha$, we can write, for any element of ${\cal G}$, 
\beq
 X=i(X_j H_j+X_{\alpha} E_\alpha)=iX_{\kappa} t_\kappa.
\eeq
The basis $\{it_\kappa\}$ is such that $(t_\kappa;t_\lambda)=\delta_{\kappa,-\lambda}$. With this compact notation, we have, for the background covariant derivative,
\beq
\label{eq:covdercan}
\bar D_\mu X  =  i(\partial_\mu-i\delta_{\mu0}g_0\bar A\cdot \kappa)X_{\kappa} t_\kappa,
\eeq
where $\bar A\cdot\kappa\equiv\bar A_j\kappa_j$. Using the Fourier convention $\partial_\mu\to-i Q_\mu$ the derivative \eqn{eq:covdercan} amounts to the multiplication of the Fourier-transformed (FT) component $X_\kappa(Q)$ by a shifted, or generalized momentum:\footnote{To avoid confusion, we reserve the set of greek letters $(\mu, \nu, \rho, \sigma)$ to denote Lorentz indices only and the other set $(\kappa, \lambda, \eta, \xi,\tau)$ to denote color states (roots or zeros) only. With this convention, $Q_\mu$ denotes the $\mu$ component of the four-vector $Q$, whereas $Q_\kappa$ refers to the shifted momentum $Q+g_0(\bar A\cdot\kappa)n$ with $n=(1,{\bf 0})$.}
\beq
(\bar D_\mu X)_\kappa\stackrel{\mbox{\tiny FT}}{\longrightarrow} -iQ^\kappa_\mu X_\kappa(Q)\,,
\eeq
where $Q^\kappa_\mu\equiv Q_\mu+\delta_{\mu0}g_0\bar A\cdot \kappa$. It is one of the many interests of the canonical bases that they allow for a simple and transparent diagonalization of the background covariant derivative. Translation invariance in space and imaginary time and the color charge symmetry mentioned above imply the conservation of such generalized momentum.
 
\begin{figure}[t!]  
\begin{center}
\epsfig{file=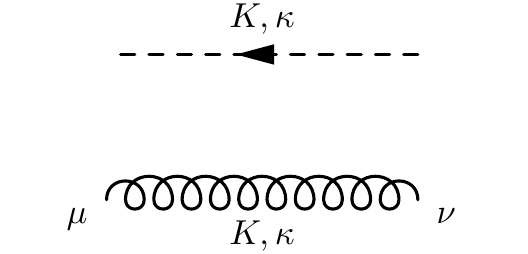,width=5cm}
 \caption{Diagrammatic representation of the propagators.}\label{fig:props}
\end{center}
\end{figure}

The typical quadratic terms appearing in the LDW action (\ref{eq:S2}) take the simple form 
\beq
 \int_x(\bar D_\mu X;\bar D_\nu Y)=\int_Q Q_\mu^\kappa Q_\nu^\kappa X^*_{\kappa}(Q)Y_\kappa(Q),
\eeq
with $X^*_\kappa(Q)=X_{-\kappa}(-Q)$ and where $\int_Q=T\sum_{n\in\mathds{Z}}\int_{\bf q}$ denotes the sum-integral over momenta $Q\equiv(\omega_n,{\bf q})$, where  $\omega_n=2n\pi T$ are the Matsubara frequencies, and $\int_{\bf q}=\int d^{d-1}q/(2\pi)^{d-1}$. One then easily obtains the ghost and gluon propagators as
\bea
\langle c^{-\kappa}(Q')\bar c^{\kappa}(Q)\rangle & = & (2\pi)^d\delta^{(d)}(Q+Q')G_0(Q_\kappa)\,,\\
\langle a_\mu^{-\kappa}(Q') a_\nu^\kappa(Q)\rangle & = & (2\pi)^d\delta^{(d)}(Q+Q')P^\perp_{\mu\nu}(Q_\kappa)G_{m_0}(Q_\kappa)\,,\nonumber\\
\eea
with
\beq
G_{m_0}(Q)\equiv\frac{1}{Q^2+m^2_0}\,.
\eeq
They are represented in Fig.~\ref{fig:props}. We note the properties $G_{m_0}(Q_\kappa)=G_{m_0}((-Q)_{-\kappa})$ and $P^\perp_{\mu\nu}(Q_\kappa)=P^\perp_{\mu\nu}((-Q)_{-\kappa})$ which follow from $(-Q)_{-\kappa}=-Q_\kappa$ and imply that the expressions of the propagators are independent of the (common) orientation of the momentum and color charge flows in their diagrammatic representation.

Let us now discuss the Feynman rules for the vertices. The cubic (derivative) interaction terms are of the form
\beq
(X;[Y,Z])=-i f_{\kappa\lambda\tau}X_{\kappa} Y_{\lambda} Z_{\tau}\,,
\eeq
where $f_{\kappa\lambda\tau}=(t_\kappa;[t_\lambda,t_\tau])$ is the tensor of structure constants of the group in the canonical basis.\footnote{This definition for $f_{\kappa\lambda\tau}$ is equivalent to $[t_\lambda,t_\tau]=f_{\lambda\tau\kappa} t_{-\kappa}$ which is to be compared to the more standard definition in the Cartesian basis: $[t_a,t_b]=if_{abc}t_c$.} Owing to Eq.~(\ref{eq:inv}), which remains true over the complexified algebra, it is completely antisymmetric. The nonvanishing elements of $f_{\kappa\lambda\tau}$ have at most one label which is a zero because $[H_j,H_k]=0$. If there is one label equal to a zero, the structure constant is equal to
\beq\label{eq:toto1}
f_{\alpha\beta0^{(j)}}=(E_{\alpha};[E_\beta,H_j])=-\beta_j(E_\alpha;E_\beta)=\alpha_j\delta_{\alpha+\beta,0}\,.
\eeq
 If all the labels are roots, the structure constant takes the form
\beq\label{eq:toto2}
f_{\alpha\beta\gamma}=(E_{\alpha};[E_\beta,E_\gamma])=N_{\beta\gamma}\delta_{\alpha+\beta+\gamma,0}\,.
\eeq
It follows that the nonvanishing elements of $f_{\kappa\lambda\tau}$ are such that $\kappa+\lambda+\tau=0$. This is nothing but the conservation of the charges associated with those color rotations that leave the background unaffected, as mentioned previously. It implies that the shifts of the momenta due to the background and, in turn, the shifted momenta themselves are conserved at the derivative vertices. For instance, in the SU($2$) case, there is only one null vector $0^{(3)}$ (neutral mode), corresponding to the color direction aligned with the background field, and a pair of roots (charged modes), which are combinations of the two orthogonal directions in color space. The nonvanishing structure constants are obtained by permutations of $f_{0+-}=1$. In the SU($3$) case, there are two neutral modes, $0^{(3)}$ and $0^{(8)}$, and three pairs of charged modes, $\pm\alpha^{(1)}$, $\pm\alpha^{(2)}$, and $\pm\alpha^{(3)}$. The nonvanishing structure constants are obtained from
\begin{align}
\label{eq:1}
f_{0^{(3)}\alpha^{(3)}(-\alpha^{(3)})}&=-1,\\
f_{0^{(3)}\alpha^{(1)}(-\alpha^{(1)})}&=f_{0^{(3)}\alpha^{(2)}(-\alpha^{(2)})}=1/2,\\
\label{eq:2}
f_{0^{(8)}\alpha^{(1)}(-\alpha^{(1)})}&=-f_{0^{(8)}\alpha^{(2)}(-\alpha^{(2)})}=-\sqrt{3}/2,
\end{align}
and
\beq
\label{eq:3}
f_{\alpha^{(1)}\alpha^{(2)}\alpha^{(3)}}=f_{(-\alpha^{(1)})(-\alpha^{(2)})(-\alpha^{(3)})}=-1/\sqrt{2}.
\eeq
The structure constants \eqn{eq:1}--\eqn{eq:2} are SU($2$)-like in that they involve a neutral mode and a pair of charged modes, whereas \Eqn{eq:3} couples charged modes together, where we recall that $\alpha^{(1)}+\alpha^{(2)}+\alpha^{(3)}=0$, see Fig.~\ref{fig:roots}.

\begin{figure}[t!]  
\begin{center}
\epsfig{file=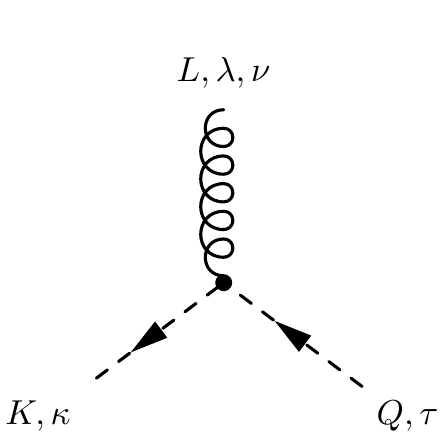,width=3.7cm}\,\,\,\,\epsfig{file=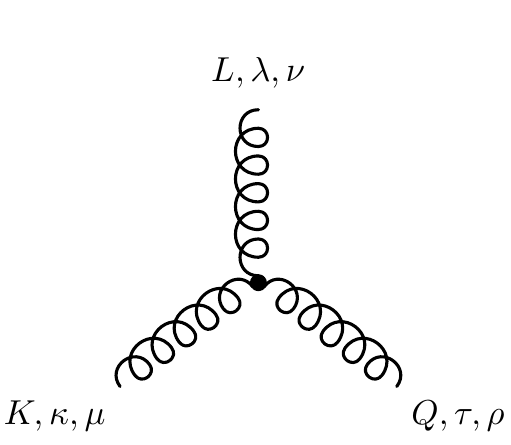,width=4.4cm}
 \caption{Diagrammatic representation of the derivative vertices.}\label{fig:der}
\end{center}
\end{figure}

In terms of the structure constants $f_{\kappa\lambda\tau}$, the ghost-gluon vertex takes the form\\
\beq
\label{eq:sdyd}
g_0f_{\kappa\lambda\tau} K_\nu^\kappa,
\eeq
where $K$ is the momentum of the outgoing antighost and $\kappa$ is the corresponding outgoing color charge. Similarly, $\lambda$ and $\tau$ are the charges of the outgoing gluon and ghost. After symmetrization, the three-gluon vertex reads
\beq
\label{eq:sdysfd}
\frac{g_0}{6}f_{\kappa\lambda\tau}\Big\{\delta_{\mu\rho}(K^\kappa_\nu-Q^\tau_\nu)+\delta_{\nu\mu}(L^\lambda_\rho-K^\kappa_\rho)+\delta_{\rho\nu}(Q^\tau_\mu-L^\lambda_\mu)\Big\},
\eeq
where $Q_\kappa$, $K_\lambda$ and $L_\tau$ denote the shifted outgoing momenta associated to the indices $\mu$, $\nu$ and $\rho$. The derivative vertices are represented in Fig.~\ref{fig:der}. 

Finally, the four-gluon interaction has the structure\\
\begin{align}
 ([X,Y];[Z,U])&=X_\kappa Y_\lambda Z_\tau U_\xi ([t_{\kappa},t_{\lambda}];[t_{\tau},t_{\xi}])\nonumber\\
& =X_\kappa Y_\lambda Z_\tau U_\xi (t_{\eta};[t_{\kappa},t_{\lambda}]) (t_{-\eta};[t_{\tau},t_{\xi}])\nonumber\\
& =f_{\kappa\lambda\eta}f_{\tau\xi(-\eta)}X_\kappa Y_\lambda Z_\tau U_\xi \,,
\end{align}
where we have used $[X,Y]=(t_{-\kappa};[X,Y])t_\kappa$.
That this vertex also conserves color charge is clear because $\kappa+\lambda+\eta=0$ and $\tau+\xi-\eta=0$ imply $\kappa+\lambda+\tau+\xi=0$. After symmetrization, the four-gluon vertex reads
\bea
\label{eq:kjdg}
& & \frac{g^2_0}{24}\sum_\eta \Big[f_{\kappa\lambda\eta}f_{\tau\xi(-\eta)}(\delta_{\mu\rho}\delta_{\nu\sigma}-\delta_{\mu\sigma}\delta_{\nu\rho})\nonumber\\
& & \hspace{1.0cm}+f_{\kappa\tau\eta}f_{\lambda\xi(-\eta)}(\delta_{\mu\nu}\delta_{\rho\sigma}-\delta_{\mu\sigma}\delta_{\nu\rho})\nonumber\\
& & \hspace{1.0cm}+f_{\kappa\xi\eta}f_{\tau\lambda(-\eta)}(\delta_{\mu\rho}\delta_{\nu\sigma}-\delta_{\mu\nu}\delta_{\sigma\rho})\Big]\,,
\eea
where $\kappa$, $\lambda$, $\tau$ and $\xi$ represent the outgoing charges. This vertex is represented in Fig.~\ref{fig:four}. We note that, using $\smash{f_{\kappa\lambda\tau}=-f^*_{(-\kappa)(-\lambda)(-\tau)}}$ (see Appendix \ref{appsec:canonical}), the momenta/charges in the vertices of Figs.~\ref{fig:der} and \ref{fig:four} can also all be considered incoming, provided one replaces Eqs.~\eqn{eq:sdyd}, \eqn{eq:sdysfd}, and \eqn{eq:kjdg} by the complex-conjugate expressions.\\

\begin{figure}[t!]  
\begin{center}
\epsfig{file=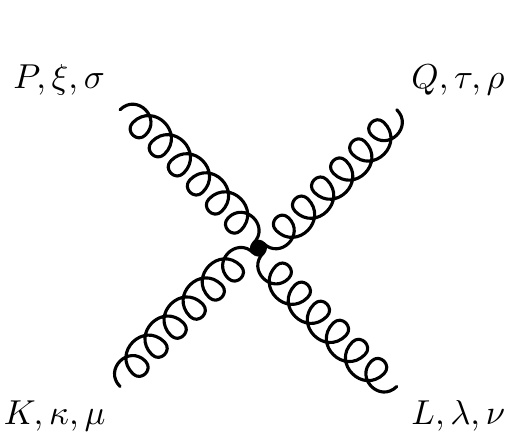,width=4.4cm}
 \caption{Diagrammatic representation of the four-gluon vertex.}\label{fig:four}
\end{center}
\end{figure}

The previous Feynman rules make clear
that it is simple to express diagrams in the LDW gauge from the
corresponding ones in the Landau gauge (that is, with $\bar A=0$). In
particular, one writes the integral in the Landau gauge without the
color factor and attributes a different charge $\kappa$ (a root or a
zero) to each line in the diagram, which amounts to a shift by
$g_0(\bar A\cdot\kappa)$ of the corresponding frequency. It is important to note that, because the shifted momenta are conserved at the vertices, one can use any manipulation based on this conservation rule, just as in the Landau gauge. One should however keep in mind that some other manipulations are not allowed. For instance, if $\kappa$ is a root, $G(Q_\kappa)\neq G((-Q)_\kappa)$ but one has instead $G(Q_\kappa)=G((-Q)_{-\kappa})$. In particular integrals such that $\int_Q \omega_\kappa G(Q_\kappa)$, with $Q_\kappa\equiv (\omega_\kappa,q)$, do not vanish in dimensional regularization unless $\kappa$ is a zero.

\section{Symmetries}\label{sec:symmetries}

The Feynman rules discussed above can be given a simple form because the gauge-fixed action \eqn{eq:S2} is invariant under the global group transformations that leave the background field unchanged. If we allow for nontrivial transformations of the latter, the action has a generalized invariance property $S_{\bar A'}[A',h',c',\bar c']=S_{\bar A}[A,h,c,\bar c]$ under a larger class of transformations. In particular, this includes the generalized gauge transformations, Eqs.~\eqn{eq:gen1} and \eqn{eq:gen2}, as well as charge conjugation, to be discussed below. This implies the corresponding invariance of the background field functional  $\tilde\Gamma[\bar A']=\tilde\Gamma[\bar A]$. 

In this section, we discuss the consequences of these symmetries for the effective potential \eqn{eq:effpot}. We consider those generalized gauge transformations that leave the background field  homogenous, along the (imaginary) time direction, and in the Cartan subalgebra. It is easy to show that these are of the form $WU(\tau)$, where $W$ is a global transformation which leaves the Cartan subalgebra globally invariant\footnote{We shall see below that these form the group of so-called Weyl transformations \cite{Zuber} of the Cartan subalgebra.} and
\beq\label{eq:39}
U(\tau)=\exp\left\{\frac{i\tau}{\beta} x_j H_j\right\}.
\eeq
As already mentioned and as we recall below, for this to be a symmetry of the theory, the transformation \eqn{eq:39} must be such that $U(\tau+\beta)=U(\tau)Z$, where $Z$ is an element of the center of $G$. This puts constraints on the real variables $x_j$ to be discussed below.  In what follows, we shall refer to the transformations \eqn{eq:39} as the winding transformations. It will be convenient to distinguish between periodic winding transformations, with $Z=\mathds{1}$, and nonperiodic ones, for which $Z\neq \mathds{1}$. Altogether, the Weyl and the periodic winding transformations generate all the standard gauge transformations that leave the background in the class considered here, i.e., homogeneous, in the time direction, and in the Cartan subalgebra.

\subsection{Winding transformations}\label{sec:center}
Under a winding transformation (\ref{eq:39}), the components of a field $X=i(X_jH_j+X_\alpha E_\alpha)$ in a given canonical basis transform as
\beq\label{eq:transfo}
X_j\to X_j\,, \quad X_\alpha\to e^{i\frac{\tau}{\beta}x\cdot\alpha} X_\alpha\,.
\eeq
For the transformed field to remain $\beta$-periodic in the imaginary time direction, we thus need to require that,
\beq\label{eq:cond3}
\frac{x\cdot\alpha}{2\pi}\in\mathds{Z}\quad\forall\alpha\,.
\eeq
It is sufficient that this condition be satisfied for a subset of roots $\{\alpha^{(j)}\}$, with $j=1,\ldots,d_C$, that generate the other roots through linear combinations with integer coefficients. In this case, the conditions (\ref{eq:cond3}) define a lattice, dual to the one generated by the roots $\alpha^{(j)}$. For later use, we introduce a basis $\{\bar\alpha^{(j)}\}$ of this dual lattice, defined as
\beq\label{eq:dual}
\alpha^{(j)}\cdot\bar\alpha^{(k)}=2\pi\delta_{jk}\,.
\eeq
The general solution to Eq.~(\ref{eq:cond3}) is then 
\beq\label{eq:y}
x=n_k\bar\alpha^{(k)},\quad{\rm with}\quad n_k\in\mathds{Z}.
\eeq 

Under a transformation of the form (\ref{eq:39}) with $x$ given by Eq.~(\ref{eq:y}), the background field (rescaled by $g_0$) transforms as
\beq
\label{eq:Abarchange}
\bar A\to\bar A+\frac{n_k\bar\alpha^{(k)}}{\beta g_0}\,.
\eeq
This implies that the potential \eqn{eq:effpot} is invariant under translations along any (multiple of) $\bar\alpha^{(k)}$:\footnote{That the potential is invariant under these transformations is also apparent from the Feynman rules. Indeed, in units of $T$, the shifts of the frequencies propagating through the lines of any diagram contributing to the potential change as  
\beq
r\cdot\alpha\to( r+\bar\alpha^{(k)})\cdot\alpha=r\cdot\alpha+2\pi m_k\,,
\eeq
with $m_k\in\mathds{Z}$ defined by $\alpha=m_j\alpha^{(j)}$. The additional frequency shifts are multiples of $2\pi T$ and, because they are conserved at the vertices of the diagram, they can be absorbed via a change of variables in the Matsubara sums.}
\beq
V(T,r)=V(T,r+\bar\alpha^{(k)})\,.
\eeq
As a consequence, one can restrict the analysis of the potential to the first Brillouin zone of the dual lattice defined above.

\begin{figure}[h]  
\vglue-15mm
\epsfig{file=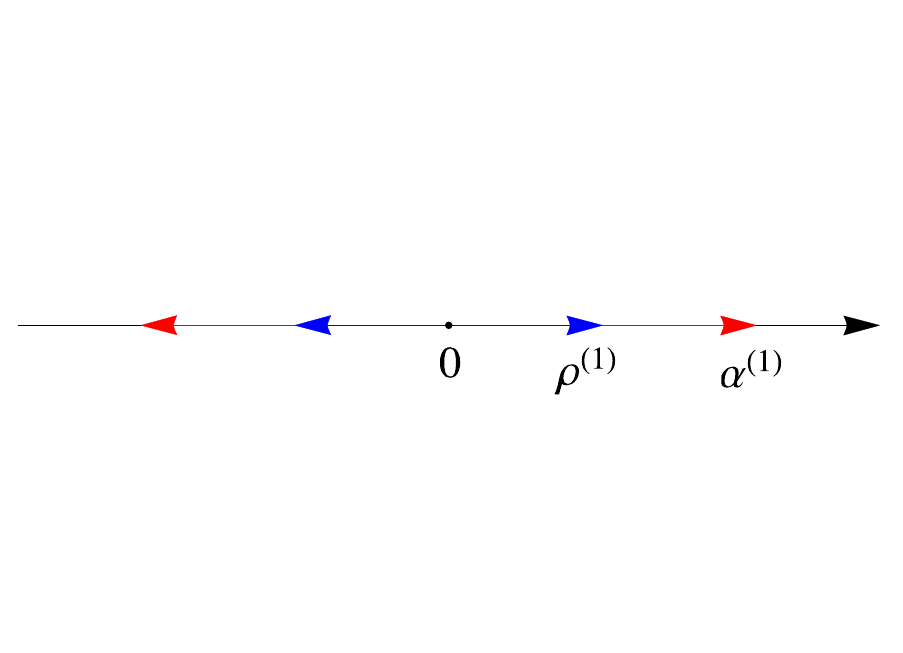,width=7cm}
\vglue-15mm
\epsfig{file=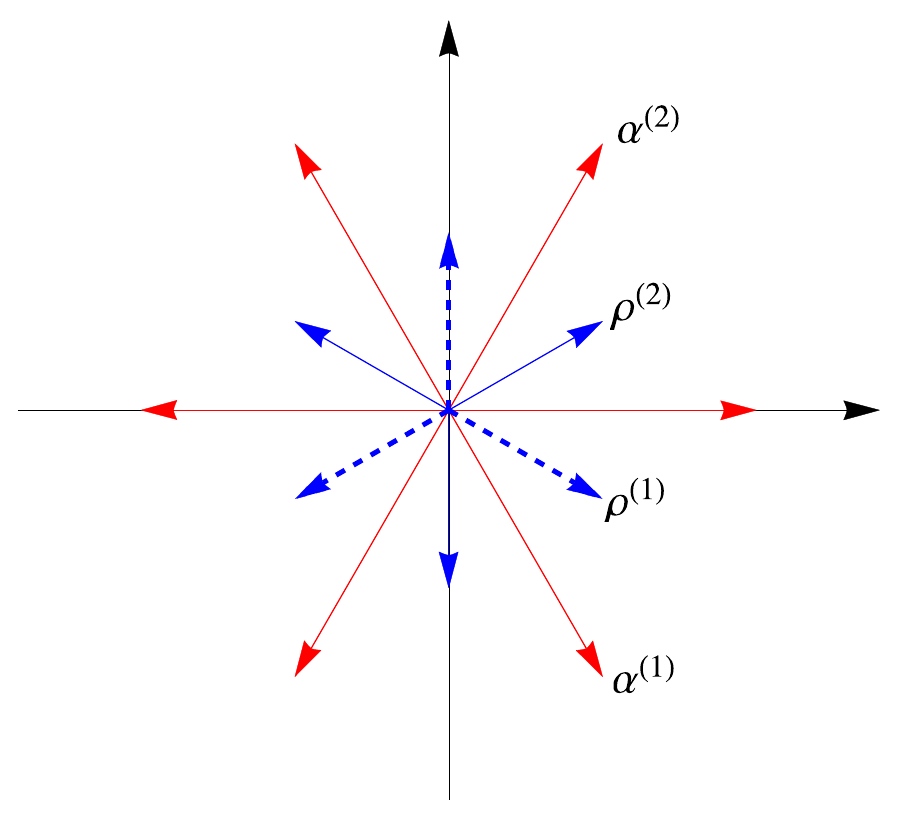,width=8cm}
 \caption{Weights $\rho$ (blue vectors) of the fundamental representations ${\bf 2}$ and ${\bf 3}$ in the SU($2$) and SU($3$) cases. In the SU($3$) case we show the weights of the contragredient fundamental representation ${\bf \bar 3}$ (dashed blue) which appear as the opposites of the weights of ${\bf 3}$ and which are convenient for the discussion in the main text. We also represent the roots $\alpha$ of each algebra (red vectors online). We have singled out certain roots (respectively weights) that generate all the other ones through linear combinations with integer coefficients.
 }\label{fig:roots_weights}
\end{figure}

Finally, let us recall how the center of the group enters the discussion here. From \Eqn{eq:39}, we have  $U(\tau+\beta)=U(\tau)Z$, where $Z=\exp\{ix_jH_j\}$. For any element $X=i(X_jH_j+X_\alpha E_\alpha)$ of the algebra, one has
\beq
\label{eq:ZZtop}
Z X Z^{-1}  =  X\,,
\eeq
where we have used Eq.~(\ref{eq:cond3}). Because the group is assumed to be compact and connex, \Eqn{eq:ZZtop} exponentiates to $ZU=UZ$ for any $U$ in $G$. This shows that, indeed, $Z$ belongs to the center of the group.

For later purposes, it is useful to identify which of the transformations \eqn{eq:Abarchange} correspond to standard (periodic) gauge transformations, with $Z=\mathds{1}$. The elements $H_j$ of the Cartan subalgebra can be viewed as a set of commuting Hermitian matrices acting on $\mathds{C}^n$ for some $n$, and can thus be diagonalized simultaneously. We have
\beq
 H_j|\rho,a\rangle=\rho_j|\rho,a\rangle,
\eeq 
where the label ``$a$'' denotes a possible degeneracy and the $\rho$'s are the weights of the fundamental representation $it_\kappa\mapsto it_\kappa$. Just like the roots, these can be represented as vectors with components $\rho_j$ in the Cartan space; see Fig.~\ref{fig:roots_weights} for the SU($2$) and SU($3$) cases. In terms of the weights $\rho$, the condition $Z=\exp\{ix_j H_j\}=\mathds{1}$ gives
\beq\label{eq:cond4}
\frac{x\cdot\rho}{2\pi}\in\mathds{Z}\quad\forall\rho.
\eeq
Repeating the previous arguments used to solve Eq.~(\ref{eq:cond3}), we find that the general solution to Eq.~(\ref{eq:cond4}) is given by $x=n_k\bar\rho^{(k)}$, with $n_k\in\mathds{Z}$ and where the $\bar\rho^{(k)}$ form a basis dual to a given basis composed of nonzero weights (or their opposites) $\{\rho^{(j)}\}$ that generate the other weights through linear combinations with integer coefficients.

\begin{figure}[t]  
\begin{center}
\epsfig{file=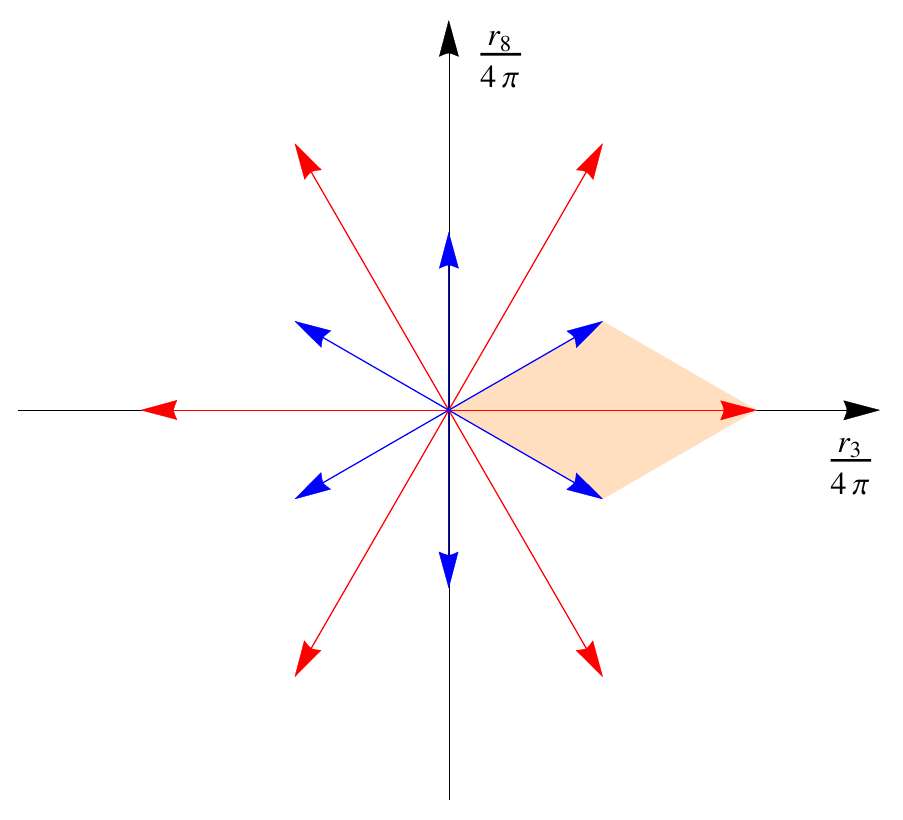,width=9cm}
 \caption{SU($3$) winding transformations. Periodic winding transformations correspond to translations along the roots (red) in the plane $(r_3/4\pi,r_8/4\pi)$, whereas nonperiodic winding transformations correspond to translation along the weights (blue). The Brillouin zone associated to winding transformations is represented in orange. In addition to the winding transformations, we have the Weyl transformations which are generated by reflections with respect to axes orthogonal to the roots, and charge conjugation which corresponds to an inversion about the origin.}\label{fig:translations}
\end{center}
\end{figure}

Let us illustrate the above discussion in the SU($2$) and SU($3$) cases. For SU($2$), the Cartan subalgebra is one dimensional. We have $\alpha^{(1)}=1$ and $\rho^{(1)}=1/2$. It follows that $\bar\alpha^{(1)}=4\pi\rho^{(1)}$ and $\bar\rho^{(1)}=4\pi\alpha^{(1)}$. Then, periodic winding transformations correspond to translations $r/4\pi\to r/4\pi\pm\alpha^{(1)}$, whereas nonperiodic winding transformations correspond to translations $r/4\pi\to r/4\pi\pm\rho^{(1)}$ (even multiples of $\rho^{(1)}$ correspond again to periodic transformations). In the SU($3$) case, the Cartan subalgebra is bidimensional and we can choose $\alpha^{(1)}$, $\alpha^{(2)}$, $\rho^{(1)}$ and $\rho^{(2)}$ as in Fig.~\ref{fig:roots_weights}. One then easily checks that $\bar\alpha^{(k)}=4\pi\rho^{(k)}$ and $\bar\rho^{(k)}=4\pi\alpha^{(k)}$, with $k=1,2$. As in the SU($2$) case, periodic winding transformations correspond to translations of the vector $r/4\pi$ along the roots $\alpha^{(j)}$ whereas nonperiodic winding transformations correspond to translations of the vector $r/4\pi$ along $\rho^{(j)}$ (again, certain combinations of the $\rho^{(j)}$ can correspond to periodic transformations); see Fig.~\ref{fig:translations}, where we also show the Brillouin zone associated to winding transformations.

\subsection{Weyl transformations}
There are also global color transformations that leave the Cartan subalgebra globally invariant. It is easy to convince oneself that these cannot be of infinitesimal form and we thus need to consider finite transformations $W=e^\theta$, where $\theta$ is an element of the algebra. These act on any other element $X$ of the algebra as
\beq
e^\theta X e^{-\theta}= e^{{\rm ad}_\theta}(X)=\sum_{n=0}^\infty \frac{1}{n!}{\rm ad}_\theta^n(X)\,,
\eeq
where ${\rm ad}_\theta(X)=[\theta,X]$. For our purposes, it is sufficient to restrict to transformations of the form $\theta=i(\theta_\alpha E_\alpha+\theta_{-\alpha}E_{-\alpha})\equiv w_\alpha$ for a given $\alpha$ (no sum over $\alpha$). A simple recursion shows that (our normalizations are such that $\theta_{-\alpha}=\theta^*_\alpha$; see Appendix~\ref{appsec:canonical})
\beq
{\rm ad}^{2p}_{w_\alpha}(H_j)=i^{2p}(2|\theta_\alpha|^2\alpha^2)^p\frac{\alpha\cdot H}{\alpha^2}\alpha_j
\eeq
for $p>0$ and
\beq
{\rm ad}^{2p+1}_{w_\alpha}(H_j)=-i^{2p+1}(2|\theta_\alpha|^2\alpha^2)^p(\theta_\alpha E_\alpha-\theta_{-\alpha} E_{-\alpha})\alpha_j
\eeq
for $p\geq 0$. It follows that an element of the Cartan subalgebra transforms as
\begin{align}
H_j & \to  H_j+\Big(\cos\big(\sqrt{2\alpha^2}|\theta_\alpha|\big)-1\Big) \frac{\alpha\cdot H}{\alpha^2}\alpha_j\nonumber\\
& - \frac{i\alpha_j}{\sqrt{2\alpha^2}|\theta_\alpha|}\sin\big(\sqrt{2\alpha^2}|\theta_\alpha|\big) (\theta_\alpha E_\alpha-\theta_{-\alpha}E_{-\alpha})
\end{align}
and choosing $|\theta_\alpha|=\pi/\sqrt{2\alpha^2}$ thus ensures that the transformation leaves the Cartan subalgebra stable. We then have
\beq
H_j \to H_j-2\frac{\alpha\cdot H}{\alpha^2}\alpha_j\,.
\eeq
Considering this as a passive transformation, the coordinates $r_j$ of a general element of the Cartan subalgebra $H=ir_jH_j$ transform as
\beq\label{eq:reflect}
r_j \to r_j-2\frac{\alpha\cdot r}{\alpha^2}\alpha_j\,.
\eeq
This corresponds to a reflection about a hyperplane orthogonal to the root $\alpha$. The potential is thus invariant under these reflections, which are nothing but the so-called Weyl reflections.\footnote{These generate the Weyl group of the Cartan subalgebra. It is well known that the latter leaves the root system invariant \cite{Zuber}.} In the SU($2$) case, there is only one nontrivial Weyl transformation corresponding to $r\to -r$. In the SU($3$) case, the Weyl transformations are generated by the reflections with respect to the axes supporting the weights of the fundamental representations ${\bf 3}$ and $\bar{\bf 3}$ which are orthogonal to the roots.

\subsection{Charge conjugation}
In the cases to be considered below the structure constants $f_{\kappa\lambda\tau}$ are real. It then follows from the Feynman rules and from the property $f_{(-\kappa)(-\lambda)(-\tau)}=-f_{\kappa\lambda\tau}^*=-f_{\kappa\lambda\tau}$ that the classical action is invariant under the simultaneous transformation of the background field and the fluctuating fields according to
\beq\label{eq_charge_conjug}
\bar A_j\to-\bar A_j\quad {\rm and}\quad X_\kappa\to -X_{-\kappa}\,.
\eeq
This symmetry corresponds to charge conjugation and implies that the background field potential \eqn{eq:effpot} is invariant under $r_j\to-r_j$:
\beq
 V(T,r)=V(T,-r).
\eeq 
In the SU($2$) case, charge conjugation coincides with the Weyl transformation, i.e., corresponds to a mere global color rotation. In contrast, in the SU($3$) case, charge conjugation is not a Weyl transformation and thus carries independent information. 

We note here that the definition of charge conjugation is not unique since it can be defined only up to other symmetries of the model, in particular the nonperiodic winding (center) transformations discussed above. This ambiguity is resolved in the presence of dynamical matter fields which break the center symmetry explicitly such as, e.g., quarks in the fundamental representation. In that case, the physical charge conjugation is the transformation (\ref{eq_charge_conjug}).

It is important to mention here that, just like center symmetry, charge-conjugation symmetry could be spontaneously broken too, in principle. However, we shall verify that this never happens in our dynamical calculations in the sense that there always exists an absolute minimum of the background field potential which is  charge-conjugation symmetric. This is important for the relation between the spontaneous breaking of the center symmetry and deconfinement since charge-conjugation invariance guarantees that the average of the Polyakov loop is real and can thus be interpreted in terms of a free energy for static sources.

\subsection{Spontaneous symmetry breaking and Weyl chambers}\label{sec:SSB}

As mentioned previously, the phase transition of pure gauge theories is associated to the spontaneous breaking of the center symmetry. In the present context, the latter can be discussed directly in terms of the background field potential $V(T,r)$, thanks to the concept of Weyl chambers. This applies to any other physical symmetry as well, such as, e.g., charge conjugation.  Although some of the notions discussed here are known (see, e.g., Ref.~\cite{Dumitru:2012fw}), this is not widespread material and we find it useful to summarize our own view on the rationale behind the use of Weyl chambers in the present context.

A given symmetry is manifest if and only if the physical state of the system (at zero sources) is invariant under the corresponding transformation. Using this criterion in the present background field approach requires some care because the potential $V(T,r)$ is invariant under standard gauge transformations that leave the background homogenous, in the time direction, and in the Cartan subalgebra. A given physical state of the system is thus not represented by a single minimum of $V(T,r)$ but rather by the gauge orbit generated from it by these transformations. Because these form a countable set (the reflections and the translations discussed above, corresponding to the Weyl and the periodic winding transformations), each gauge orbit is an infinite lattice of {\em physically equivalent} points. Thus, a symmetry is manifest in a given physical state if and only if the lattice associated to this state is globally invariant under the corresponding transformation.

Equivalently, one can restrict the analysis to an elementary cell of this lattice, called a Weyl chamber and defined as the minimal region that allows one to cover the whole Cartan space using standard gauge transformations.\footnote{The usual definition of Weyl chambers in the mathematical literature differs somewhat from the one used here. There, they correspond to the minimal regions that allow one to cover the whole Cartan space using Weyl transformations only \cite{Zuber}.} The various points of a Weyl chamber are thus associated to {\em physically distinct} states and the problem reduces to analyzing the symmetry group of the Weyl chamber. 
The relation between these geometrical symmetries and the physical symmetries of the problem is easily established as follows. The transformations of interest (e.g., center or charge-conjugation transformations) typically displace the Weyl chamber in the Cartan space, but one can bring it back to its original location by means of either Weyl reflections or translations associated to periodic winding transformations---which correspond to (unbroken) standard gauge transformations. Only the points of the Weyl chamber which are unaffected by this mapping (fixed points) correspond to symmetric states. Conversely, points corresponding to broken symmetry states are the images of one another under this mapping. 

Let us illustrate the above considerations in the cases of the SU($2$) and SU($3$) groups. Other groups will be discussed in Sec.~\ref{sec:su4}. The Cartan space of SU($2$) is one dimensional and the Weyl chambers are intervals of length $2\pi$. This is because periodic winding transformations correspond to translations $r\to r+4\pi$ and the only Weyl transformation is a reflection about the origin: $r\to-r$. The symmetry group of these Weyl chambers is $Z_2$, which corresponds to reversing the segment on itself, e.g., $[0,2\pi]\to[2\pi,0]$. This actually corresponds to the center symmetry of SU($2$). To see this, consider the chamber $[0,2\pi]$, which, as discussed above (see \Fig{fig:roots_weights}), is displaced by $2\pi$ under a nonperiodic winding transformation. The displaced chamber can be brought back to its original location by means of a Weyl reflection about the origin and a translation by $4\pi$. This results in the mapping $r\to2\pi-r$, which corresponds to the $Z_2$ symmetry of the Weyl chamber described above. Clearly, the only center-symmetric state corresponds to the point $r=\pi$ in that chamber. All other points are center-breaking and come in pairs $(r,2\pi-r)$. As already mentioned, charge conjugation corresponds to $r\to-r$ and is thus nothing but a Weyl reflection. It follows that all points in a given Weyl chamber (hence, on the whole Cartan line) are charge-conjugation invariant in this case.

\begin{figure}[t]  
\begin{center}
\epsfig{file=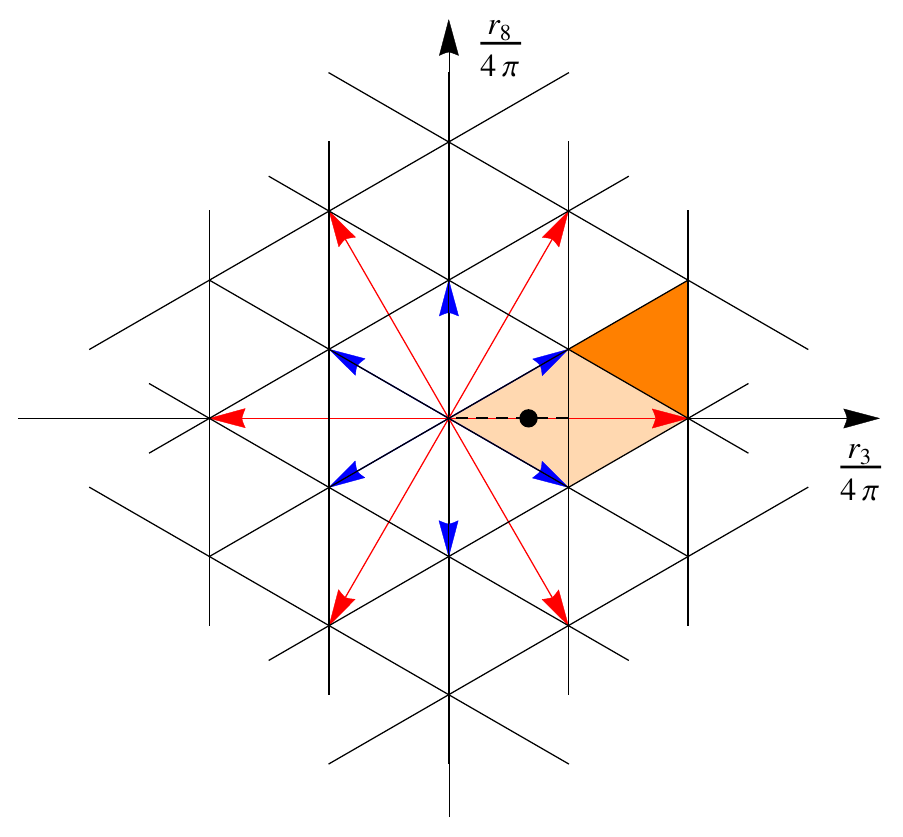,width=8.5cm}
 \caption{Weyl chambers of the SU($3$) Cartan plane. The Brillouin zone is made of two (clearest colored) Weyl chambers. The darker colored chamber is the image of the original one (leftmost clearest colored) via a nonperiodic winding transformation. To bring this chamber back into the original one, one uses two reflections but in the process the Weyl chamber rotates by an angle $-2\pi/3$ leaving room for only one center-symmetric point in the Weyl chamber: the center of the triangle.}\label{fig:Weyl_chambers}
\end{center}
\end{figure}

The Weyl chambers of the two-dimensional Cartan space of SU($3$) are the equilateral triangles shown in \Fig{fig:Weyl_chambers}.\footnote{By combining a translation by a vector $4\pi\alpha$ (periodic winding transformation) and a reflection with respect to the axis passing through the origin and orthogonal to $\alpha$ (Weyl reflection), one obtains a reflection with respect to a new axis, parallel to the first one and translated by the vector $2\pi\alpha$. The Weyl chambers appear as the regions delimitated by all these reflection axes.} Their symmetry group is the dihedral group $D_3$, whose six elements are rotations by $n\pi/3$, with $n=0,1,2$, and reflections about the medians. The former correspond to the physical center ($Z_3$) transformations,\footnote{To see this, we proceed as in the SU($2$) case. In the plane $(r_3/4\pi,r_8/4\pi)$ of \Fig{fig:Weyl_chambers}, we translate a given Weyl chamber along the vectors $\bar\alpha^{(1)}$, $\bar\alpha^{(2)}$, or any linear combination with integer coefficients (this corresponds to winding transformations) and we bring it back to its original location using Weyl reflections and/or periodic winding transformations. The so-obtained chamber has been rotated with respect to the original one by a multiple of $\pi/3$ about its center.} from which we conclude that the only center-symmetric point in a given chamber is the center of the triangle. All other points are center-breaking and come in triplets. As for the other symmetries of the Weyl chamber, namely, the reflections about the medians, they correspond to the physical charge conjugation and its combination with $Z_3$ center transformations.\footnote{As mentioned previously, the definition of charge-conjugation symmetry is not unique. In the pure gauge SU($3$) theory, any of the reflections of the $D_3$ symmetry group of the Weyl chambers can be equivalently defined as charge conjugation due to the $Z_3$ symmetry of the theory. This degeneracy is lifted in the presence of matter in a representation which breaks the $Z_3$ symmetry explicitly (e.g., quarks in the fundamental representation) and there remains a unique charge-conjugation symmetry.} Consider, for instance, the colored triangle touching the origin in \Fig{fig:Weyl_chambers}. The charge-conjugation transformation $(r_3,r_8)\to(-r_3,-r_8)$ combined with the Weyl reflection $(r_3,r_8)\to(-r_3,r_8)$ maps this chamber onto itself up to a reflection about the $r_3$ axis. Incidentally, this implies that all the points of the median $r_8=0$ of this triangle correspond to charge-conjugation-invariant states (this includes the center-symmetric point). In particular, we note that in cases of broken center symmetry but unbroken charge-conjugation symmetry, the triplet of center-breaking points must lie on the medians of the triangle, such that one of them is always charge-conjugation symmetric whereas the other two are charge conjugates of one another. We shall check below that this is always the situation which is dynamically realized in practice, i.e., the charge-conjugation symmetry is always manifest.\\

The above considerations show that the minimum of the background field potential $V(T,r)$ is an order parameter both for center symmetry (see also Refs.~\cite{Braun:2007bx,Marhauser:2008fz}) and for charge-conjugation invariance.

\subsection{Miscellaneous consequences of the background field dependence}

We end this section with general remarks concerning some consequences of the background field gauge symmetry. First, the invariance of the classical action under the gauge transformations \eqn{eq:two} implies the same symmetry for the effective action (this is because the symmetry transformation is at most linear in all the fields \cite{WeinbergBook}), i.e., $\Gamma[\bar A,X]=\Gamma[\bar A',X']$, with $X=(a_\mu,h,c,\bar c)$. In particular, restricting to homogeneous background fields, in the time direction, and in the Cartan subalgebra, the effective action is invariant under the winding transformation of parameter $\bar\alpha^{(k)}$:
\beq
 r'_j=r_j+\bar\alpha^{(k)}_j\,, \quad X_\kappa'=e^{i\frac{\tau}{\beta}\bar\alpha^{(k)}\cdot\kappa}X_\kappa\,,
\eeq
with $r=\beta g_0\bar A$. It follows that the vertex and correlation functions in the background $\bar A'=\bar A+\bar\alpha^{(k)}/(\beta g_0)$ only differ from those in the background $\bar A$ by appropriate phase factors and are thus essentially the same functions. 

We can also conclude from the previous observation that the zero-temperature potential at fixed background, $\hat V(\smash{T=0},\bar A)$, does not depend on $\bar A$. Indeed, from the previous symmetry considerations, we have, for a given~$k$, $\hat V(T,\bar A)=\hat V(T,\bar A+n_k\bar\alpha^{(k)}T/g_0)$, with $n_k\in\mathds{Z}$. If we choose $T=g_0\delta \bar A/n_k$ and we take the limit $n_k\to\infty$, we obtain $\hat V(\smash{T=0},\bar A)=\hat V(\smash{T=0},\bar A+\delta \bar A\bar\alpha^{(k)})\,\,\forall\,\delta \bar A$, and hence, the announced result. This defines the vacuum contribution $V_{\rm vac}=\hat V(\smash{T=0},\bar A)$ in \Eqn{eq:effpot}. A similar conclusion holds for the correlation functions at zero temperature: the latter only depend on the background field through trivial phase factors multiplying the correlation functions in the Landau gauge (i.e., at vanishing background). 

An important consequence is that the counterterms of the Landau gauge are enough to renormalize the potential and the correlation functions in the LDW gauge. Moreover, the background field gauge symmetry \eqn{eq:gen1}--\eqn{eq:gen2} implies that the product $g_0\bar A$ is finite \cite{WeinbergBook}. In the following, we consider a renormalization scheme such that $g_0\bar A=g\bar A_{\rm R}$, where $g$ and $\bar A_{\rm R}$ are the renormalized coupling and background field, respectively \cite{Reinosa:2014zta}. We omit the index R for simplicity and write $r=\beta g\bar A$.

\section{The background field potential at NLO}\label{sec:pot}

We now come to the explicit calculation of $V(T,r)$ in a straightforward expansion in the coupling $g$ with $g\bar A\sim {\cal O}(1)$; see Ref.~\cite{Reinosa:2014zta}. Here, we determine the LO (one-loop) and NLO (two-loop) contributions, extending our previous calculation for SU($2$) \cite{Reinosa:2014ooa,Reinosa:2014zta} to any compact Lie group with a simple Lie algebra; see also Refs.~\cite{Dumitru:2012fw,Dumitru:2013xna} for related work in the massless theory. To this order, we can replace $g_0$ by $g$ in the Feynman rules given above. We can also replace the bare mass $m_0$ by the renormalized mass $m$ provided we take into account a one-loop diagram involving the renormalization factor $\delta Z_{m^2}$, as in Ref.~\cite{Reinosa:2014zta}.

\subsection{One loop}

The first nontrivial contribution to the effective potential occurs at one-loop order. It has been computed in Ref.~\cite{Reinosa:2014ooa} for SU($2$) and SU($3$), and generalized to the class of gauge groups considered here in Ref.~\cite{Reinosa:2015oua}. Here, we simply recall the relevant formulas. 

Introducing the function (we recall that $d_C$ is the dimension of the Cartan subalgebra and we define $\varepsilon_q=\sqrt{q^2+m^2}$)
\begin{align}\label{eq:Weiss0}
 {\cal F}_m(T,r)&\equiv  T\sum_\kappa\int_{\bf q}\ln\left[1+e^{-2\beta\varepsilon_{q}}-2e^{-\beta\varepsilon_{q}}\cos (r\cdot\kappa)\right]\nn
 &=\frac{T}{\pi^2}\int_0^\infty \!\!dq\,q^2\Big\{d_C \ln\left(1-e^{-\beta\varepsilon_q}\right)\nn
 &\quad+\frac{1}{2}\sum_\alpha \ln\left[1+e^{-2\beta\varepsilon_{q}}-2e^{-\beta\varepsilon_{q}}\cos (r\cdot\alpha)\right]\!\Big\},
\end{align}
which is such that
\bea\label{eq:Weiss}
{\cal F}_0(T,r) = -\frac{\pi^2T^4}{45}d_C & + & \frac{T^4}{12}\sum_\alpha\left[\frac{(\{r\cdot\alpha\}-\pi)^4}{2\pi^2}\right.\nonumber\\
& - & \left.(\{r\cdot\alpha\}-\pi)^2+\frac{7\pi^2}{30}\right],\nonumber\\
\eea
with $\{r\cdot\alpha\}$ the remainder in the (Euclidean) division of $r\cdot\alpha$ by $2\pi$, the one-loop background field potential can be written as
\beq
\label{eq:onelooppot}
 V^{(1)}(T,r)=\frac{3}{2}{\cal F}_m(T,r)-\frac{1}{2}{\cal F}_0(T,r),
\eeq
where the first term on the right-hand side is the contribution from the three massive gluon modes per color state and the second one is due to the incomplete cancellation between the contributions of the remaining massless gluonic and ghost-antighost degrees of freedom.\\

\subsection{Two loop}

From the considerations of Sec.~\ref{sec:Frules}, it is clear that the calculation of the two-loop correction to the background field potential is almost identical to that in the SU($2$) case \cite{Reinosa:2014zta}. The essential difference is a change in some constants which depend
on the gauge group. More precisely, the tensor
$\varepsilon_{\kappa\lambda\tau}$ is replaced by the
structure constant $f_{\kappa\lambda\tau}$ and the labels $\kappa$, $\lambda$
and $\tau$ now take values among the zeros and roots of the algebra
${\cal G}$.
As already emphasized, an important point is that the nonvanishing elements of $f_{\kappa\lambda\tau}$ are such that $\kappa+\lambda+\tau=0$. The effective momentum flowing through the lines of the diagram is then conserved at the vertices and we can use the same manipulations of the Feynman integrals as in the SU($2$) case.  The net result is that we can take the formula (D3) of Ref.~\cite{Reinosa:2014zta} and just replace ${\cal C}_{\kappa\lambda\tau}=(\varepsilon_{\kappa\lambda\tau})^2\to|f_{\kappa\lambda\tau}|^2$ and the constant $C_N$ by a new one $C_G$ to be given below. We obtain, using the notations of Ref.~\cite{Reinosa:2014zta},
\begin{align}\label{eq:pot2loop}
V^{(2)}(T,r) & = m^2 C_G \sum_\kappa J_m^\kappa(1n)\nonumber\\
& +  \frac{3g^2}{8}\sum_{\kappa,\lambda,\tau}{\cal C}_{\kappa\lambda\tau}\left[\frac{5}{2}U^\kappa V^\lambda- \frac{7}{m^2}\tilde U^\kappa\tilde V^\lambda\right]\nonumber\\
& +  \frac{g^2m^2}{16}\sum_{\kappa,\lambda,\tau}{\cal C}_{\kappa\lambda\tau}\left[33S_{mmm}^{\kappa\lambda\tau}(2n)+S_{m00}^{\kappa\lambda\tau}(2n)\right].
\end{align}
Here, 
\begin{align}
 U^\kappa&=J_m^\kappa(1n)+\frac{1}{3}J^\kappa_0(1n)\,,\\
 V^\kappa&=J_m^\kappa(1n)-\frac{1}{5}J^\kappa_0(1n)\,,\\
 \tilde U^\kappa&=\tilde J_{m}^\kappa-\tilde J_{0}^\kappa\,,\\
 \tilde V^\kappa&=\tilde J_{m}^\kappa-\frac{5}{21}\tilde J_{0}^\kappa\,,
\end{align}
with the tadpole-type integrals
\bea
J^\kappa_m(1n) & = & \int_{\bf q}\, {\rm Re}\,\frac{n_{\varepsilon_{q}-i\hat r\cdot\kappa}}{\varepsilon_{q}}\,,\\
\tilde J^\kappa_m & = & \int_{\bf q} \,{\rm Im}\,n_{\varepsilon_{q}-i\hat r\cdot\kappa}\,,
\eea
where  $n_z=(e^z-1)^{-1}$ is the Bose-Einstein distribution function and where we defined $\hat r= r T=g\bar A$. Equation~\eqn{eq:pot2loop} also involves the sunset-type integrals
\begin{widetext}
\begin{align}
S_{m_1m_2m_3}^{\kappa\lambda\tau}(2n) & = \frac{1}{32\pi^4}\int_0^\infty \!\!dq\,q\int_0^\infty \!\!dk\,k\, {\rm Re}\,\frac{n_{\varepsilon_{m_1,q}-i\hat r\cdot\kappa}\,n_{\varepsilon_{m_2,k}-i\hat r\cdot\lambda}}{\varepsilon_{m_1,q}\,\varepsilon_{m_2,k}}{\rm Re}\,\ln\frac{(\varepsilon_{m_1,q}+\varepsilon_{m_2,k}+i0^+)^2-(\varepsilon_{m_3,k+q})^2}{(\varepsilon_{m_1,q}+\varepsilon_{m_2,k}+i0^+)^2-(\varepsilon_{m_3,k-q})^2}\nonumber\\
& + \frac{1}{32\pi^4}\int_0^\infty \!\!dq\,q\int_0^\infty \!\!dk\,k \,{\rm Re}\,\frac{n_{\varepsilon_{m_1,q}-i\hat r\cdot\kappa}\,n_{\varepsilon_{m_2,k}+i\hat r\cdot\lambda}}{\varepsilon_{m_1,q}\,\varepsilon_{m_2,k}}\,{\rm Re}\,\ln\frac{(\varepsilon_{m_1,q}-\varepsilon_{m_2,k}+i0^+)^2-(\varepsilon_{m_3,k+q})^2}{(\varepsilon_{m_1,q}-\varepsilon_{m_2,k}+i0^+)^2-(\varepsilon_{m_3,k-q})^2}\nonumber\\
\label{appeq:S2n}
&+{\rm perm.}\,,
\end{align}
\end{widetext}
where we used the explicit notation $\varepsilon_{m,q}\equiv\sqrt{q^2+m^2}$ and where ``perm.'' denotes circular permutations of $(m_1,\kappa)$, $(m_2,\lambda)$, and $(m_3,\tau)$. Finally, the constant $C_G$ is given by
\beq\label{eq:pot2loop_fin}
C_G= \frac{g^2C_{\rm ad}}{128\pi^2}\left(z_f+35\ln\frac{\bar\mu^2}{m^2}+\frac{313}{3}-\frac{99\pi}{2\sqrt{3}}\right),
\eeq
where $\bar\mu^2=4\pi e^{-\gamma}\mu^2$, where $\gamma$ is the Euler constant and $\mu$ is the renormalization scale of dimensional regularization, and $z_f$ is a finite constant which encodes the chosen renormalization conditions but which is independent of the gauge group $G$. Here, we use the same renormalization scheme as in Ref.~\cite{Reinosa:2014zta} and $z_f$ is given by Eq.~(D2) of that reference. Finally, $C_{\rm ad}$ is the Casimir of the adjoint representation given by (see Appendix \ref{appsec:casimir})
\beq
C_{\rm ad}=\sum_{\lambda\tau}{\cal C}_{\kappa\lambda\tau}=\frac{\sum_\alpha \alpha^2}{d_C}=\sum_\alpha \alpha^2_j \quad \forall j\,.
\eeq

The last two lines of Eq.~(\ref{eq:pot2loop}) contain two types of terms: those for which one of the indices of ${\cal C}_{\kappa\lambda\tau}$ is a zero and those for which the three indices are roots. The former are of the same type as the ones computed in the SU($2$) case. Using the fact that ${\cal C}_{\alpha(-\alpha)0^{(j)}}=\alpha_j^2$ and that the integral multiplying this factor does not depend on $j$, these can be obtained from the SU($2$) calculation with the replacement $r\to r\cdot\alpha$ and by multiplying the result by $\sum_j\alpha_j^2=\alpha^2$. For the other types of terms, we remark that ${\cal C}_{\alpha\beta\gamma}={\cal C}_{(-\alpha)(-\beta)(-\gamma)}$ so that, using $U^{-\alpha}=U^\alpha$, $V^{-\alpha}=V^\alpha$, $\tilde U^{-\alpha}=-\tilde U^\alpha$, $\tilde V^{-\alpha}=-\tilde V^\alpha$, and $S_{m_1m_2m_3}^{(-\alpha)(-\beta)(-\gamma)}(2n)=S_{m_1m_2m_3}^{\alpha\beta\gamma}(2n)$, we can combine these two types of contributions, with an effective factor $2{\cal C}_{\alpha\beta\gamma}=2|N_{\alpha\beta}|^2$, whose expression is given in \Eqn{eq:sadeft}. 

The final, complete expression of the background potential at two-loop order is given explicitly in Appendix~\ref{appsec:final} in terms of one- and two-dimensional radial momentum integrals that can be easily computed numerically.

\section{The Polyakov loop at NLO}\label{sec:pol}

Here, we evaluate the Polyakov loop for an arbitrary finite representation $it_\kappa\mapsto it^R_\kappa$ of a compact and connex Lie group with a simple Lie algebra in a perturbative expansion in the presence of the nontrivial background field. Specifically, we compute the following function of the background \cite{Reinosa:2014zta}:
\beq
\label{eq:decadix}
\ell_R(r)=\frac{1}{d_R}{\rm tr}\,\left\langle P \exp\left(ir^jt^R_{0^{(j)}}+ig_0\int_0^\beta \!\!d\tau \,a_0^\kappa t^R_\kappa\right)\right\rangle\,,
\eeq
where it is understood that $\langle a_0\rangle=0$ on the right-hand side. The physical Polyakov loop is obtained by evaluating $\ell_R(r)$ at the minimum of the effective potential. 

It is easy to show that, under the symmetry transformations of the Weyl chambers which correspond to center transformations of the gauge group $G$, the function \eqn{eq:decadix} transforms just as the Polyakov loop \eqn{eq_pol_trace}. For instance, we have, for a representation of $N$-ality $p$,
\beq
\ell_R(r)=z^p\ell_R(r'),
\eeq
for any pair $(r,r')$ of points in a Weyl chamber related by a center transformation $z\mathds{1}$. At a center-symmetric point, $r_{\rm sym}=r'_{\rm sym}$ and $\ell_R(r_{\rm sym})=0$  for representations of nontrivial $N$-ality.

\subsection{Tree level}

The LO expression of the function \eqn{eq:decadix} is ${\cal O}(g^0)$ and reads 
\beq
\ell_R^{(0)}(r)=\frac{1}{d_R}{\rm tr}\,\exp\Big\{ir_j t^R_{0^{(j)}}\Big\}\,.
\eeq
In order to evaluate the color trace on the right-hand side, it is convenient to introduce the weights of the representation $R$, just as we did above for the fundamental representation $it_\kappa\mapsto it_\kappa$. For a unitary representation,\footnote{Note that for a compact group, as considered here, any finite representation is equivalent to a unitary one.} the $t^R_{0^{(j)}}$ are Hermitian. Because they commute, they can be diagonalized simultaneously:
\beq\label{eq:wei}
t^R_{0^{(j)}}|\mu,a\rangle = \mu_j|\mu,a\rangle\,,
\eeq
where the $\mu$'s are nonzero, real-valued vectors with components $\mu_j$, called the weights of the representation, which can be represented on the root diagram. The set of eigenvalues represented by a given weight can be degenerate, as denoted by the label ``$a$'' in $|\mu,a\rangle$. In terms of the weights, we have
\beq\label{eq:l1l}
\ell_R^{(0)}(r)=\frac{1}{d_R}\sum_\mu {\rm mul}(\mu)\,e^{ir\cdot\mu}\,,
\eeq
where ${\rm mul}(\mu)$ denotes the degeneracy of the weight $\mu$ and $r\cdot\mu=r_j\mu_j$. 

\subsection{One loop}

In Ref.~\cite{Reinosa:2014zta}, we obtained the general expression of the NLO contribution to the function \eqn{eq:decadix} for any representation of the group SU($N$) in terms of the gluon propagator $G^{\lambda\kappa}_{\mu\nu}(\tau,{\bf 0})\equiv\langle a^\lambda_\mu(\tau,{\bf 0})a^\kappa_\nu(0,{\bf 0})\rangle$ and of the color trace ${\rm tr}\left\{t_\kappa^R\bar A^q t_\lambda^R\bar A^{n-q}\right\}$, with $n,q\in\mathds{N}$. The generalization to the present case is straightforward and we obtain, for the ${\cal O}(g^2)$ correction,
\beq\label{appeq:PL}
\ell_R^{(1)}(r)=-\frac{g^2\beta}{2d_R}\int_0^\beta \!\!d\tau \,G^{\lambda\kappa}_{00}(\tau,{\bf 0})\,\tr\left\{t^\kappa_Re^{(\beta-\tau)g\bar A} t^\lambda_Re^{\tau g\bar A}\right\}\!,
\eeq
with $\bar A=i\bar A_j t^R_{0^{(j)}}$. Using the weights $\mu$ of the representation to evaluate the trace, we get (see Appendix \ref{appsec:PL})
\begin{widetext}
\beq\label{eq:PL_res}
\ell_R(r)=\frac{1}{d_R}\sum_\mu {\rm mul}(\mu)e^{ir\cdot\mu}\left\{1+ \frac{g^2m}{T}\left[\frac{C_R}{8\pi}+\sum_{\alpha,a,b}\frac{\left|\langle\mu+\alpha,b|t^R_\alpha|\mu,a\rangle\right|^2}{{\rm mul}(\mu)}\sin^2\left(\frac{r\cdot\alpha}{2}\right)\frac{a(T,r\cdot\alpha)}{2\pi^2}\right]\right\}+{\cal O}(g^4)\,,
\eeq
where the state $|\mu+\alpha\rangle\equiv 0$ if $\mu+\alpha$ is not a weight, $C_R$ is the Casimir of the representation, and
\beq
\label{eq:ATX}
a(T,x)\equiv\int_0^\infty \frac{q^2dq}{m^3}\left(\frac{1}{\cosh (q/T)-\cos x}-\frac{q^2}{\varepsilon^2_q}\frac{1}{\cosh(\varepsilon_q/T)-\cos x}\right).
\eeq
\end{widetext}
Equation~\eqn{eq:PL_res} must be evaluated at the minimum of the effective potential computed at the same approximation order.

\section{The SU($3$) theory}\label{sec:su3}
 We now specialize to the physically relevant SU(3) case. For completeness, we briefly review the previous LO results \cite{Reinosa:2014ooa} and then present our findings at NLO.

\subsection{Roots and weights}
We consider the basis $\{i\lambda_a/2\}$ where the $\lambda_a$ denote the Gell-Mann matrices. These satisfy $(\lambda_a/2;\lambda_b/2)=\delta_{ab}$ and $\sum_a (\lambda_a/2)^2=(4/3)\mathds{1}$, so that $C_R=4/3$ in the fundamental representation. Using $f_{123}=1$, $f_{345}=-f_{367}=1/2$, $f_{845}=f_{867}=\sqrt{3}/2$, one checks that 
\begin{align}
t_{\pm\alpha^{(1)}}&=\frac{\lambda_6\mp i\lambda_7}{2\sqrt{2}},\nn 
t_{\pm\alpha^{(2)}}&=\frac{\lambda_4\pm i\lambda_5}{2\sqrt{2}},\\ 
t_{\pm\alpha^{(3)}}&=\frac{\lambda_1\mp i\lambda_2}{2\sqrt{2}},\nonumber
\end{align}
form an orthonormal Cartan-Weyl basis. The root diagram is represented in Fig.~\ref{fig:roots_weights}. For the calculation of the potential (see Appendix \ref{appsec:final}), we only need to know that $(\pm\alpha^{(j)})^2=1$ and $2|N_{\alpha\beta}|^2=1$ if $\alpha+\beta$ is a root and vanishes otherwise. From the expressions of $\lambda_3$ and $\lambda_8$, one finds that the weights of the fundamental representation ${\bf 3}$ are $\mu^{(1)}=(1,1/\sqrt{3})^{\rm t}/2$, $\mu^{(2)}=(-1,1/\sqrt{3})^{\rm t}/2$, $\mu^{(3)}=(0,-1/\sqrt{3})^{\rm t}$, corresponding respectively to $\rho^{(2)}$, $-\rho^{(1)}$ and $\rho^{(1)}-\rho^{(2)}$ in the notations of Fig.~\ref{fig:roots_weights}. They are nondegenerate [${\rm mul}(\mu^{(j)})=1$] and we have
\begin{align}
& t_{\alpha^{(1)}}|\mu^{(2)}\rangle=\frac{1}{\sqrt{2}}|\mu^{(3)}\rangle\,,  \quad t_{-\alpha^{(3)}}|\mu^{(2)}\rangle=\frac{1}{\sqrt{2}}|\mu^{(1)}\rangle\,,\nonumber\\
& t_{\alpha^{(2)}}|\mu^{(3)}\rangle=\frac{1}{\sqrt{2}}|\mu^{(1)}\rangle\,,  \quad t_{-\alpha^{(1)}}|\mu^{(3)}\rangle=\frac{1}{\sqrt{2}}|\mu^{(2)}\rangle\,,\nonumber\\
& t_{\alpha^{(3)}}|\mu^{(1)}\rangle=\frac{1}{\sqrt{2}}|\mu^{(2)}\rangle\,,  \quad t_{-\alpha^{(2)}}|\mu^{(1)}\rangle=\frac{1}{\sqrt{2}}|\mu^{(3)}\rangle\,,
\end{align}
and all other contributions vanish. Then, the Polyakov loop function in the fundamental representation $R={\bf 3}$ reads
\begin{widetext}
\beq\label{eq:pl3}
\ell_{\bf 3}(r)=\frac{1}{3}\sum_\mu e^{ir\cdot\mu}\left\{1+g^2\frac{m}{T}\left(\frac{1}{6\pi}+\frac{1}{4\pi^2}\sum_{\mu'\neq\mu}\sin^2\left[\frac{r\cdot(\mu'-\mu)}{2}\right]a\Big[T,r\cdot(\mu'-\mu)\Big]\right)\right\}+{\cal O}(g^4).
\eeq
\end{widetext}
We could equally well consider the conjugate fundamental representation $\bar{\bf 3}$, for which $\ell_{\bar{\bf 3}}(r)=\ell_{\bf 3} ^*(r)$. This trivially follows from the fact that the weights of $\bar{\bf 3}$ are the opposite of those of ${\bf 3}$. It is easy to check that, at LO,  $\ell_{\bf 3}^{(0)}(r)=0$ only at the center-symmetric point so that confinement in the fundamental representation implies that center symmetry is manifest and thus that all sources with nonzero $N$-ality are confined. We have checked that this remains true at NLO.

\subsection{LO results}

\begin{figure}[h]  
\begin{center}
\epsfig{file=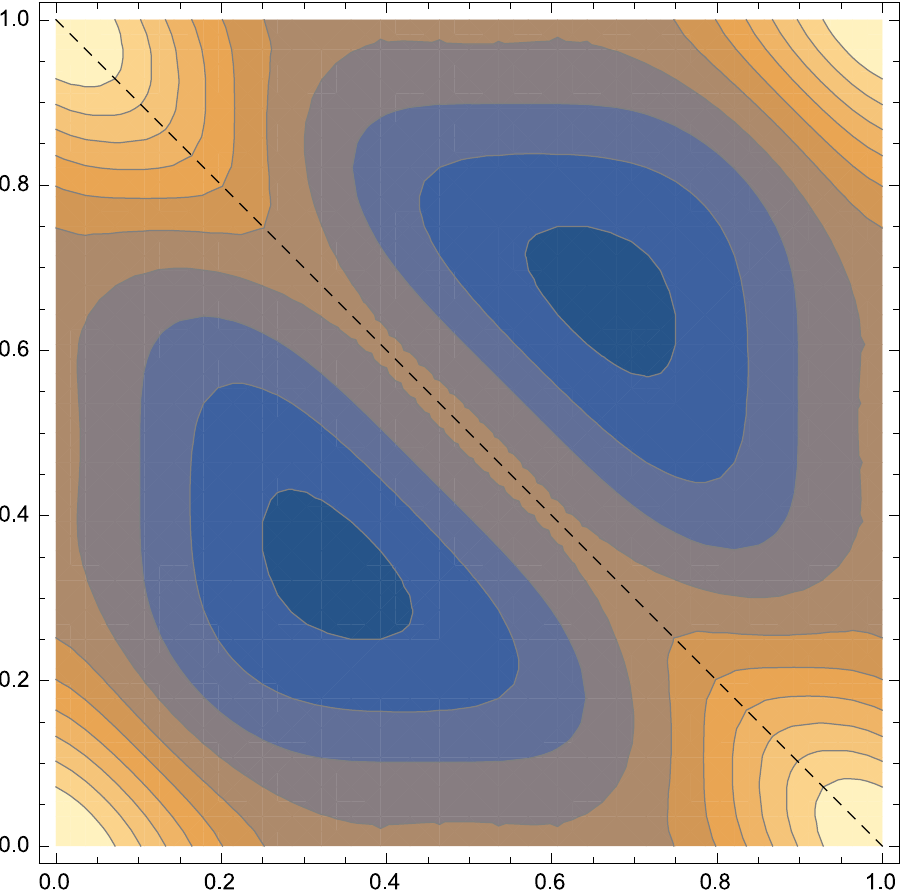,width=2.5cm}\quad\epsfig{file=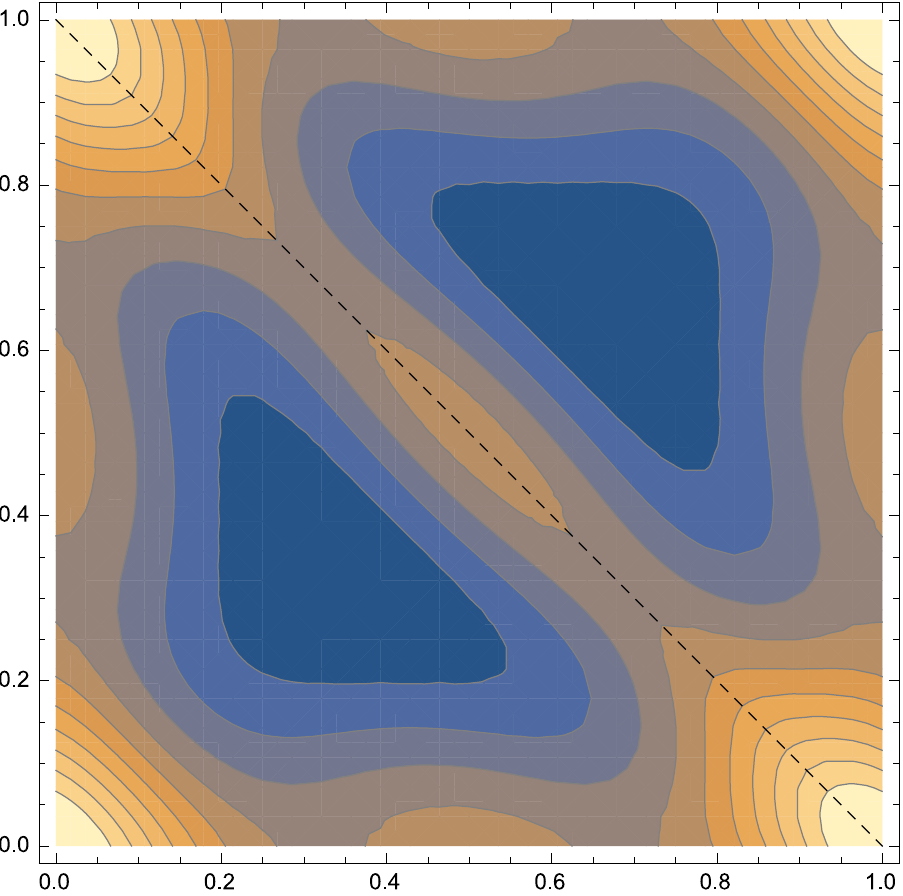,width=2.5cm}\quad\epsfig{file=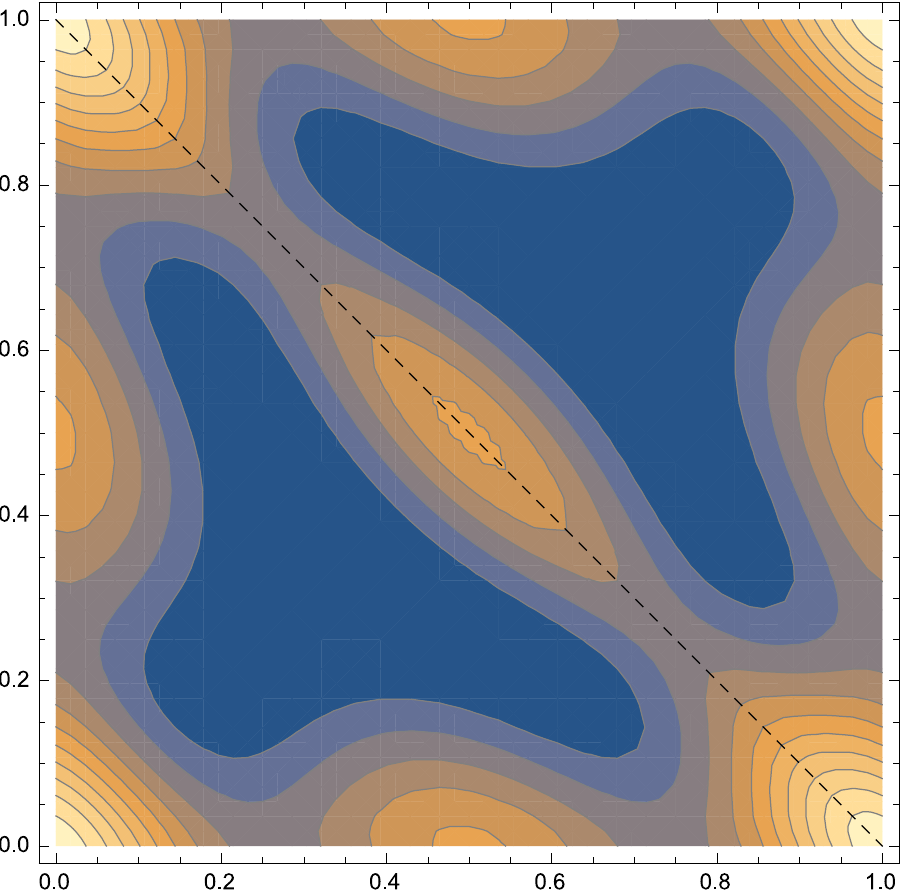,width=2.5cm}\\
\vspace{0.6cm}
\epsfig{file=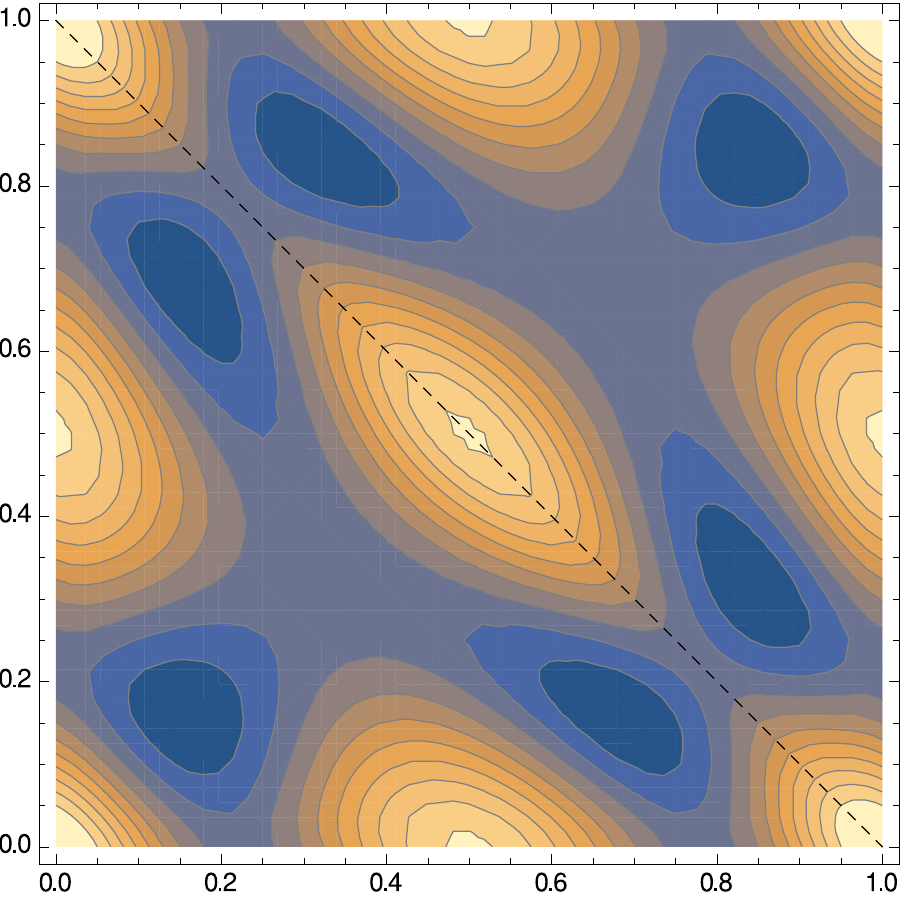,width=2.5cm}\quad\epsfig{file=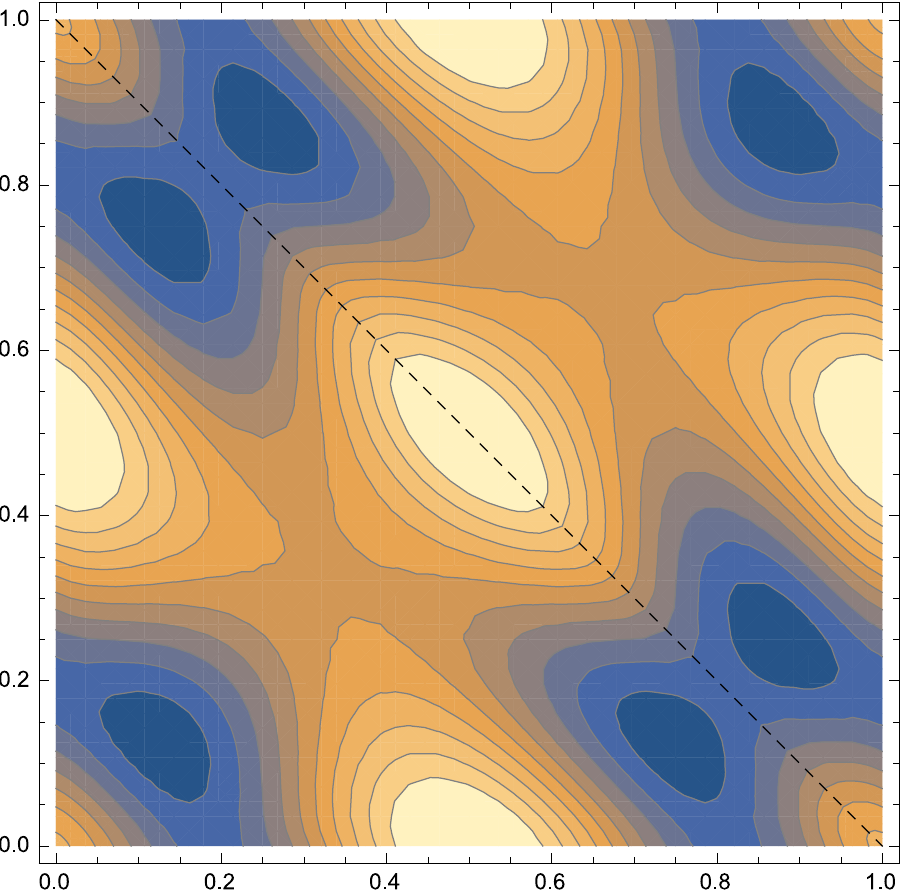,width=2.5cm}\quad\epsfig{file=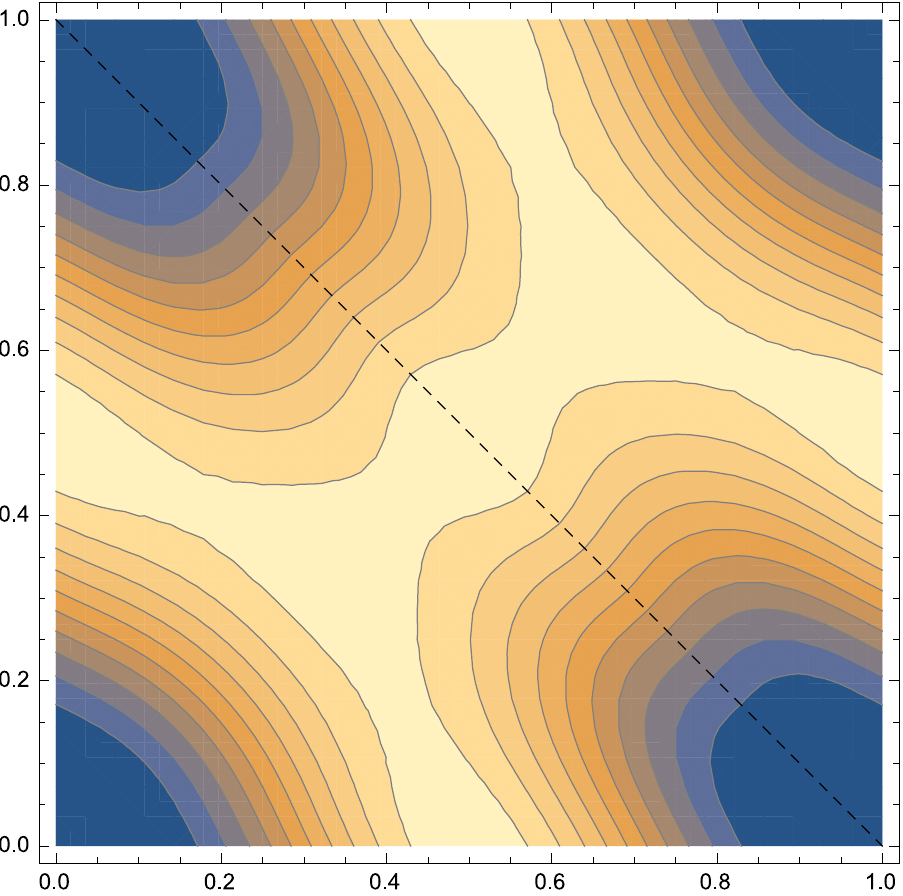,width=2.5cm}
 \caption{Contour plot of the SU($3$) LO potential in the first Brillouin zone $(\bar r_1,\bar r_2)\in[0,1]^2$ for increasing values of the temperature. The dashed line separates two equivalent Weyl chambers. In the one closer to the origin, the center-symmetric point is located at $(\bar r_1,\bar r_2)=(1/3,1/3)$. The upper (respectively lower) plots correspond to temperatures below (respectively above) the transition temperature. Darker colors correspond to regions where the potential is the deepest.}\label{fig:su3}
\end{center}
\end{figure}

Let us briefly revisit the one-loop results \cite{Reinosa:2014ooa} for completeness. It is convenient  to decompose $r$ in the dual basis $\{\bar\alpha^{(1)},\bar\alpha^{(2)}\}$ depicted in Fig.~\ref{fig:translations}:
\beq
 r=\bar r_k\bar\alpha^{(k)},
\eeq
so that the first Brillouin zone is $(\bar r_1,\bar r_2)\in [0,1]^2$; see Fig.~\ref{fig:su3}. The LO potential is given by \Eqn{eq:onelooppot}. Using the roots $\pm\alpha^{(1)}$, $\pm\alpha^{(2)}$, and $\pm(\alpha^{(1)}+\alpha^{(2)})$ and the relation $r\cdot\alpha^{(k)}=2\pi \bar r_k$, we get, for the function \eqn{eq:Weiss0},
\begin{align}\label{eq:pot3}
 {\cal F}_m(T,r)&\equiv\frac{T}{\pi^2}\int_0^\infty \!\!dq\,q^2\Big\{3\ln\left[1-e^{-\beta\varepsilon_q}\right]\nn
 &+ \ln\left[1+e^{-2\beta\varepsilon_q}-2e^{-\beta\varepsilon_q}\cos(2\pi \bar r_1)\right]\nn
  &+ \ln\left[1+e^{-2\beta\varepsilon_q}-2e^{-\beta\varepsilon_q}\cos(2\pi \bar r_2)\right]\nn
   &+ \ln\left[1+e^{-2\beta\varepsilon_q}-2e^{-\beta\varepsilon_q}\cos(2\pi (\bar r_1+\bar r_2))\right]\!\Big\}.
 \end{align}

Figure \ref{fig:su3} shows a contour plot of the one-loop SU($3$) potential for increasing values of the temperature in the first Brillouin zone. As discussed previously, the latter contains two Weyl chambers, which are symmetric to one another. At low temperatures, we have a confining phase with a minimum of the potential located at the center-symmetric points, e.g.,  $\bar r_1=\bar r_2=1/3$ in the Weyl chamber closer to the origin. At the temperature $(T_c/m)^{\rm LO}\simeq 0.363$ a first-order deconfinement phase transition occurs, with the appearance of a triplet of center-breaking minima degenerate with the center-symmetric one. Using the value $m_{\rm LO}\simeq 510\,{\rm MeV}$ obtained by fitting tree-level propagators to lattice data in the Landau gauge at zero temperature\footnote{ As explained in Ref.~\cite{Reinosa:2014ooa}, this is a valid procedure
because, at zero temperature, correlation functions of the fields
computed in the LDW gauge (with a constant, temporal, background) and
in the Landau gauge coincide.} yields $T_c^{\rm LO}\simeq 185\,{\rm MeV}$. Above the transition temperature, we have three degenerate minima related by center transformations, which signals the spontaneous breaking of the $Z_3$ symmetry, and hence, a deconfined phase.

As mentioned in the Introduction, a somewhat different model with effectively massive gluons has been considered in Ref.~\cite{Meisinger:2001cq}, where no low-temperature confined phase is found at one-loop order. This can be traced back to the fact  that only two (transverse) gluon degrees of freedom per momentum are given a mass. This is an important difference with the present approach, where the mass term affects three gluon degrees of freedom per mode, resulting in an only partial cancellation of the ghost degrees of freedom. In particular, this yields a ghost-dominated potential at low temperatures with an inverted shape as compared to the high-temperature potential, and thus a confined phase, in agreement with the scenario of Refs.~\cite{Braun:2007bx,Fister:2013bh}.

Finally, we emphasize that the background field effective potential computed here (or in similar background field approaches) is not directly the effective potential for the Polyakov loop. The latter can be obtained in a standard way, either by imposing a constraint or as a Legendre transform with respect to sources linearly coupled to the Polyakov loop, and it is a gauge-independent quantity by construction \cite{Dumitru:2013xna}.  Here, we can access this physical potential for the Polyakov loop by expressing the background field potential in terms of the Polyakov loop (and its conjugate) in a loop expansion.\footnote{We defer the study of the precise relation between the background field potential and the gauge-independent Polyakov loop potential to a future work. We have explicitly checked that the procedure described here reproduces, at high temperatures, the results of Ref.~\cite{Dumitru:2013xna} in the SU($2$) case at two-loop order.} For instance, at LO, the conjugate Polyakov loop variables read, in terms of the backgrounds $r_3$ and $r_8$:
\begin{align}
\ell = & \frac{1}{3}\Big[e^{-i\frac{r_8}{\sqrt{3}}}+2e^{i\frac{r_8}{\sqrt{3}}}\cos(r_3/2)\Big]\,,\label{eq:ell}\\
\bar\ell = & \frac{1}{3}\Big[e^{i\frac{r_8}{\sqrt{3}}}+2e^{-i\frac{r_8}{\sqrt{3}}}\cos(r_3/2)\Big]\,.\label{eq:ellbar}
\end{align}
and we obtain,  for $\tilde{\cal F}_m(\ell,\bar\ell\,)\equiv{\cal F}_m(T,r_3,r_8)$,
\begin{align}
\label{eq:lsfjfj}
\tilde{\cal F}_m(\ell,\bar\ell\,) &= \frac{T}{\pi^2}\!\!\int_0^\infty \!\!\!\!\!dq\,q^2\ln\Big[1+e^{-8\beta\varepsilon_q}\nn
&-\big(9\ell\bar\ell-1\big)\big(e^{-\beta\varepsilon_q}+e^{-7\beta\varepsilon_q}\big)\nonumber\\
& -\,\big(81\ell^2\bar\ell^2-27\ell\bar\ell+2\big)\big(e^{-3\beta\varepsilon_q}+e^{-5\beta\varepsilon_q}\big)\nonumber\\
& +\,\big(27\ell^3+27\bar\ell^3-27\ell\bar\ell+1\big)\big(e^{-2\beta\varepsilon_q}+e^{-6\beta\varepsilon_q}\big)\nonumber\\
& +\,\big(162\ell^2\bar\ell^2-54\ell^3-54\bar\ell^3+18\ell\bar\ell-2\big)e^{-4\beta\varepsilon_q}\Big],\nonumber\\
\end{align}
which coincides with the expression for $\Omega_g$ in Ref.~\cite{Sasaki:2012bi}. This procedure can be systematically implemented at higher orders as well.
 
\subsection{NLO results}

\begin{figure}[t!]  
\begin{center}
\epsfig{file=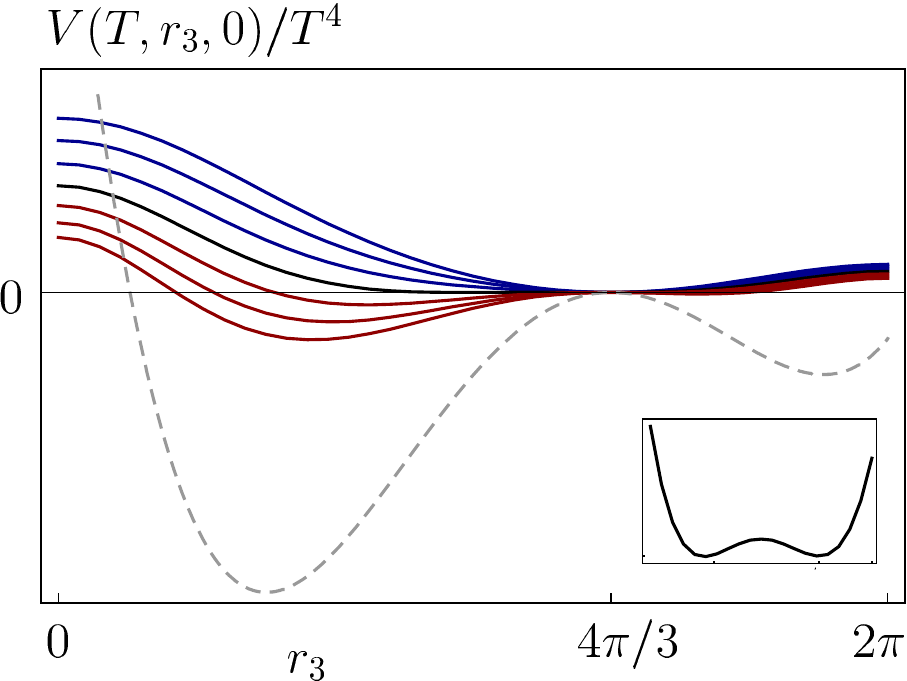,width=7cm}
\caption{Rescaled two-loop background field potential $V(T,r_3,r_8=0)/T^4$ for various temperatures, below (three upper curves, blue) and above (three lower curves, red) the transition temperature (middle curve, black). All curves have been shifted by their respective values at $r=4\pi/3$ for clarity. The dashed curve corresponds to the asymptotic high-temperature limit $v_\infty(r_3)$; see \Eqn{eq:asymptotic}. The inset plot shows a close up view at $T=T_c$ revealing the first-order nature of the transition.}\label{fig:pot}
\end{center}
\end{figure}

As announced earlier we see that, at LO, charge-conjugation invariance is manifest in the sense that there is always one minimum compatible with this symmetry (in the Weyl chamber under consideration, this corresponds to $r_8=0\Leftrightarrow\bar r_1=\bar r_2$). We expect this to be true to all orders and we have explicitly checked it at NLO. For our present purposes, it is thus sufficient to plot the potential on the axis $r_8=0$, as shown in \Fig{fig:pot}. The transition remains first order and we obtain a transition temperature $(T_c/m)^{\rm NLO}\simeq 0.471$. Using the value $m_{\rm NLO}\simeq 540\,{\rm MeV}$ obtained by fitting one-loop propagators (with $\mu=1\,{\rm Gev}$ and $g=4.9$) to lattice data in the Landau gauge at zero temperature \cite{Tissier_11}, we get $T_c^{\rm NLO}\simeq 254\,{\rm MeV}$. Although the comparison to the known lattice result $T_c^{\rm latt}=270\,{\rm MeV}$ must be taken with a pinch of salt due to the ambiguity in fixing the scale, we observe that, indeed, NLO corrections seem to quantitatively improve the previous LO estimate.

Figure~\ref{fig:pot} shows the NLO background potential on the charge-conserving axis $r_8=0$ as a function of $r_3$ for various temperatures. We show in Appendix \ref{appsec:final} that the rescaled potential in the high-temperature limit, $v_\infty(r_3)=\lim_{T\to\infty} V(T,r_3,r_8=0)/T^4$, is a polynomial in the range $[0,2\pi]$:
\begin{align}\label{eq:asymptotic}
v_\infty(r_3) & =  \frac{135r_3^4-600\pi r_3^3+720 \pi^2r_3^2-256\pi^4}{1440\pi^2}\nn
& +  g^2\frac{63r_3^4-316\pi r_3^3+552\pi^2 r_3^2-384\pi^3 r_3+80\pi^4}{384\pi^4}\,,
\end{align}
whose absolute minimum is located at 
\begin{align}
\label{eq:rinf}
r_{\infty}&=\frac{4 \pi }{3}\!\!\left[1-\frac{8 \pi ^2+5 g^2+3 \sqrt{65 g^4+144 \pi ^2 g^2+64 \pi ^4}}{8\left(7 g^2+4 \pi ^2\right)}\right]\nonumber\\
&=\frac{g^2}\pi+\mathcal O(g^4).
\end{align}

\begin{figure}[t!]  
\begin{center}
\epsfig{file=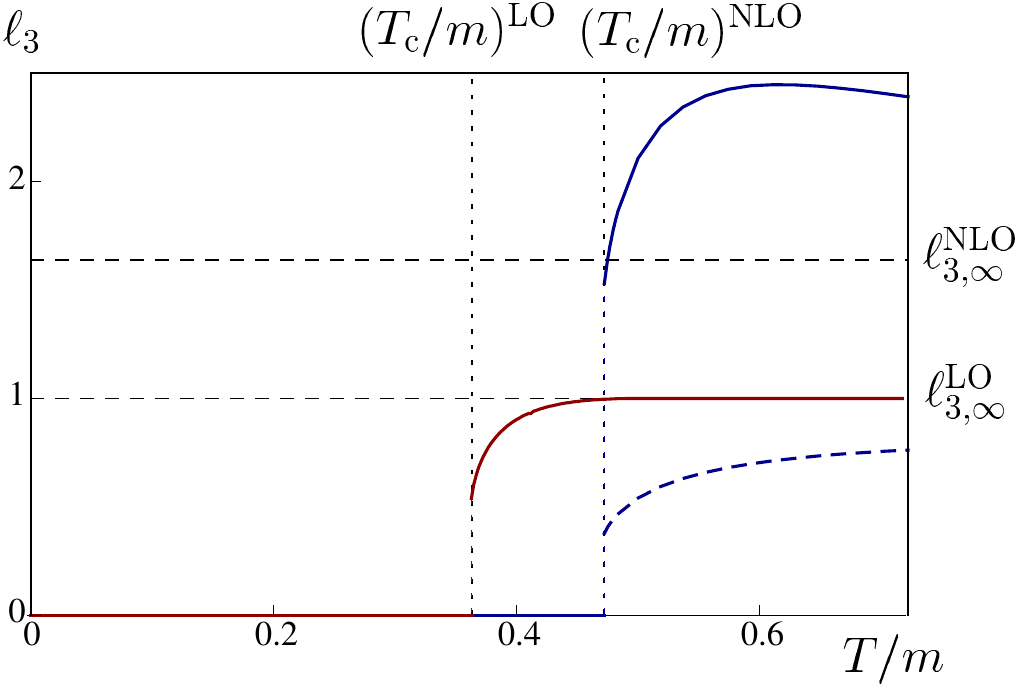,width=8cm}
 \caption{Temperature dependence of the Polyakov loop in the fundamental representation ${\bf 3}$ at LO (red) and NLO (blue). The horizontal dashed lines denote the corresponding asymptotic values at high temperature, denoted here by $\ell_{{\bf 3},\infty}^{\rm LO}=1$ and $\ell_{{\bf 3},\infty}^{\rm NLO}$, respectively. The dashed blue curve shows the LO (or mean-field) Polyakov loop \eqn{eq:l1l} evaluated at $r=r_{\rm min}^{\rm NLO}$ as considered, e.g., in Refs.~\cite{Braun:2007bx,Fister:2013bh,Braun:2010cy,Reinhardt:2012qe}. The respective LO and NLO critical temperatures are indicated by vertical dashed lines.}\label{fig:pl}
\end{center}
\end{figure}

In Fig.~\ref{fig:pl}, we represent the Polyakov loop at LO and NLO. We observe that, as was already the case for the SU($2$) group, the LO Polyakov loop exactly reaches its limiting value at a finite temperature $T_\star/m\approx 0.5$, where the minimum of the potential reaches exactly $r=0$. This generates an additional spurious singularity in the thermodynamical quantities.\footnote{Similar singularities are found in other approaches as well \cite{Reinhardt:2015kxa}.} This unphysical feature disappears at NLO, where the Polyakov loop reaches its limiting value
\begin{align}
\label{eq:above}
\ell^{\rm NLO}_{{\bf 3},\infty}&=\frac{1+2\cos(r_{\infty}/2)}{3} +\frac{3 g^2 }{16 \pi ^2}\!\left(\frac{4 \pi }{3}-r_{\infty}\right)\sin(r_{\infty}/2)\nonumber\\
&=1+\mathcal O(g^4)
\end{align}
only as $T\to\infty$, after first overshooting it at some finite temperature. 
We mention that the high-temperature Polyakov loop has been computed to order $g^4$ in the (standard) Landau gauge in Ref.~\cite{Gava:1981qd}. Even though the present high-temperature limit is purely academic since it does not take into account the necessary (hard thermal loops) resummations for temperatures $T\gg m$ \cite{Braaten:1989kk}, or the running of the coupling with the temperature, $g^2\propto 1/\ln T$, it is interesting that it approaches its asymptotic value from above as in Ref.~\cite{Gava:1981qd}. We postpone a more quantitative comparison for a future work. Similar issues have been recently discussed in the context of a FRG calculation of the Polyakov loop in Ref.~\cite{Herbst:2015ona}.

\begin{figure}[t]  
\begin{center}
\epsfig{file=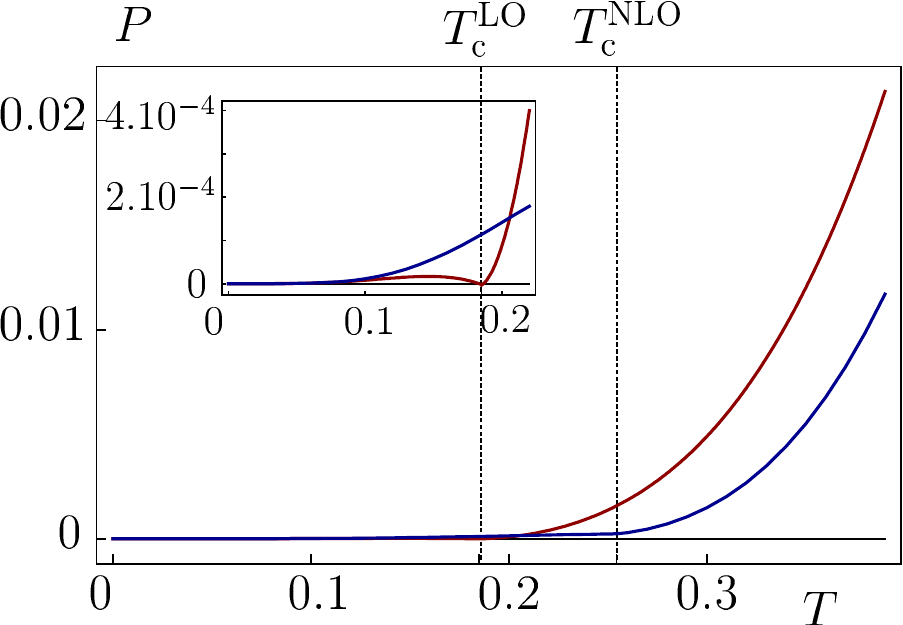,width=8.0cm}
 \caption{Thermodynamic pressure at one- (red) and two-loop (blue) orders, obtained from the minimum of the background field potential as a function of the temperature (in GeV). The plot in the inset is a zoom on the low-temperature region. The respective LO and NLO transition temperatures are indicated by vertical dashed lines.}\label{fig:pressure}
\end{center}
\end{figure}

\begin{figure}[t]  
\begin{center}
\epsfig{file=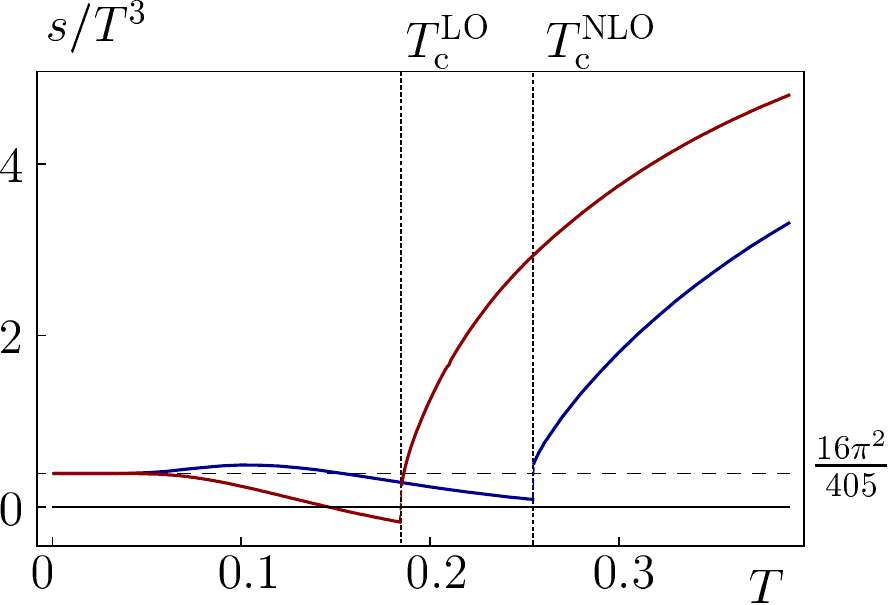,width=8.0cm}
 \caption{Entropy density normalized by $T^3$ at one- (red) and two-loop (blue) orders, obtained from $s=-dP/dT$ as a function of the temperature (in GeV). The respective LO and NLO transition temperatures are indicated by vertical dashed lines.}\label{fig:entropy}
\end{center}
\end{figure}

The pressure and the entropy density are shown respectively in Figs.~\ref{fig:pressure} and \ref{fig:entropy}. We observe that both the LO and NLO results show a positive entropy (increasing pressure) at small temperatures. In this regime the entropy is dominated by the LO result and, despite the fact that only the ghost modes play a role, those ghost modes which feel the presence of the background have a modified statistics, contributing positively to the entropy as in the SU($2$) case \cite{Reinosa:2014zta}. Indeed, one obtains, for the entropy density, in the center-symmetric phase,
\begin{align}
\frac{s}{4T^3} & = 2\int_{\bf q}\ln\Big(1-e^{-q}\Big)+3\int_{\bf q}\ln\Big(1+e^{-q}+e^{-2q}\Big)\nonumber\\
& = -\int_{\bf q}\ln\Big(1-e^{-q}\Big)+3\int_{\bf q}\ln\Big(1-e^{-3q}\Big)\nonumber\\
& = -\frac{8}{9}\int_{\bf q}\ln\Big(1-e^{-q}\Big)=\frac{4\pi^2}{405}\,.
\end{align}
This is to be compared with the corresponding result at vanishing background
\beq
\frac{s_{\bar A=0}}{4T^3}= 8\int_{\bf q}\ln\Big(1-e^{-q}\Big)=-\frac{4\pi^2}{45}\,.
\eeq
At larger temperatures, in particular close to the transition, the LO entropy becomes slightly negative. This is because, as the temperature increases, gluonic modes start playing a role but, as long as the system is in the confined phase, the charged gluons contribute with negative distribution functions, e.g., 
\bea
{\rm Re}\, n_{\varepsilon-i(4\pi/3)T}=-\frac{e^{\beta\varepsilon}/2+1}{e^{2\beta\varepsilon}+e^{\beta\varepsilon}+1}.
\eea 
Interestingly, similar pathologies are observed in other approaches \cite{Sasaki:2012bi,Canfora:2015yia,Kondo:2015noa} which, we believe, have the same origin as the one described here. In the  present case, these spurious features disappear at NLO, where we obtain a positive entropy---a monotonously increasing pressure---at all temperatures;\footnote{Despite these satisfying aspects, our results still present some unwanted artifacts. The low-temperature pressure (respectively entropy density) behave as $T^4$ (respectively $T^3$) as $T\to 0$, in contradiction with lattice results. We believe that this is a generic issue in approaches where the (massless) ghost modes dominates over the (massive) gluonic modes at small temperatures \cite{Braun:2007bx,Canfora:2015yia,Kondo:2015noa}.} see Figs.~\ref{fig:pressure} and \ref{fig:entropy}. Finally, the entropy density is discontinuous at the transition, as it should for a first-order phase transition. We obtain, for the latent heat, $(L/T^4_c)_{\rm NLO}\approx 0.41$ (vs. $0.43$ at LO),  that is only $30$ percent of the lattice value $1.40$ \cite{Beinlich:1996xg}. We stress, however, that the present calculation disregards a possible temperature dependence of the parameters which could arise, e.g., through renormalization group improvement effects or from our assumption that the gluon mass parameter is related to the presence of Gribov copies which depend themselves on the temperature. Such effects may be important for a quantitative agreement with lattice simulations; see, e.g., Ref.~\cite{Reinosa:2013twa}, or Ref.~\cite{Fukushima:2013xsa} for an example in the Gribov-Zwanziger approach. We defer the investigation of such effects to future work.

\begin{figure}[t!]  
\begin{center}
\epsfig{file=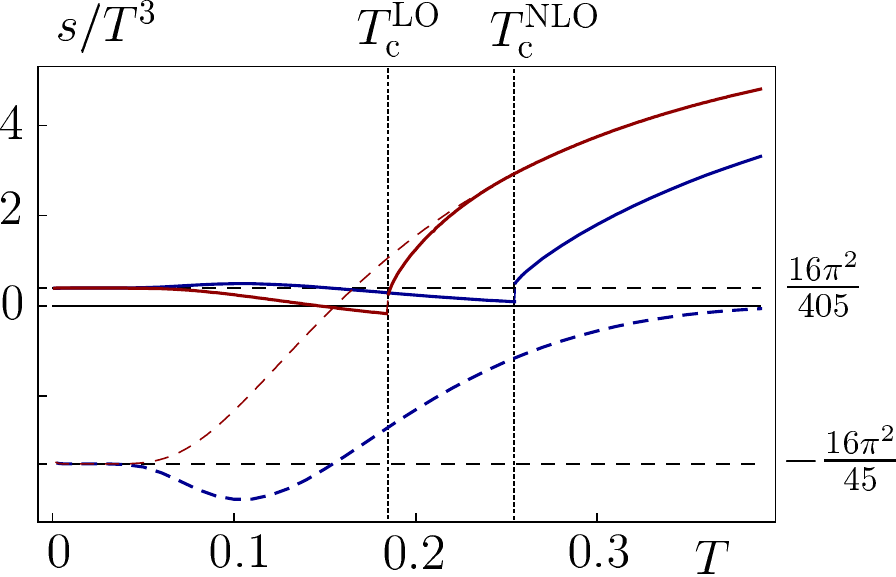,width=8.0cm}
 \caption{Comparison of the perturbative expansions in the (massive) Landau and LDW gauges. Temperatures are given in GeV.}\label{fig:Landau}
\end{center}
\end{figure}

As a last remark, we compare in Fig.~\ref{fig:Landau} the present result in the LDW gauge to the ones in the Landau gauge, which corresponds to setting $\bar A=0$. As we observed already in the SU($2$) case, the results are dramatically different: in the Landau gauge, the LO thermodynamics is inconsistent (negative entropy) in a wide range of temperatures including the confined phase and the NLO result is thermodynamically inconsistent at all temperatures explored here. At first sight, such differences might seem surprising because the behavior of thermodynamical observables should not depend on the gauge. However, we stress that finite orders of a given expansion scheme usually do depend on the gauge. To illustrate this point further, we note that the results in the Landau gauge can be read off from the LDW effective potential at the origin $\bar A=0$. However, as shown in Fig.~\ref{fig:pot}, at two-loop order, this corresponds to a maximum at all temperatures and the correct physics can certainly not be accessed from this point by perturbative means. We note that a similar remark applies to other functional approaches such as the FRG \cite{Braun:2007bx} where the origin of the LDW potential is also a maximum in the relevant range of temperatures. 

\section{Other groups}\label{sec:su4}

It is interesting to extend the previous calculations to other gauge groups \cite{Braun:2010cy,Dumitru:2012fw}. In this section, we perform a LO analysis for the groups Sp(2) and
SU(4), for which lattice simulations predict a first-order phase transition in $d=4$ \cite{Moriarty:1981wc,Green:1984ks,Batrouni:1984vd,Wheater:1984wd,Holland:2003kg}. The Sp($2$) group has the same center as SU($2$). However, its Cartan subalgebra is two-dimensional and the background field potential has a richer structure. Another curiosity of Sp($2$) is that there exists a whole segment of center-symmetric points in each Weyl chamber which are actually explored by the dynamics. The group SU($4$) is interesting because there exist center-breaking points in the Weyl chamber where some Polyakov loops vanish but not others, which would correspond to partial deconfinement (this extends to the SU($N\ge4$) groups, as discussed in Appendix~\ref{appsec:ploopcounting}; see also Ref.~\cite{Dumitru:2012fw}). However, we shall verify that these points never actually correspond to an absolute minimum of the background potential at LO. 

\subsection{Sp($2$)}

\begin{figure}[h]  
\epsfig{file=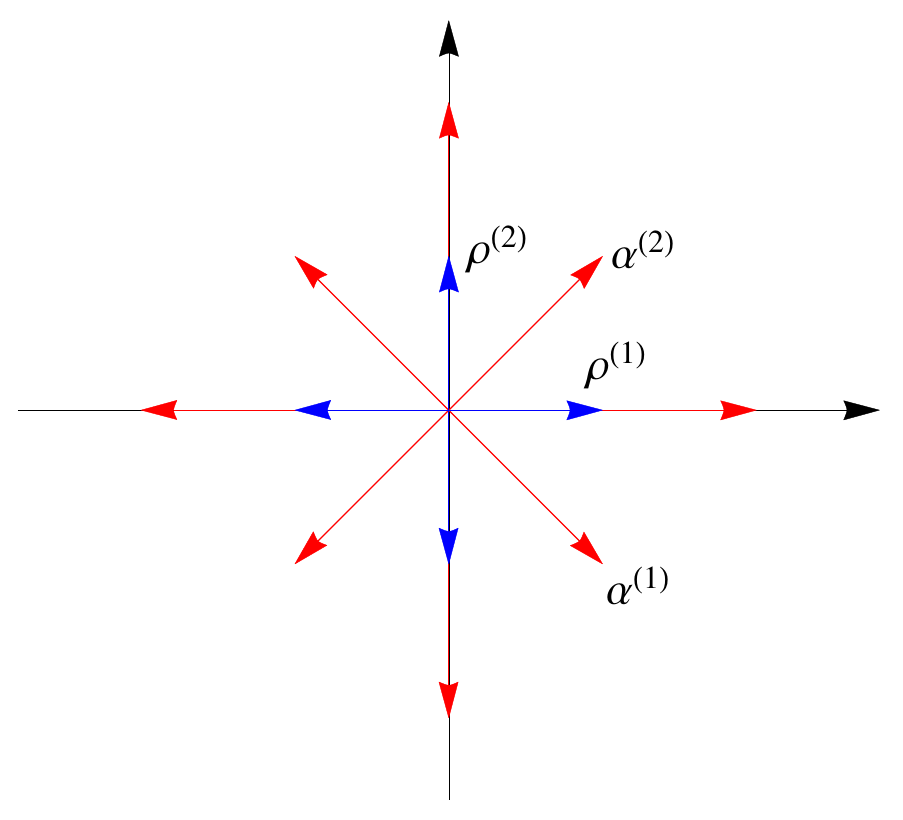,width=8cm}
 \caption{The system of roots $\alpha$ (red) for the Sp($2$) algebra. We also show the weights $\rho$ of the fundamental representation (blue). We have $\alpha^{(1)}=(1/2,-1/2)$, $\alpha^{(2)}=(1/2,1/2)$, $\rho^{(1)}=(1/2,0)$ and $\rho^{(2)}=(0,1/2)$.}\label{fig:roots_sp2}
\end{figure}

\begin{figure}[t]  
\begin{center}
\epsfig{file=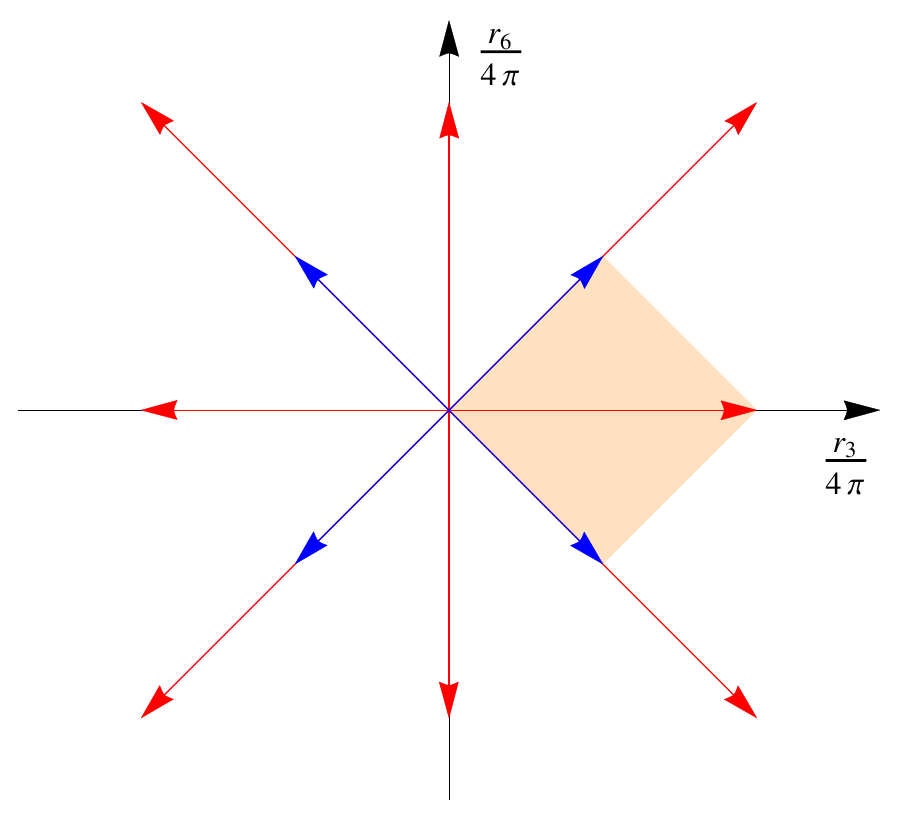,width=8cm}
 \caption{The Sp($2$) Brillouin zone. Periodic winding transformations correspond to translations along certain roots (red) in the plane $(r_3/4\pi,r_6/4\pi)$, whereas nonperiodic winding transformations correspond to translation along the remaining roots (blue); see text. The Brillouin zone (associated to periodic winding transformations) is represented in orange. In addition to the winding transformations, we have the Weyl transformations which are generated by reflections with respect to axes orthogonal to the roots. charge conjugation corresponds to an inversion about the origin and is thus a Weyl transformation.\label{fig:translations_sp2}}
\end{center}
\end{figure}

The algebra of Sp($2$) is ten dimensional with a Cartan subalgebra of dimension two. The root diagram is represented in Fig.~\ref{fig:roots_sp2}, together with the weight diagram of the fundamental representation. We can choose the roots $\pm\alpha^{(1)}=\pm(1/2,-1/2)$ and $\pm\alpha^{(2)}=\pm(1/2,1/2)$ in terms of which the other roots can be obtained using linear combinations with integer coefficients: $\pm(\alpha^{(1)}+\alpha^{(2)})$ and $\pm(\alpha^{(1)}-\alpha^{(2)})$. Similarly, we can choose the weights $\rho^{(1)}=(1/2,0)$ and $\rho^{(2)}=(0,1/2)$. 
As discussed in Sec.~\ref{sec:center}, the lattice dual to the one generated by the weights $\rho^{(j)}$ gives translations that correspond to periodic winding transformations. This corresponds to $\bar\rho^{(1)}=4\pi(\alpha^{(1)}+\alpha^{(2)})$ and $\bar\rho^{(2)}=4\pi(\alpha^{(2)}-\alpha^{(1)})$. The lattice dual to the one generated by the roots $\alpha^{(j)}$ gives the translations that correspond to general winding transformations. This corresponds to $\bar\alpha^{(1)}=4\pi\alpha^{(1)}$ and $\bar\alpha^{(2)}=4\pi\alpha^{(2)}$. So contrary to the SU($2$) and SU($3$) cases, in the plane $(r_3/4\pi,r_6/4\pi)$, only certain roots are associated to periodic winding transformations, namely $\pm(\alpha^{(1)}+\alpha^{(2)})$ and $\pm(\alpha^{(1)}-\alpha^{(2)})$. The other roots $\pm\alpha^{(1)}$ and $\pm\alpha^{(2)}$ are associated to nonperiodic winding transformations. We mention, however, that $2\alpha^{(1)}$ and $2\alpha^{(2)}$ correspond again to periodic winding transformations. The various relevant translations in the Cartan plane $(r_3/4\pi,r_6/4\pi)$ are represented in Fig.~\ref{fig:translations_sp2}.

\begin{figure}[t]  
\begin{center}
\epsfig{file=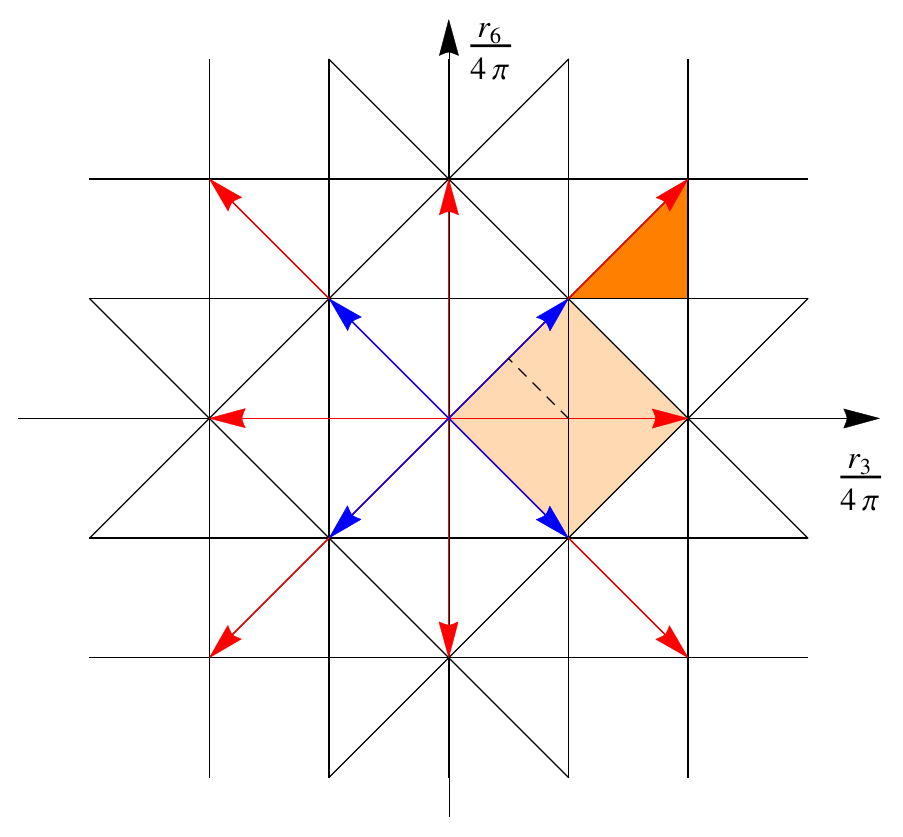,width=7cm}
 \caption{The Weyl chambers of Sp($2$). The reflection-symmetry lines resulting from combinations of periodic winding transformations and Weyl reflections are shown in black. These separate the Brillouin zone in four square triangles (light orange) which thus correspond to four equivalent Weyl chambers. The $Z_2$ symmetry group of the Weyl chamber corresponds to the center of Sp($2$). Indeed, a nonperiodic winding transformation translates the Weyl chamber along one of the short (blue) vectors, as illustrated here as an example. The translated chamber (dark orange) can be brought back to its original location using three reflections, after which one obtains its mirror image about its short median (dashed line). It follows that the center-symmetric points are all the points of the median.}\label{fig:Weyl_chambers_sp2}
\end{center}
\end{figure}

 The Weyl chambers are obtained as before, i.e., by exploiting the reflections and the translations corresponding to the standard gauge transformations. This is described in \Fig{fig:Weyl_chambers_sp2}. The Brillouin square is made of four equivalent Weyl chambers, which are square triangles. Their symmetry group is $Z_2$, made of the identity and the reflection about the short median of the triangle. This clearly corresponds to the center symmetry of the theory, as explained in \Fig{fig:Weyl_chambers_sp2}. It follows that all the points on the short median are center symmetric. As for SU($2$), charge conjugation is nothing but a Weyl transformation and all the points in the Weyl chamber are charge-conjugation invariant.

\begin{figure}[t!]  
\begin{center}
\epsfig{file=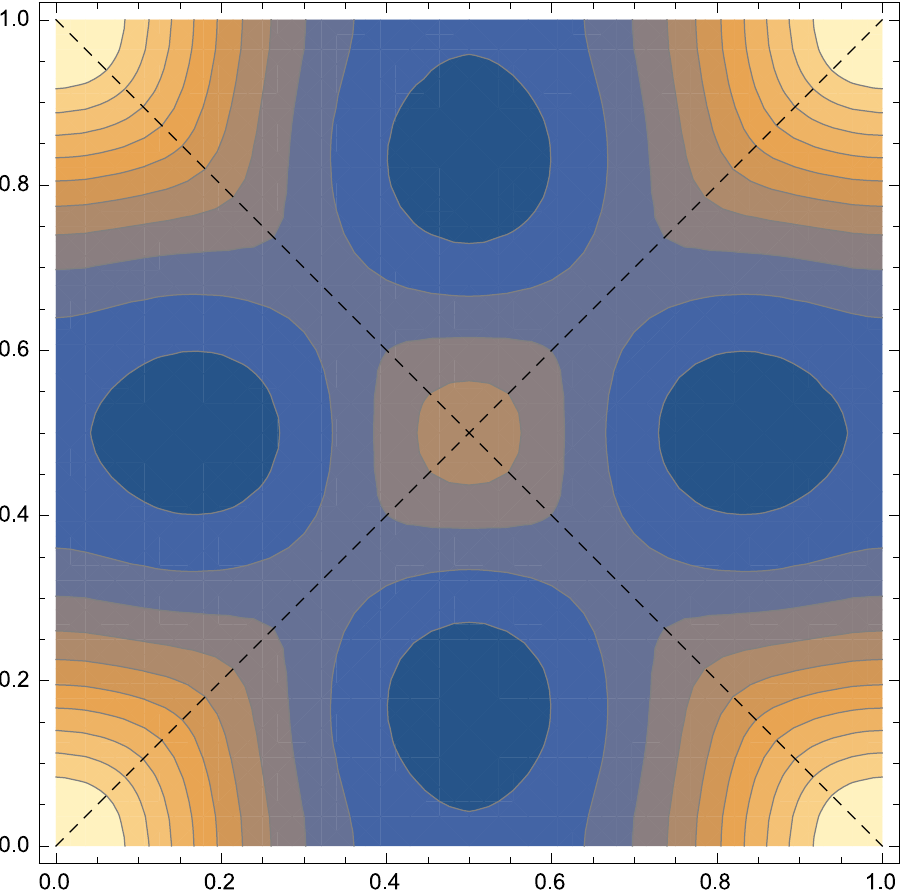,width=2.5cm}\quad\epsfig{file=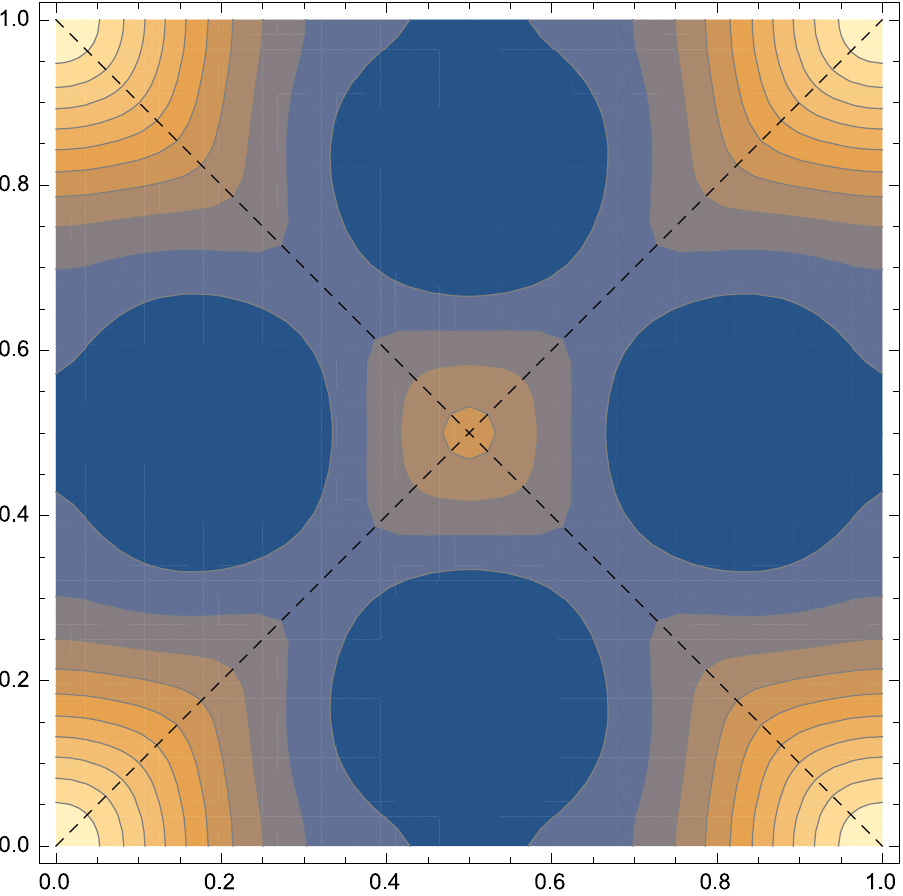,width=2.5cm}\quad\epsfig{file=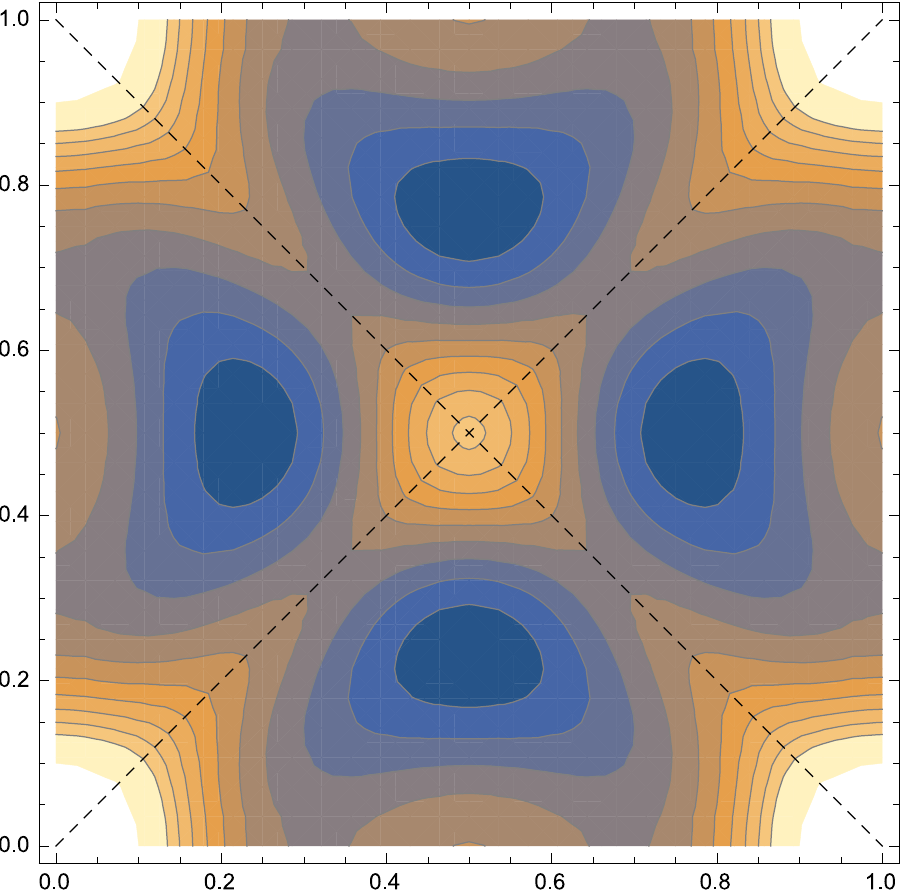,width=2.5cm}\\
\vspace{0.6cm}
\epsfig{file=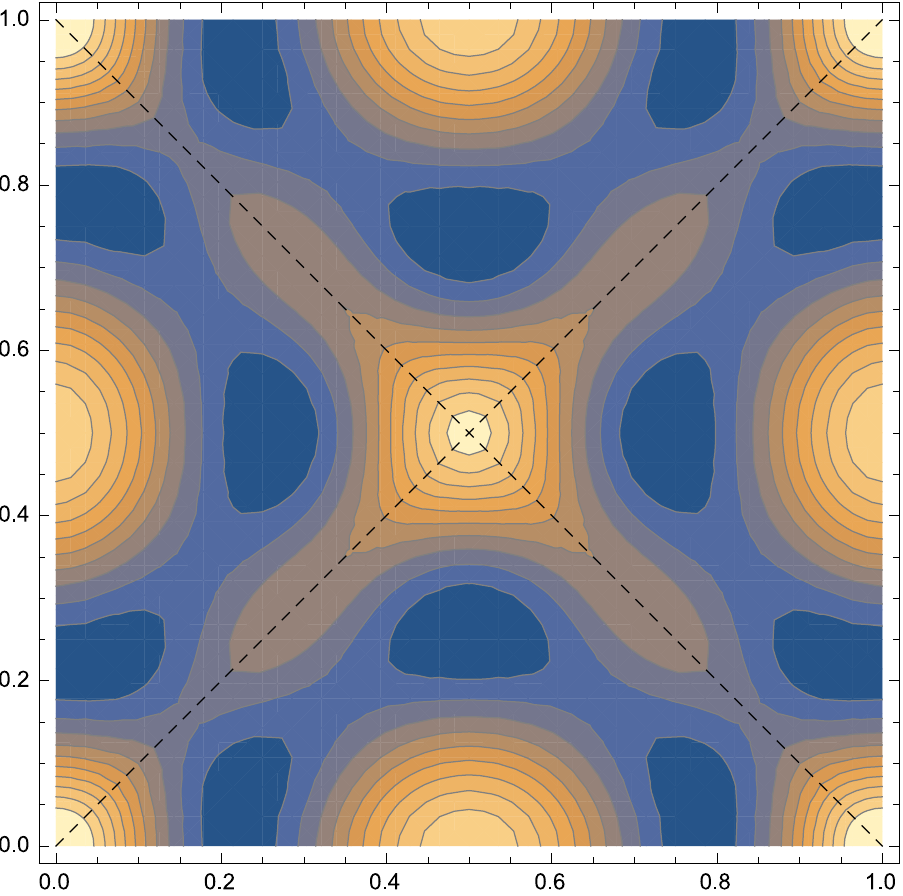,width=2.5cm}\quad\epsfig{file=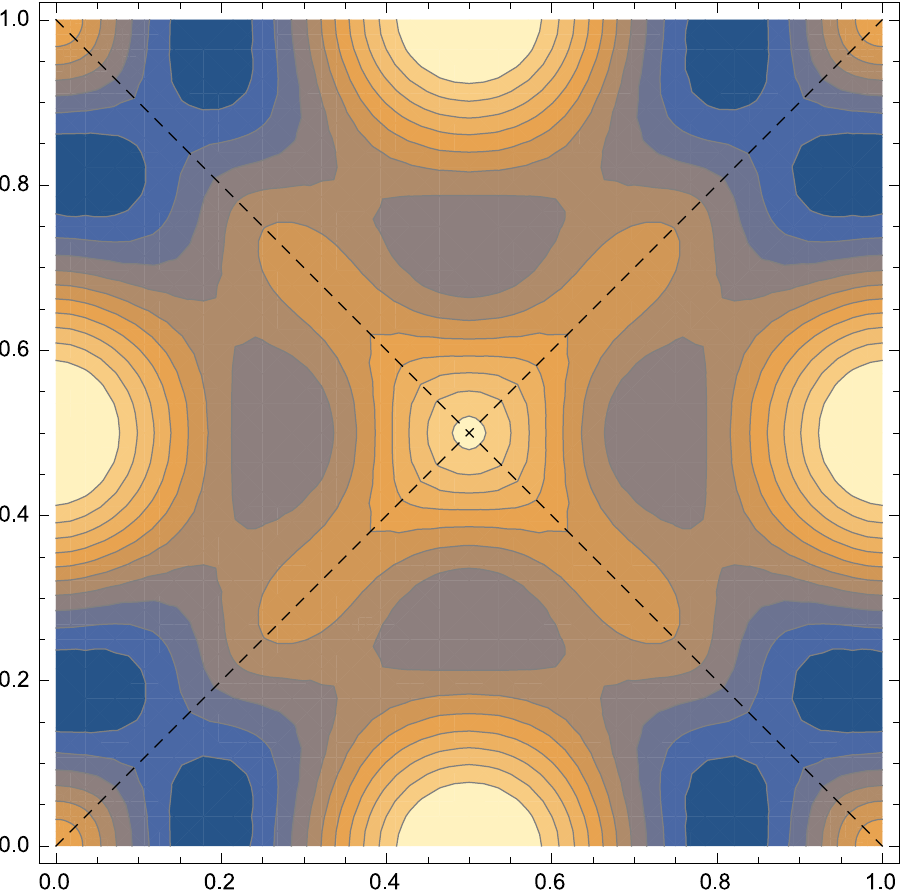,width=2.5cm}\quad\epsfig{file=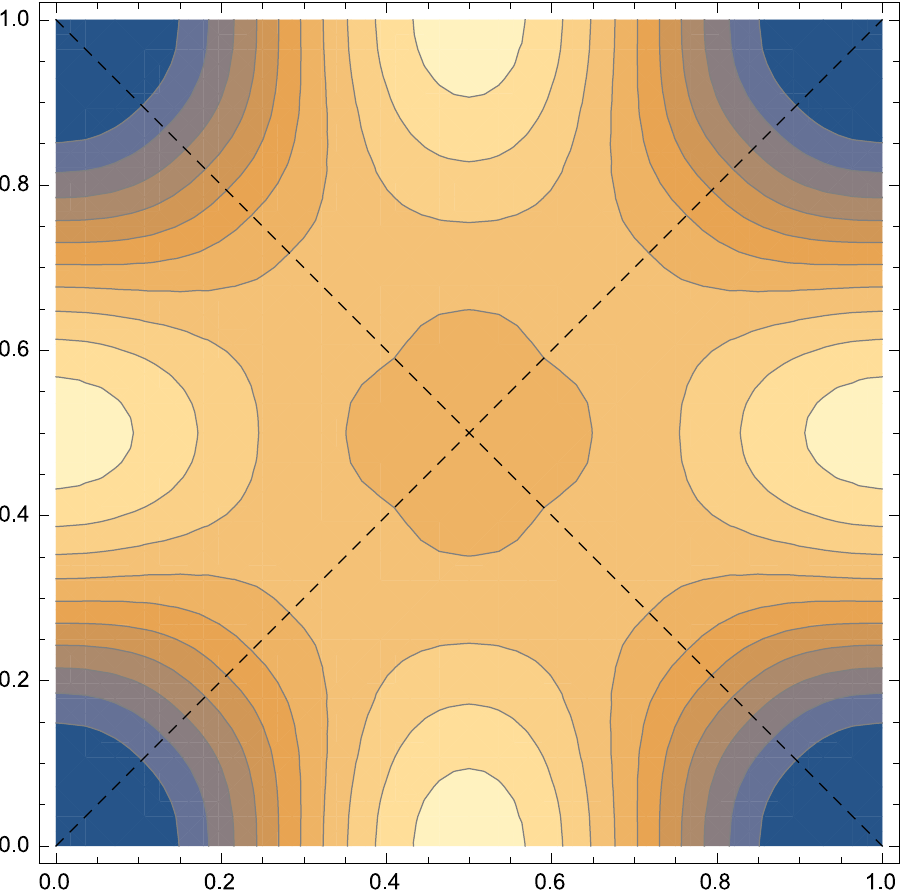,width=2.5cm}
 \caption{Contour plot of the Sp($2$) one-loop potential for increasing values of the temperature. The dashed lines separate the four Weyl chambers that form the Brillouin zone. The center-symmetric points are such that $\bar r_1=1/2$ or $\bar r_2=1/2$ depending on the choice of Weyl chamber. The upper (respectively lower) plots correspond to temperatures below (respectively above) the transition temperature.  One observes the appearance of a $Z_2$ pair of degenerate minima above the (first-order) transition, signaling the spontaneous breaking of the center symmetry. Darker colors correspond to regions where the potential is the deepest.}\label{fig:sp2}
\end{center}
\end{figure}

We now study the LO potential. As before, it is convenient to decompose $r=\bar r_k\bar\alpha^{(k)}$ in the dual basis $\{\bar\alpha^{(1)},\bar\alpha^{(2)}\}$, where the Brillouin zone is $(\bar r_1,\bar r_2)\in [0,1]^2$. The center-symmetric segments are located at $\bar r_1=1/2$ or $\bar r_2=1/2$ depending on which Weyl chamber one chooses. Using the one-loop formula \eqn{eq:onelooppot} with the roots $\pm\alpha^{(1)}$, $\pm\alpha^{(2)}$, $\pm(\alpha^{(1)}+\alpha^{(2)})$ and $\pm(\alpha^{(1)}-\alpha^{(2)})$, we get, for the function \eqn{eq:Weiss0},
\begin{align}
 {\cal F}_m(T,r)&\equiv\frac{T}{\pi^2}\int_0^\infty \!\!dq\,q^2\Big\{2\ln\left[1-e^{-\beta\varepsilon_q}\right]\nn
 &+ \ln\left[1+e^{-2\beta\varepsilon_q}-2e^{-\beta\varepsilon_q}\cos(2\pi \bar r_1)\right]\nn
  &+ \ln\left[1+e^{-2\beta\varepsilon_q}-2e^{-\beta\varepsilon_q}\cos(2\pi \bar r_2)\right]\nn
 &+ \ln\left[1+e^{-2\beta\varepsilon_q}-2e^{-\beta\varepsilon_q}\cos(2\pi(\bar r_1+\bar r_2))\right]\nn
 &+ \ln\left[1+e^{-2\beta\varepsilon_q}-2e^{-\beta\varepsilon_q}\cos(2\pi(\bar r_1-\bar r_2))\right]\!\Big\}.\nn
\end{align}
Figure~\ref{fig:sp2} shows a contour plot of the potential for increasing values of the temperature. At low temperatures, there is a single absolute minimum in each chamber, located at a center-symmetric point that moves along the center-symmetric segment as the temperature is varied. We observe a first-order phase transition at a temperature $T_c^{\rm LO}/m\simeq 0.37$, close to the corresponding LO value in the SU($3$) theory. 

Using Eq.~(\ref{eq:l1l}) with the weights $\pm\rho^{(1)}=\pm(\alpha^{(1)}+\alpha^{(2)})/2$, $\pm\rho^{(2)}=\pm(\alpha^{(2)}-\alpha^{(1)})/2$, we obtain, for the Polyakov loop in the fundamental representation ${\bf 4}$ at LO,
\beq
\ell_{{\bf 4}}^{(0)}(r)=  \cos(\pi \bar r_1)\cos(\pi \bar r_2),
\eeq
which vanishes if and only if $\bar r_1=\frac{1}{2}\,({\rm mod}\,1)$ or  $\bar r_2=\frac{1}{2}\,({\rm mod}\,1)$, that is, precisely on the center-symmetric segments.

\subsection{SU($4$)}

The algebra of SU($4$) is $15$ dimensional with a Cartan subalgebra of dimension three. From the generalization of the Gell-Mann matrices, one obtains the roots $\pm\alpha^{(1)}$, $\pm\alpha^{(2)}$, $\pm\alpha^{(3)}$, $\pm\alpha^{(4)}=\pm(\alpha^{(1)}-\alpha^{(2)})$, $\pm\alpha^{(5)}=\pm(\alpha^{(2)}-\alpha^{(3)})$, and $\pm\alpha^{(6)}=\pm(\alpha^{(3)}-\alpha^{(1)})$, with
\beq
\alpha^{(1)}=\left(
\begin{array}{c}
1\\
0\\
0
\end{array}\right)\!, \,\,
\alpha^{(2)}=\frac{1}{2}\left(
\begin{array}{c}
1\\
\sqrt{3}\\
0
\end{array}\right)\!, \,\, 
\alpha^{(3)}=\frac{1}{2}\left(
\begin{array}{c}
1\\
\sqrt{1/3}\\
\sqrt{8/3}
\end{array}\right)\!.
\eeq
The weights of the fundamental representation ${\bf 4}$ are found to be
\bea\label{eq:we}
\mu^{(1)}&=&\frac{1}{2}\left(
\begin{array}{c}
-1\\
\sqrt{1/3}\\
\sqrt{1/6}
\end{array}\right)\!, \quad 
\mu^{(2)}=\frac{1}{2}\left(
\begin{array}{c}
0\\
-\sqrt{4/3}\\
\sqrt{1/6}
\end{array}\right)
\!, \quad\nonumber\\ 
\mu^{(3)}&=&\frac{1}{2}\left(
\begin{array}{c}
0\\
0\\
- \sqrt{3/2}
\end{array}\right)\!, \quad
\mu^{(4)}=\frac{1}{2}\left(
\begin{array}{c}
1\\
\sqrt{1/3}\\
\sqrt{1/6}
\end{array}\right)\!.
\eea
For later use, it will be helpful to obtain the Brillouin zone $\{\bar\alpha^{(1)},\bar\alpha^{(2)},\bar\alpha^{(3)}\}$. A straightforward calculation shows that $\bar\alpha^{(j)}=-4\pi\mu^{(j)}$, from which it also follows that $\bar\mu^{(j)}=-4\pi\alpha^{(j)}$. Then, according to the general discussion of Sec.~\ref{sec:center}, periodic winding transformations correspond to translations of the background by vectors equal to $4\pi$ times any of the roots, while general winding transformations correspond to translations of the background by vectors equal to $4\pi$ times any of the weights (\ref{eq:we}) as was already the case for SU($2$) and SU($3$).

\begin{figure}[t]  
\begin{center}
\epsfig{file=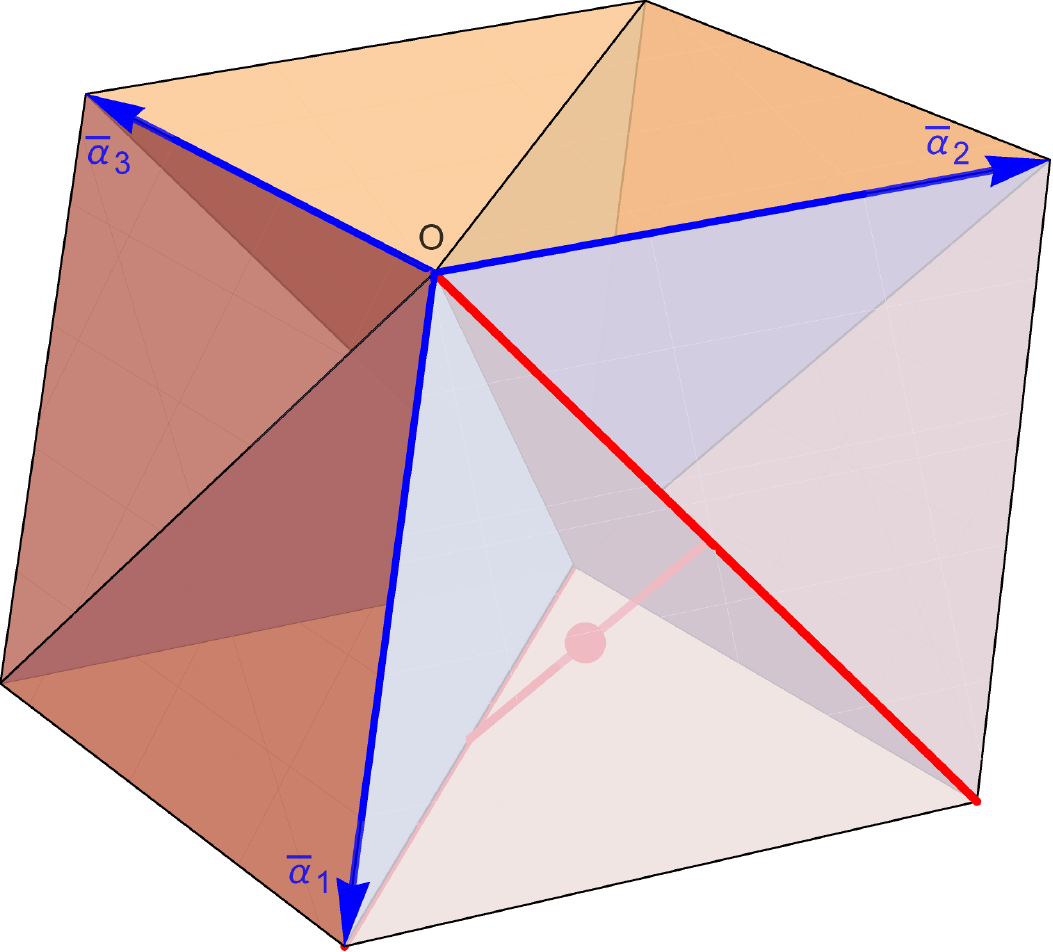,width=5.5cm}
 \caption{The Brillouin zone and the Weyl chambers of SU($4$). The vectors $\bar\alpha^{(1)}$, $\bar\alpha^{(2)}$, $\bar\alpha^{(3)}$ (blue) define the Brillouin zone. The latter is composed of six tetrahedral Weyl chambers:  $\{O,\bar\alpha^{(1)},\bar\alpha^{(1)}+\bar\alpha^{(2)},\bar\alpha^{(1)}+\bar\alpha^{(2)}+\bar\alpha^{(3)}\}$, $\{O,\bar\alpha^{(2)},\bar\alpha^{(1)}+\bar\alpha^{(2)},\bar\alpha^{(1)}+\bar\alpha^{(2)}+\bar\alpha^{(3)}\}$ (blue), $\{O,\bar\alpha^{(1)},\bar\alpha^{(1)}+\bar\alpha^{(3)},\bar\alpha^{(1)}+\bar\alpha^{(2)}+\bar\alpha^{(3)}\}$, $\{O,\bar\alpha^{(3)},\bar\alpha^{(1)}+\bar\alpha^{(3)},\bar\alpha^{(1)}+\bar\alpha^{(2)}+\bar\alpha^{(3)}\}$ (red), $\{O,\bar\alpha^{(2)},\bar\alpha^{(2)}+\bar\alpha^{(3)},\bar\alpha^{(1)}+\bar\alpha^{(2)}+\bar\alpha^{(3)}\}$, $\{O,\bar\alpha^{(3)},\bar\alpha^{(2)}+\bar\alpha^{(3)},\bar\alpha^{(1)}+\bar\alpha^{(2)}+\bar\alpha^{(3)}\}$ (yellow). Each chamber has two edges longer than the other four, namely the one connecting the origin to the sum of two $\bar\alpha$'s and the one connecting one $\bar\alpha$ to the sum of the three $\bar\alpha$'s, as illustrated in the figure (red lines). These longer edges have length $\sqrt{2}$ while the others have length $\sqrt{3/2}$. The symmetry group of the Weyl chamber is $D_{2d}$ and there is a single center-symmetric point at the barycenter of the tetrahedron such as, e.g., the one (red dot) located at $(3/4,1/2,1/4)$ in the basis of the Brillouin zone.}\label{fig:Weyl_chambers_su4}
\end{center}
\end{figure}

The Weyl chambers are identified following the same procedure as before. Combining Weyl transformations (reflections about planes orthogonal to the roots) and periodic winding transformations (translations by $4\pi$ along the roots), one obtains a family of reflection planes shifted by $2\pi$ along the roots, which divide the Cartan space in Weyl chambers. Let us proceed in two steps. First, one easily shows\footnote{To see this, consider the cell delimited by the planes containing the origin and orthogonal to $\alpha^{(1)}$, $\alpha^{(2)}$, and $\alpha^{(3)}$, and their respective translations by $2\pi\alpha^{(1)}$, $2\pi\alpha^{(2)}$, $2\pi\alpha^{(3)}$. The vertices $r$ connected to the origin obey the equations $(r-2\pi\alpha^{(j)})\cdot\alpha^{(j)}=0$, $r\cdot\alpha^{(j+1)}=0$ and $r\cdot\alpha^{(j+2)}=0$, whose solutions are nothing but the $\bar\alpha^{(j)}$ [we use $(\alpha^{(j)})^2=1$].} that the family of reflection planes along the roots $\alpha^{(1)}$, $\alpha^{(2)}$ and $\alpha^{(3)}$ actually defines cells equal to the Brillouin zones generated by the triad $(\bar\alpha^{(1)},\bar\alpha^{(2)},\bar\alpha^{(3)})$. In order to understand how the Brillouin zone is further divided, we need to examine its intersections with the remaining families of planes, associated to the roots $\smash{\alpha^{(1)}-\alpha^{(2)}}$, $\smash{\alpha^{(2)}-\alpha^{(3)}}$, and $\smash{\alpha^{(3)}-\alpha^{(1)}}$. Consider for instance the hyperplane orthogonal to $\alpha^{(1)}-\alpha^{(2)}$. It contains $\bar\alpha^{(3)}$ as well as $\bar\alpha^{(1)}+\bar\alpha^{(2)}$. We obtain two other similar planes by permuting $\bar\alpha^{(1)}$, $\bar\alpha^{(2)}$, and $\bar\alpha^{(3)}$, as shown in Fig.~\ref{fig:Weyl_chambers_su4}. Altogether they divide the Brillouin zone into six tetrahedra. A simple calculation shows that these tetrahedra have two nonadjacent edges that are longer than the other four, of equal length, by a factor of $2/\sqrt{3}$. 

Such an irregular tetrahedron, made of four identical isoceles triangles, is called a tetragonal disphenoid, whose symmetry group is the dihedral group $D_{2d}$, with eight elements. These are the identity, the three rotations by $\pi$ around any of the axes which relate the midpoints of nonadjacent edges, the reflections about the two planes perpendicular to one of the long edges and containing the other, and the remaining two elements can be obtained by combining these. As before, these correspond to either elements of the center $Z_4=\{\mathds{1},i\mathds{1},-\mathds{1},-i\mathds{1}\}$ of SU($4$), charge conjugation, or any combination of these. The elements corresponding to nontrivial center transformations are those which result in cyclic permutations of the vertices\footnote{This can be seen from the fact that the LO Polyakov loop function $\ell^{(0)}(r)$ [see \Eqn{eq:cla} below] takes the values $1$, $i$, $-1$, and $-i$, respectively, at these vertices. The nontrivial elements of $Z_4$ are multiplications by $i$, $-1$, or $-i$, which correspond to cyclic permutations of the values of $\ell^{(0)}(r)$ mentioned here.} $(0,0,0)$, $(1,0,0)$, $(1,1,0)$, and $(1,1,1)$. It is not difficult to convince oneself\footnote{This can be done either by using the method described in Sec.~\ref{sec:SSB} or, more simply, by evaluating simple quantities sensitive to these symmetries at various points in the Weyl chamber, such as, for instance, the LO Polyakov loop at the vertices of the tetrahedron for the center symmetry. These values can also be used to determine the symmetry plane corresponding to charge conjugation because, in the pure Yang-Mills theory, the Polyakov loops at points related by charge conjugation are complex conjugates of each other.} that the element $-\mathds{1}$ corresponds to the rotation by $\pi$ around the axis crossing the two longer edges whereas $\pm i\mathds{1}$ are obtained by combining the two other rotations with any of the reflection planes described above. Finally, charge conjugation corresponds to one of these reflection planes---the other one being a combination of a nontrivial center transformation and charge conjugation. 

Clearly, the only center-symmetric point is the intersection of the three rotation axes mentioned above, that is the barycenter of the tetrahedron. In the Weyl chambers containing the origin, represented on \Fig{fig:Weyl_chambers_su4}, it is located at the coordinates $(3/4,1/2,1/4)$ (dot in the figure) or any permutation of these, depending on the Weyl chamber one chooses.

\begin{figure}[t]  
\begin{center}
\epsfig{file=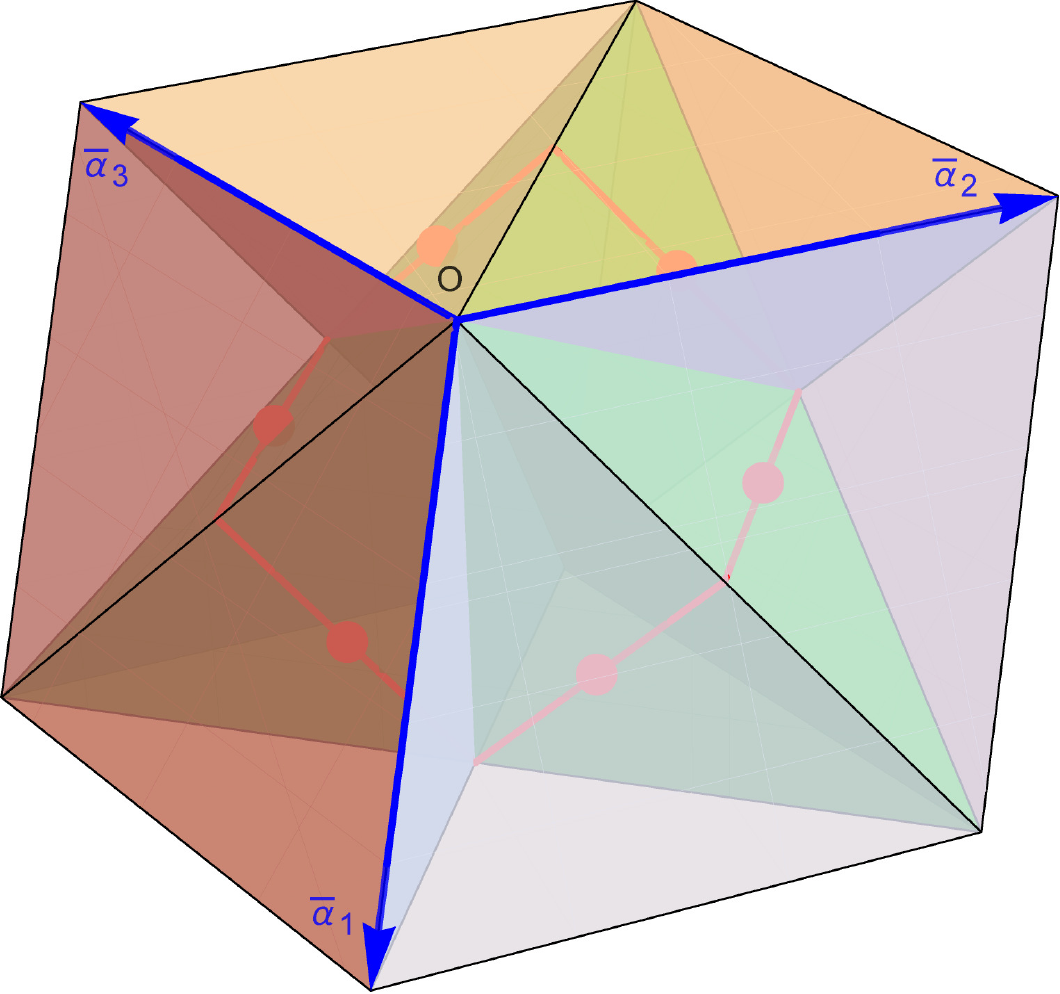,width=5.5cm}
 \caption{Reflection-symmetry planes of the Weyl chambers corresponding to charge conjugation. These contain one of the longer edges of the tetrahedron and are orthogonal to the other one, intersecting it in its center. Each chamber contains two such reflection-symmetry planes, one of which (green) corresponds to charge conjugation, and the other (not shown) is its image under the element $-\mathds{1}$ of the center $Z_4$. The transformed versions of the charge conjugation reflection under the elements $\pm i\mathds{1}$ of the center correspond to rotations by $\pi$ around the two axes relating the centers of pairs of opposite short edges of the tetrahedron.}\label{fig:Weyl_chambers_su4_C}
\end{center}
\end{figure}

\begin{figure}[t!]  
\begin{center}
\epsfig{file=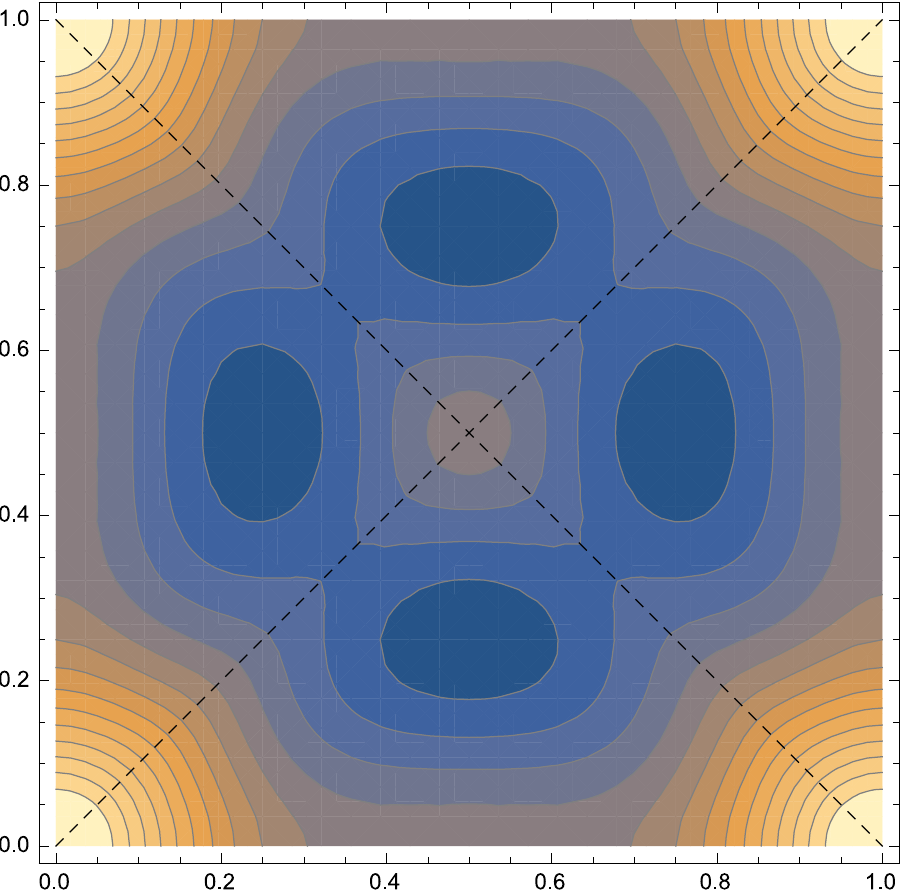,width=2.5cm}\quad\epsfig{file=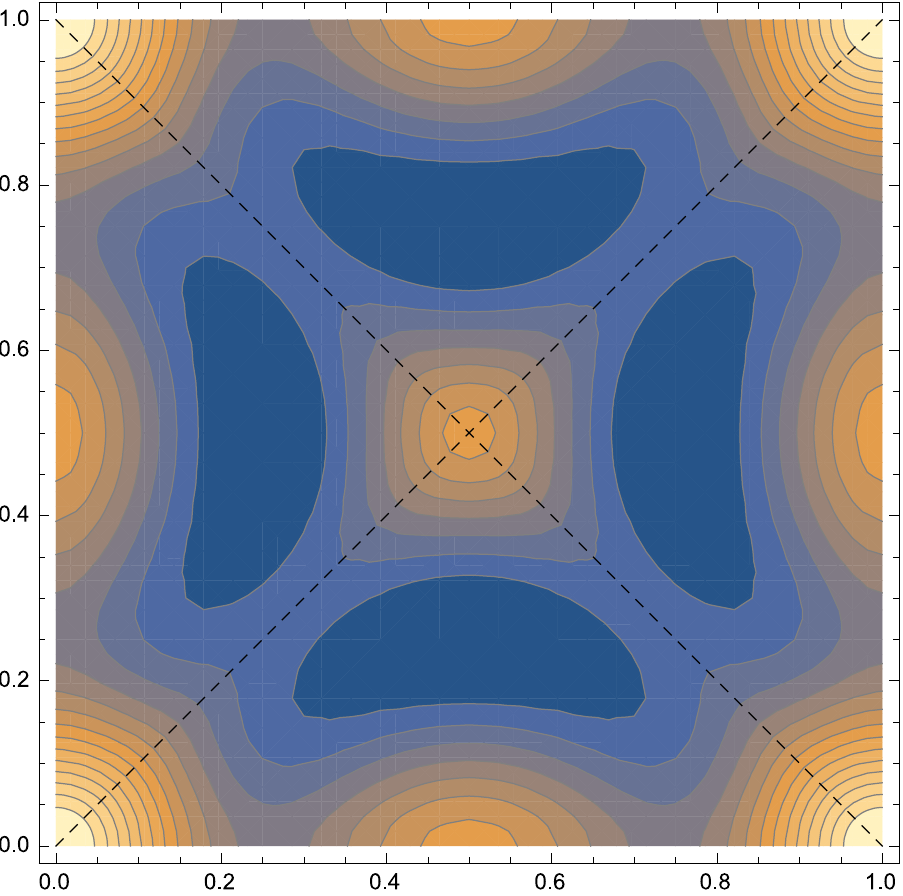,width=2.5cm}\quad\epsfig{file=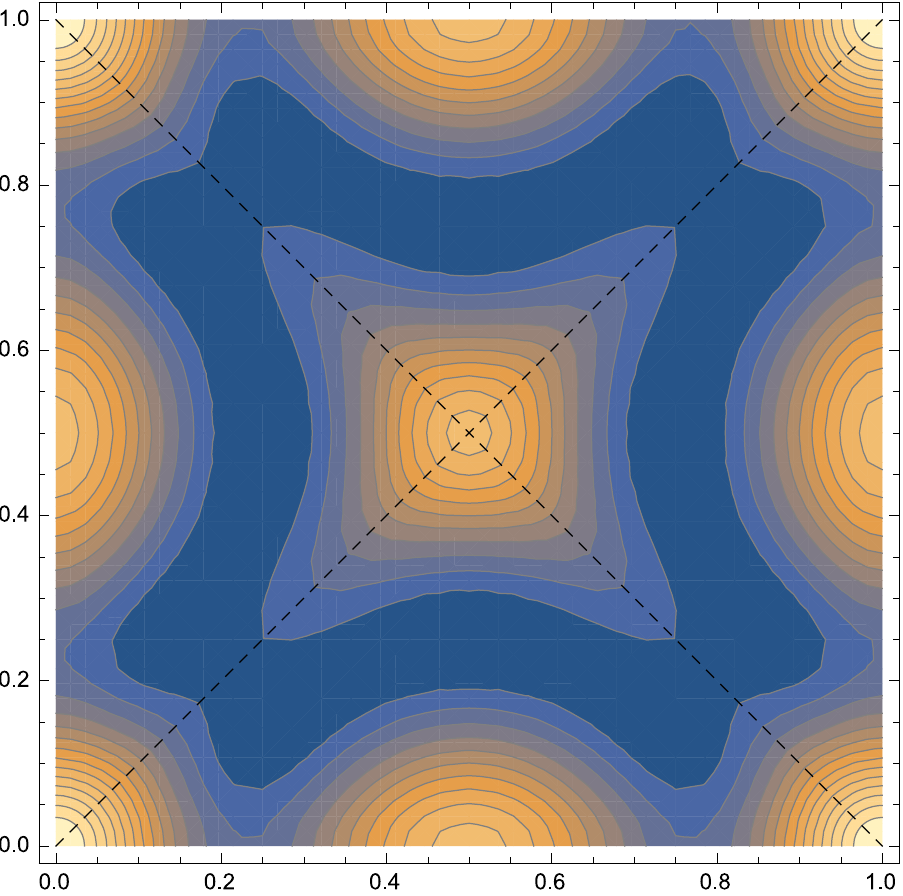,width=2.5cm}\\
\vspace{0.6cm}
\epsfig{file=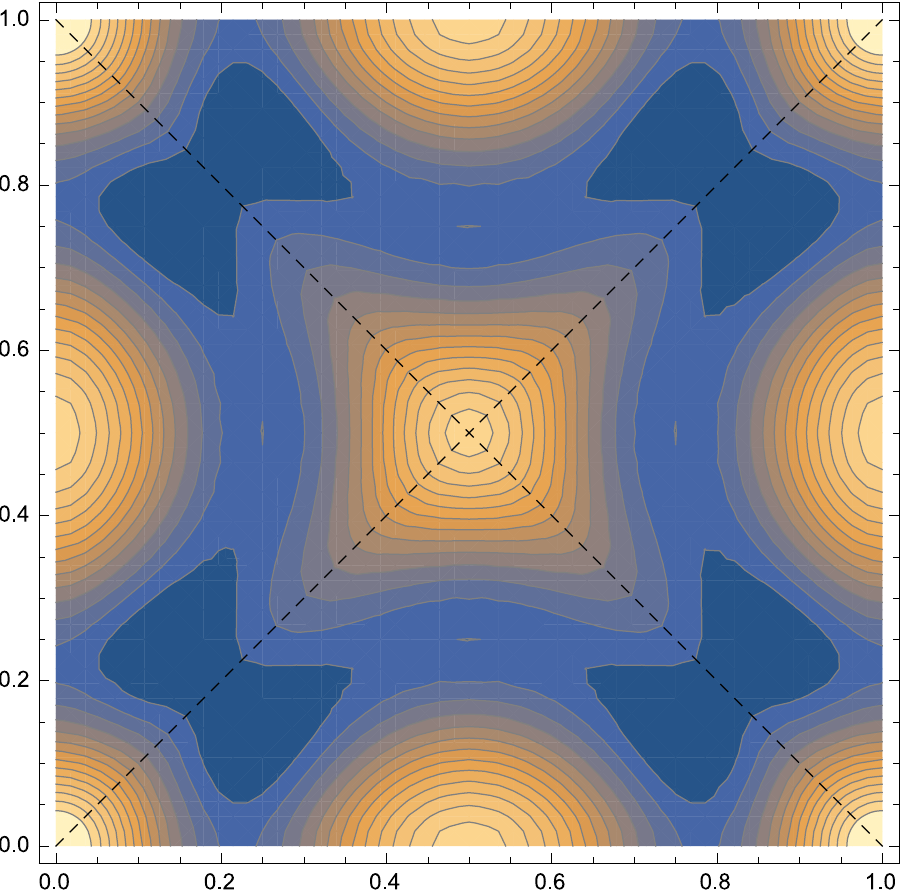,width=2.5cm}\quad\epsfig{file=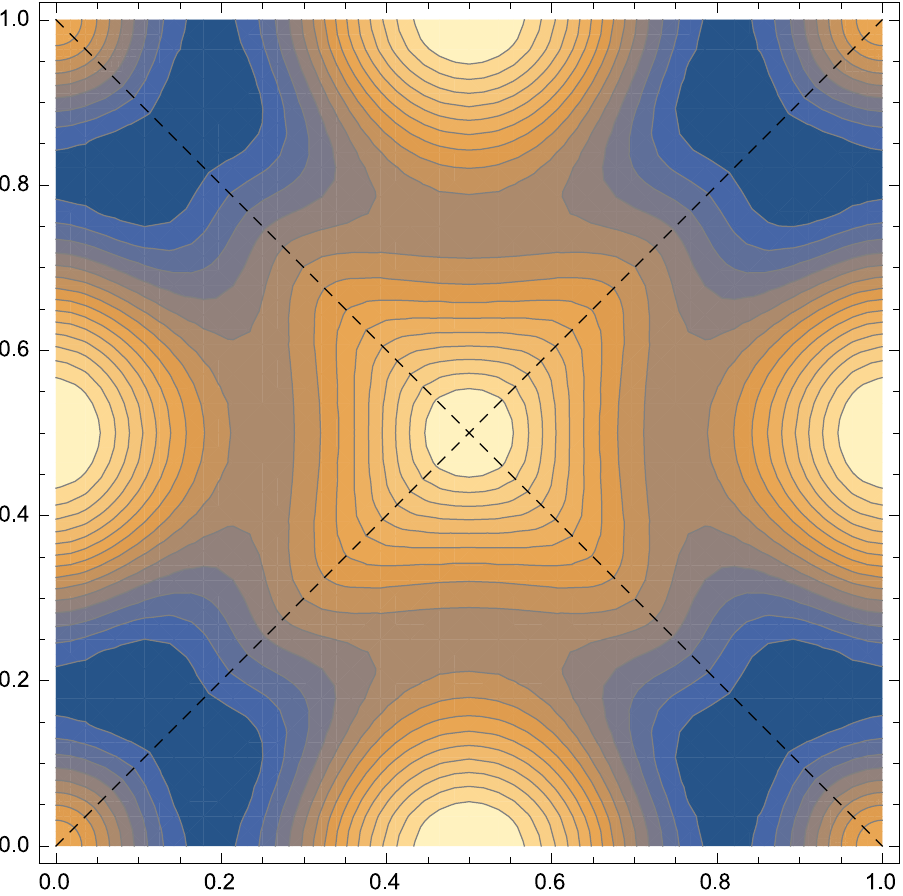,width=2.5cm}\quad\epsfig{file=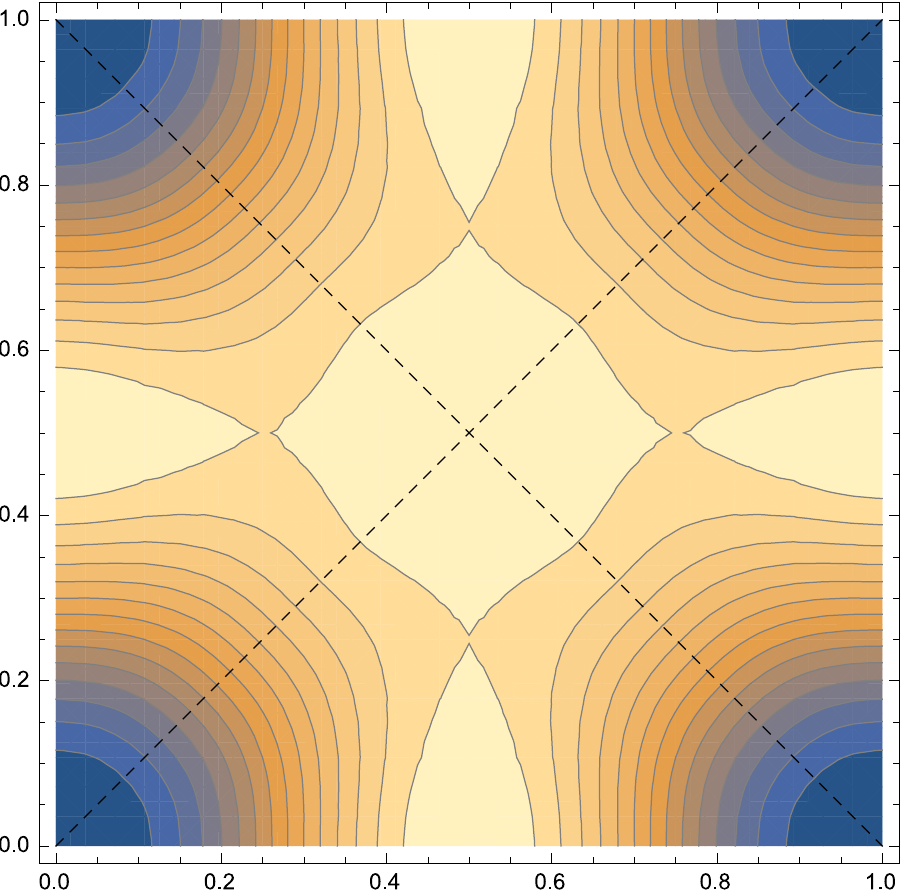,width=2.5cm}
 \caption{Contour plots of the SU($4$) potential at LO in a charge-conjugation-invariant plane. In this figure, the latter intersects four Weyl chambers separated by dashed lines. The upper (respectively lower) plots correspond to temperatures below (respectively above) the transition temperature. Darker colors correspond to regions where the potential is the deepest.}\label{fig:su4}
\end{center}
\end{figure}

As before, we decompose $r=\bar r_k\bar\alpha^{(k)}$, and we study the LO potential in the Brillouin zone $(\bar r_1,\bar r_2,\bar r_3)\in[0,1]^3$. We have 
\begin{align}
 {\cal F}_m(T,r)&=\frac{T}{\pi^2}\int_0^\infty \!\!dq\,q^2\Big\{3\ln\left(1-e^{-\beta\varepsilon_q}\right)\nn
 &+ \ln\left[1+e^{-2\beta\varepsilon_q}-2e^{-\beta\varepsilon_q}\cos(2\pi \bar r_1)\right]\nn
  &+ \ln\left[1+e^{-2\beta\varepsilon_q}-2e^{-\beta\varepsilon_q}\cos(2\pi \bar r_2)\right]\nn
    &+ \ln\left[1+e^{-2\beta\varepsilon_q}-2e^{-\beta\varepsilon_q}\cos(2\pi \bar r_3)\right]\nn
 &+ \ln\left[1+e^{-2\beta\varepsilon_q}-2e^{-\beta\varepsilon_q}\cos(2\pi(\bar r_1-\bar r_2))\right]\nn
  &+ \ln\left[1+e^{-2\beta\varepsilon_q}-2e^{-\beta\varepsilon_q}\cos(2\pi(\bar r_2-\bar r_3))\right]\nn
 &+ \ln\left[1+e^{-2\beta\varepsilon_q}-2e^{-\beta\varepsilon_q}\cos(2\pi(\bar r_3-\bar r_1))\right]\!\Big\}.
\end{align}
To simplify the presentation, we restrict the analysis to the charge-conjugation-invariant plane mentioned above, which contains one of the long edges and is orthogonal to the other one, as represented in \Fig{fig:Weyl_chambers_su4_C}. Indeed, we do not expect charge-conjugation symmetry to be spontaneously broken so that at least one of the absolute minima of the potential should belong to that plane. We have explicitly checked that this is, indeed, the case. Figure~\ref{fig:su4} shows contour plots of the potential in this plane. At low temperatures, the minimum of the potential is at the center-symmetric point (upper panels), whereas we find a $Z_4$ quadruplet of degenerate minima at high temperatures (lower panels), located pairwise in the two reflection-symmetry planes of the tetrahedron. There is thus a pair of states in which the charge-conjugation symmetry remains unbroken.\footnote{As discussed previously in the SU($3$) case, there are various {\it a priori} equivalent definitions of charge conjugation in the pure gauge theory which are the center-transformed versions of one another. In the present case, the $Z_4$-transformed versions of the charge-conjugation reflection-symmetry are the reflection itself (obtained by applying the element $-\mathds{1}$ of $Z_4$) and the other reflection-symmetry of the Weyl chamber with respect to a plane (obtained by applying the elements $\pm i\mathds{1}$ of $Z_4$). At LO, the absolute minima of the potential in the $Z_4$ broken phase sit pairwise in the two reflection-symmetry planes.} We have checked that the transition is first order with $T_c^{\rm LO}/m\approx 0.367$, again, close to the value obtained for SU($3$) at LO.\\

We compute the LO  Polyakov loop in the fundamental representation ${\bf 4}$ using \Eqn{eq:l1l}. Observing that $\mu^{(j)}-\mu^{(4)}=-\alpha^{(j)}$ for $j=1,2,3$ and using $r\cdot\alpha^{(j)}=2\pi \bar r_j$, we have
\beq
\label{eq:cla}
\ell_{\bf 4}^{(0)}(r) =\frac{e^{i\frac{\pi}{2}(\bar r_1+\bar r_2+\bar r_3)}}{4}\left(1+e^{-i2\pi \bar r_1}+e^{-i2\pi \bar r_2}+e^{-i2\pi \bar r_3}\right)\!.
\eeq
We can interpret the vanishing of $\ell^{(0)}_{\bf 4}(r)$ as the closure of a rhombus with external angles $2\pi \bar r_1$, $2\pi(\bar r_2-\bar r_1)$, $2\pi(\bar r_3-\bar r_2)$ and $2\pi(\bar r_1-\bar r_3)$ $({\rm mod}\,2\pi)$ up to permutations of $\bar r_1$, $\bar r_2$, and $\bar r_3$. This implies $2\pi \bar r_1+2\pi(\bar r_2-\bar r_1)=\pi ({\rm mod}\,2\pi)$ and $2\pi \bar r_1=2\pi(\bar r_3-\bar r_2)+({\rm mod}\,2\pi)$, that is,
\beq
\label{eq_zero_pol_1}
\bar r_2=\frac{1}{2}\,({\rm mod}\,1)\quad {\rm and} \quad \bar r_3=\bar r_1-\frac{1}{2}\,({\rm mod}\,1).
\eeq
In a given Weyl chamber, this corresponds to the segment joining the centers of the long edges of the tetrahedron, to which belong, in particular, the center-symmetric point, as exemplified in \Fig{fig:Weyl_chambers_su4}. The corresponding segments in the other chambers of \Fig{fig:Weyl_chambers_su4} are obtained by permutations of $\bar r_1$, $\bar r_2$ and $\bar r_3$. 

The fact that $\ell_{\bf 4}(r)$ vanishes on these segments is not an accident. It is because this function transforms nontrivially under center transformations which leave these segments invariant, namely, the element $-\mathds{1}$ of $Z_4$, which corresponds to a rotation of the Weyl chambers by $\pi$ around these axes. It follows that $\ell_{\bf 4}(r)=0$ on these segments to all orders. As announced, we conclude that there exist center-breaking points where some Polyakov loops still vanish \cite{Dumitru:2012fw}. Clearly, this concerns all Polyakov loops in representations of $N$-ality 1 or 3, such as ${\bf 4}$ or $\bar{\bf 4}$, which are sensitive to the element $-\mathds{1}$ of $Z_4$. Remarkably, we have found, in the explicit LO calculation given above, that such points are never absolute minima of the potential and are thus not physically realized.

Still, it is interesting to remark that, contrary to the cases of the SU($2$), SU($3$), and Sp($2$) theories, the vanishing of the Polyakov loops in the first pair of fundamental representations---here, ${\bf 4}$ and $\bar{\bf 4}$---is not sufficient {\it a priori} to conclude that center-symmetry is manifest and, thus, that all sources with nonvanishing $N$-ality are confined, even at LO. This has to do with the fact that these representations do not exhaust all possible $N$-alities in the SU($4$) case. In fact, there is another fundamental (real) representation, denoted ${\bf 6}$, with $N$-ality 2.\footnote{In general, there are $d_C$ fundamental representations \cite{Zuber}.} Clearly, the Polyakov loop function in this representation, $\ell_{\bf 6}(r)$, is invariant under the element $-\mathds{1}$ of the center and it has no reason to vanish on the above-mentioned segments. In contrast $\ell_{\bf 6}(r)$ directly probes the elements $\pm i\mathds{1}$ of $Z_4$. It reads, at LO, 
\beq
\ell_{\bf 6}^{(0)}(r)=\frac{1}{3}{\rm Re}\,(e^{i\nu^{(1)}\cdot r}+e^{i\nu^{(2)}\cdot r}+e^{i\nu^{(3)}\cdot r}),
\eeq
where $\pm\nu^{(j=1,2,3)}$ are the weights of ${\bf 6}$ with
\beq
\nu^{(1)}\!=\!\left(\begin{array}{c}
0\\
\frac{1}{\sqrt{3}}\\
\frac{1}{\sqrt{6}}
\end{array}\right)\!,\,\,
\nu^{(2)}\!=\!\frac{1}{2}\!\left(\begin{array}{c}
1\\
-\frac{1}{\sqrt{3}}\\
\frac{2}{\sqrt{6}}
\end{array}\right)\!,\,\,
\nu^{(3)}\!=\!\frac{1}{2}\!\left(\begin{array}{c}
1\\
\frac{1}{\sqrt{3}}\\
-\frac{2}{\sqrt{6}}
\end{array}\right)\!.
\eeq
After some algebra, we get
\begin{align}
\ell_{\bf 6}^{(0)}(r) & = 8\cos(\pi\bar r_1)\cos(\pi\bar r_2)\cos(\pi\bar r_3)\nonumber\\
&  -2\cos[\pi(\bar r_1+\bar r_2+\bar r_3)]\,.
\end{align}
The vanishing of the real function $\ell_{\bf 6}^{(0)}(r)$ yields one constraint and thus defines a two-dimensional surface in the Cartan space. The intersections with the lines $\ell^{(0)}_{{\bf 4}}(r)=0$, e.g., $\bar r_2=1/2$ and $\bar r_3=\bar r_1-1/2$, with $\bar r_1\in [1/2,1]$ for the Weyl chamber exemplified in \Fig{fig:Weyl_chambers_su4}, yield the equation
\beq
\cos(2\pi\bar r_1)=0.
\eeq
In the considered Weyl chamber, the unique solution $\bar r_1=3/4$ is the center-symmetric point. We conclude that the conditions $\ell^{(0)}_{{\bf 4}}(r)=\ell^{(0)}_{\bar{\bf 4}}(r)=\ell^{(0)}_{{\bf 6}}(r)=0$ uniquely specify the center-symmetric points at LO and thus the vanishing of all Polyakov loops in representations of nonzero $N$-ality. In other words, at this approximation order, confinement of static sources in the ensemble of fundamental representations (in fact in any set of representations that span all possible $N$-alities) implies confinement in all representations of nonzero $N$-ality. 

Finally, we mention that one can alternatively consider other observables that probe the various elements of the center in order to fully characterize the possible center-symmetric points \cite{Dumitru:2012fw,Pisarski:2002ji}. For instance, in the SU($4$) case, $\langle \tr\,L_{\bf 4}^2\rangle$ is invariant under the element $-\mathds{1}$ of the center but probes the elements $\pm i\mathds{1}$. This observable has the advantage on $\langle\tr\,L_{\bf 6}\rangle$, considered above, that it is simpler to compute, at least at LO, where it simply amounts to the replacement $r\to2r$:
\beq
\langle\tr\,L_{\bf 4}^2\rangle=\ell^{(0)}_{\bf 4}(2r) + {\cal O}(g^2),
\eeq
where it it understood that the right-hand side must be evaluated at the minimum of the potential. The disadvantage of this observable is that its interpretation in terms of confinement is not obvious; see, however, Ref.~\cite{Pisarski:2002ji}. Nevertheless, the condition that this observable vanishes trivially yields
\beq
\label{eq_zero_pol_2}
2\bar r_2=\frac{1}{2}\,({\rm mod}\,1)\quad {\rm and} \quad 2\bar
r_3=2\bar r_1-\frac{1}{2}\,({\rm mod}\,1)\,.  
\eeq 
up to permutations of the $\bar r_i$'s. As expected, the sets of lines \eqn{eq_zero_pol_1} and \eqn{eq_zero_pol_2} intersect only at the center-symmetric points. We provide a generalization of this discussion to the SU($N$) groups in Appendix~\ref{appsec:ploopcounting}.


\section{Conclusions}
\label{sec:conc}

We have investigated the phase structure and some aspects of the thermodynamics of non-Abelian gauge theories in a perturbative framework based on a simple massive extension of the LDW gauge-fixed action. In particular, we have computed the NLO contributions to both the background-field effective potential and the Polyakov loop in an arbitrary unitary representation for any compact and connex Lie group with a simple Lie algebra. This generalizes our previous work \cite{Reinosa:2014zta} for the SU($2$) theory. For the SU($3$) theory, we have shown that NLO corrections lead to a quantitative improvement of the value of the transition temperature---although one has to keep in mind the issue of scale setting---and to dramatic qualitative improvements in the behavior of thermodynamic observables. Both the negative pressure and entropy and the presence of a spurious singularity where the modulus of the Polyakov loop becomes exactly $1$ disappear when NLO corrections are included. This was already observed in the SU($2$) case. 

Finally, we have studied the phase structure of the Sp($2$) and SU($4$) theories, for which the relation between confinement and center symmetry is not as simple as in the SU($2$) and SU($3$) cases. Our massive description correctly predicts first-order phase transitions in $d=4$. It is worth emphasizing that, although the transition may become second order in lower dimensions, e.g., for SU($3$) and Sp($2$) in $d=3$ \cite{Holland:2003kg}, our LO results still predict a first-order transition. However, unlike the SU($2$) case, where the second-order nature of the transition (in any dimension) is simple to understand at the LO level, the fact that the first-order transition of SU($3$) and Sp($2$) in $d=4$ becomes second-order in $d=3$ is a nontrivial effect hardly describable by perturbative means. An instructive analogy can be made with the Potts model in statistical physics.\footnote{The second-order transition of the SU($3$) theory in $d=3$ is in the same universality class as the three-states Potts model in $d=2$ \cite{Svetitsky:1982gs}.} The three-states Potts model exhibits a first-order transition in $d\ge3$, which becomes continuous (second order) in $d=2$ \cite{Baxter:2000ez}. This change in the order of the phase transition cannot be captured by
a simple mean-field or perturbative analysis.

The present results corroborate our previous findings \cite{Reinosa:2014ooa,Reinosa:2014zta}, namely, that various qualitative and quantitative aspects of the confinement-deconfinement transition in YM theories can be accurately captured by a simple massive extension of the LDW gauge. The latter is motivated by the issue of Gribov ambiguities in the standard FP quantization and appears to be a better starting point than the standard FP action for an efficient perturbative description. Of course, this does not solve all the problems. Related to the fundamental issue of understanding the dynamical generation of the gluon mass is the question of $T^4$ contributions to the pressure at low temperatures. These are due to the presence of massless degrees of freedom, precisely those which are responsible for the very existence of a confined phase in the present approach; see also Ref.~\cite{Fister:2013bh}. This is a serious issue which requires further investigation. 

Another feature of the present NLO result, already present in the SU($2$) case \cite{Reinosa:2014zta}, is the rapid rise of the Polyakov loop above the transition temperature, at odds with lattice data, which exhibit a much slower increase \cite{Kaczmarek:2002mc,Dumitru:2003hp,Gupta:2007ax,Mykkanen:2012ri}. We expect that this could be resolved by taking into account renormalization group improvement and by including hard thermal loop effects at high temperature. A recent study of these aspects in the FRG approach has been performed in Ref.~\cite{Herbst:2015ona}, which yields a good comparison with the renormalized Polyakov loop computed on the lattice.

\section*{Acknowledgements}

We acknowledge interesting discussions with J.~M.~Pawlowski. We also thank PEDECIBA for supporting the visits of U. Reinosa, J. Serreau, and M. Tissier to Montevideo where part of this work was realized. 
Finally, N.~Wschebor thanks the laboratory APC of the Paris Diderot University for support during the last stages of this work.

\appendix

\section{Background field (in)dependence of the partition function}\label{appsec:mass}

The partition function $Z$ is an observable and should thus be independent of the background field, which only enters through the gauge-fixing condition; see \Eqn{eq:LDW}. In the FP quantization, this simply follows from the local nilpotent BRST symmetry of the gauge-fixed action. It is a question whether this property remains valid in a quantization scheme  where Gribov ambiguities are properly taken into account, which usually leads to an explicit (soft) breaking of the BRST symmetry. Here, we show how this property is altered by the soft breaking of the (nilpotent) BRST symmetry due to a bare gluon mass. It would be of interest to investigate this question in the (refined) Gribov-Zwanziger quantization as well.

Consider the generating functional
\beq
\label{appeq:partition}
 Z[J,\bar A]=\int{\cal D}X\, e^{-S[\bar A,X]-\int_x (J_\mu; A_\mu)}\,,
\eeq
with $X=(a_\mu,c,\cb,h)$ and $A_\mu=\bar A_\mu+a_\mu$. For a given background $\bar A_\mu$, the gauge-fixed action $S$, given by \Eqn{eq:intrinsic}, is invariant under a non-nilpotent BRST-like symmetry, whose action on the fluctuating fields is
\beq
 sa_\mu=D_\mu c\,,\quad sc=ig_0c^2,
\eeq
and 
\beq
 s\bar c=ih\,,\quad sih=m_0^2c.
\eeq
One sees that $s^2a_\mu=s^2c=0$, whereas $s^2\propto m_0^2$ in the sector $(\cb,ih)$. 
In the following it is useful to explicitly separate the gluon mass term in the action. We write
\beq
 S=S_{\rm YM}+S_{\rm FP} + S_{m_0^2},
\eeq
where $S_{\rm YM}$ is the Yang-Mills action \eqn{eq:YM}, 
\beq
 S_{\rm FP}=-\int_x\Big\{\!\left(\bar D_\mu\cb\,; D_\mu c\right)+\left(ih\,;\bar D_\mu a_\mu\right)\!\Big\}
\eeq 
is the FP action corresponding to the LDW gauge and 
\beq
 S_{m_0^2}=-\frac{m_0^2}{2}\int_x\left(a_\mu;a_\mu\right)
\eeq
is the bare mass term [here, we employ rescaled fields, i.e., $X\to g_0X$, as compared to \Eqn{eq:intrinsic}].

Let us analyze $Z[J,\bar A+\alpha]$ for an arbitrary variation $\alpha_\mu$ of the background field. Under the functional integral \eqn{appeq:partition}, we change $a_\mu\to a_\mu-\alpha_\mu$ so that $A_\mu$ remains unchanged. The source term and $S_{\rm YM}$, which only depend on $A_\mu=\bar A_\mu+a_\mu$, are unchanged. The variation of the FP contribution $S_{\rm FP}$ is best written by noticing that
\beq
 S_{\rm FP}=s\int_x\left(a_\mu\,; D_\mu\cb\right),
\eeq
with $s$ is the BRST symmetry operation defined above. The covariant derivative on the right-hand side only involves $A_\mu$ and the variation of $S_{\rm FP}$ under the change $\bar A_\mu\to\bar A_\mu+\alpha_\mu$ is, thus, simply
\beq
 S_{\rm FP}\to S_{\rm FP}-\int_x\left(\alpha_\mu \,; sD_\mu\cb\right).
\eeq
Finally, the mass term is changed to
\beq
 S_{m_0^2}\to S_{m_0^2}+m_0^2\int_x\left(\alpha_\mu\,; a_\mu\right)-\frac{m_0^2}{2}\int_x\left(\alpha_\mu\,;\alpha_\mu\right).
\eeq
Altogether, we have
\beq
 \frac{Z[J,\bar A+\alpha]}{Z[J,\bar A]}=\left<\exp\int_x\left(\alpha_\mu; sD_\mu\cb-m_0^2 a_\mu+\frac{m_0^2}{2}\alpha_\mu\right)\right>_{\!J},
\eeq
where the bracket denotes an average in the presence of the source $J_\mu$, that is, with the action $S+\int_x (J_\mu\,; A_\mu)$. Considering an infinitesimal variation, we get the symmetry identity (in the following, we work in a standard Cartesian basis $\{it^a\}$)
\beq
 \frac{\delta \ln Z[J,\bar A]}{\delta \bar A_\mu^a}=\left<s(D_\mu \cb)^a+m_0^2 a_\mu^a\right>_{\!J}\,.
\eeq

The partition function is obtained from \Eqn{appeq:partition} at vanishing source:
\beq
 Z[\bar A]=Z[J=0,\bar A].
\eeq
Because $s$ is a symmetry of the theory at vanishing source, we have $\left<sD_\mu \cb\right>_{\!J=0}=0$ and, thus,\footnote{Notice that, using the LDW gauge condition $\bar D_\mu a_\mu=0$, one recovers from \Eqn{appeq:funcder} that $\bar D_\mu(\delta\ln Z[\bar A]/\delta\bar A_\mu)=0$, which is nothing but the statement of the invariance of $Z[\bar A]$ under the formal gauge transformation $\bar A\to \bar A^U$.
}
\beq
\label{appeq:funcder}
 \frac{\delta \ln Z[\bar A]}{\delta \bar A_\mu^a}=m_0^2\left< a_\mu^a\right>_{\!J=0}
\eeq
where, by definition, $\left< a_\mu^a\right>_{\!J=0}=A_{{\rm min},\mu}^a(\bar A)-\bar A_\mu^a$, with the notations of Sec.~\ref{subsec:LdW}. We recover the fact that, in the FP quantization, which ignores the issue of Gribov ambiguities and which corresponds to taking $m_0=0$, the partition function is independent of the background, as it should be. In the present massive theory, we conclude that the functional derivative \eqn{appeq:funcder} vanishes when evaluated at a self-consistent background field $\bar A=\bar A_s$, defined as $A_{\rm min}(\bar A_s)=\bar A_s$:
\beq
\label{appeq:egalzero}
 \left.\frac{\delta \ln Z[\bar A]}{\delta \bar A_\mu^a}\right|_{\bar A_s}=0.
\eeq

We see that, although the soft BRST breaking mass term spoils the exact background field independence of the partition function,\footnote{Interestingly, in the context of the gauge-fixing procedure put forward in Ref.~\cite{Serreau:2012cg} the bare mass must be sent to zero together with the bare coupling in the continuum limit, hence, possibly implying the background field independence of the partition function. This needs to be investigated further.}, the latter is recovered locally for self-consistent backgrounds. We emphasize though that this is not yet enough as certain properties used in background field techniques rely on the exact background field independence of the partition function; see the discussion in Sec.~\ref{subsec:LdW}. These aspects require further investigation.

Incidentally, \Eqn{appeq:egalzero} implies that a self-consistent background, if it exists, is necessarily an extremum of the functional $\tilde \Gamma[\bar A]$, as announced in Sec.~\ref{subsec:LdW}. The proof goes as follows. The effective action in the presence of the background is defined as the usual Legendre transform 
\beq
 \Gamma[A,\bar A]=-\ln Z[J(A,\bar A),\bar A]-\int_x\left(J_\mu(A,\bar A)\,;A_\mu\right),
\eeq 
with the current $J_\mu(A,\bar A)$ defined by the implicit relation
\beq 
 \left.\frac{\delta\ln Z[J,\bar A]}{\delta J^a_\mu}\right|_{J(A,\bar A)}=A^a_\mu.
\eeq 
It follows that the functional $\tilde\Gamma[\bar A]=\Gamma[\bar A,\bar A]$ satisfies
\beq
\label{appeq:from}
 \frac{\delta \tilde \Gamma[\bar A]}{\delta \bar A_\mu^a}=J_\mu^a(\bar A,\bar A)-\left.\frac{\delta\ln Z[J,\bar A]}{\delta \bar A^a_\mu}\right|_{J(\bar A,\bar A)}.
\eeq
Now, from the relation
\beq
 \frac{\delta\Gamma[A,\bar A]}{\delta A^a_\mu}=J^a_\mu(A,\bar A),
\eeq
one has, by definition, 
\beq
 J^a_\mu(A_{\rm min}(\bar A),\bar A)=0\,,\,\,\forall\,\bar A.
\eeq
For a self-consistent background $\bar A_s$, defined as $A_{\rm min}(\bar A_s)=\bar A_s$, one thus has $J(\bar A_s,\bar A_s)=J(A_{\rm min}(\bar A_s),\bar A_s)=0$, and, using Eqs.~\eqn{appeq:from} and \eqn{appeq:egalzero},
\beq 
 \left.\frac{\delta \tilde \Gamma[\bar A]}{\delta \bar A_\mu^a}\right|_{\bar A_s}=-\left.\frac{\delta\ln Z[J=0,\bar A]}{\delta \bar A^a_\mu}\right|_{\bar A_s}=0.
\eeq
As announced, a self-consistent background is always an extremum of $\tilde \Gamma[\bar A]$.\\

\section{Complexified algebra\\ and canonical bases}\label{appsec:canonical}

For completeness, we here gather some known material on how the canonical or Cartan-Weyl bases of the algebra are introduced. We refer for instance to Ref.~\cite{Zuber} for more details.

The original algebra ${\cal G}$ is real. We can extend it to a complex algebra by considering ${\cal G}\times {\cal G}\equiv{\cal G}\oplus i{\cal G}$. A general element of the complexified algebra can be written as $Z=(X,Y)\equiv X+iY$, where $X,Y\in{\cal G}$, and we define its complex conjugate as $Z^*=X-iY$ for later use. An element of ${\cal G}$ can be seen as an element of the complexified algebra through the identification $(X,0)\equiv X$.  The elements of ${\cal G}$ are thus real in the sense that $X^*=X$, whereas those of $i{\cal G}$ are purely imaginary in the sense that $(iY)^*=-iY$. 

The Lie bracket is extended on the complexified algebra as
\beq
[X+iU,Y+iV]=[X,Y]+i[X,V]+i[U,Y]-[U,V]
\eeq
and the invariant form as
\beq
(X+iU;Y+iV)=(X;Y)+i(X;V)+i(U;Y)-(U;V)\,.
\eeq
By construction
\beq
[Z_1,Z_2]^*=[Z_1^*,Z_2^*]
\eeq
and
\beq
(Z_1;Z_2)^*=(Z_1^*;Z_2^*)\,.
\eeq
These extensions have the same properties as the original definitions on the real algebra ${\cal G}$. In particular, the extended form is bilinear symmetric, that is $(Z_1;Z_2)=(Z_2;Z_1)$, and obeys Eq.~(\ref{eq:cyclic}). Even though it is convenient to work with this particular extension of the invariant form, we shall also consider the sesquilinear symmetric form
\beq
\langle Z_1;Z_2\rangle=(Z_1^*;Z_2)\,,
\eeq
which obeys $\langle Z_1;Z_2\rangle=\langle Z_2;Z_1\rangle^*=\langle Z_2^*;Z_1^*\rangle$.  An important difference with respect to the bilinear symmetric extension $(\,\,;\,)$ is that\footnote{We note also that the sesquilinear form $\langle\,\,;\,\rangle$ is nondegenerate and negative definite, whereas the bilinear symmetric form $(\,\,;\,)$ is only non-degenerate but has no definite sign on the complexified algebra. In particular, it is positive definite on $i{\cal G}$.}
\beq
\langle Z_1;[Z_2,Z_3]\rangle=\langle Z_2^*;[Z_3,Z_1^*]\rangle=\langle Z_3^*;[Z_1^*,Z_2]\rangle\,.
\eeq
It follows in particular that, for a purely imaginary element $iY\in i{\cal G}$ of the complexified algebra,
\beq\langle Z_1;[iY,Z_3]\rangle =\langle [iY,Z_1];Z_3\rangle\,,
\eeq
and thus the operator ${\rm ad}_{iY}\equiv [iY,\,]$, acting on ${\cal G}\oplus i{\cal G}$, is Hermitian with respect to the sesquilinear symmetric form $\langle\,\,;\,\rangle$.\\

Let us now use these considerations to solve the set of equations ${\rm ad}_{iH_j}(X)\equiv[iH_j,X]=\lambda_{j,X} X$, where $\{iH_j\}$ is a basis of a Cartan subalgebra of the (real) algebra ${\cal G}$. From the above discussion we know that the operators ${\rm ad}_{H_j}$ are Hermitian with respect to $\langle\,;\,\rangle$. They can thus be diagonalized. Moreover, since $[{\rm ad}_{H_j},{\rm ad}_{H_k}]={\rm ad}_{[H_j,H_k]}=0$, one can find a common basis of diagonalization. Since $[H_j,H_k]=0$, each $H_j$ is a common eigenvector to all the ${\rm ad}_{H_k}$ with corresponding eigenvalue $0$. The common basis of diagonalization can thus be chosen of the form $\{iH_j,iE_\alpha\}$, such that $\forall j$,
\beq
[H_j,E_\alpha]=\alpha_j E_\alpha\,.
\eeq 
Each label $\alpha$ denotes a nonzero real-valued vector called a root. Its number of components is equal to the dimension of the Cartan subalgebra. It is useful to think of $j$ as labeling a quantum number corresponding to the ``observable'' (Hermitian operator) ${\rm ad}_{H_j}\cdot=[H_j,\cdot]$. Each (color) state $E_\alpha$ of the system is characterized by a collection of quantum numbers $\alpha_j$ that can be represented by a vector $\alpha$ in the Cartan space. The states $H_j$ are particular in the sense that their quantum numbers are all equal to zero and are thus represented by the null vector.

It can be shown that the eigenspace associated to a given root $\alpha$ is one-dimensional \cite{Zuber}. This has various important consequences. First of all, by considering the complex conjugate of the above commutation relation and using that $H_j\in i{\cal G}$, we obtain $[H_j,E_\alpha^*]=-\alpha_j E_\alpha^*$. It follows that, for any root $\alpha$, $-\alpha$ is another root and $E_{-\alpha}$ is proportional to $E^*_\alpha$. In what follows we split the roots into two subsets related by $\alpha\to -\alpha$ and we choose normalizations such that $E_{-\alpha}=-E^*_\alpha$.\footnote{In the case of a compact Lie group, we can assume that the elements of ${\cal G}$ are anti-Hermitian matrices. If $E_\alpha=X+iY$, we have $E_\alpha^\dagger=-X+iY=-E_\alpha^*=E_{-\alpha}$.} We also have
\bea
\label{appeq:appeq}
[H_j,[E_\alpha,E_\beta]] & = & [E_\alpha,[H_j,E_\beta]]-[E_\beta,[H_j,E_\alpha]]\nonumber\\
& = & (\alpha+\beta)_j [E_{\alpha},E_{\beta}]\,.
\eea
If $\alpha+\beta$ is a root, we conclude that $[E_\alpha,E_\beta]=N_{\alpha\beta}E_{\alpha+\beta}$ for some coefficients $N_{\alpha\beta}$. If $\alpha+\beta=0$, \Eqn{appeq:appeq} implies $[E_\alpha,E_{-\alpha}]=u_j H_j$ for some coefficients $u_j$. Finally, if $\alpha+\beta$ is neither a root nor a zero, we have $[E_\alpha,E_\beta]=0$. 

Without loss of generality, we can choose the $H_j$ such that $(H_j;H_k)=-\langle H_j;H_k\rangle=\delta_{jk}$ (recall that the form $\langle\,\,;\,\rangle$ is negative definite). Then
\bea
u_j & = & (H_j;[E_\alpha,E_{-\alpha}])\nonumber\\
& = & (E_{-\alpha};[H_j,E_\alpha])=\alpha_j(E_{-\alpha};E_\alpha)\,.
\eea
With the normalization $(E_{-\alpha};E_\alpha)=-\langle E_{\alpha};E_\alpha\rangle=1$, we have $u_j=\alpha_j$. The values of the coefficients $N_{\alpha\beta}$ can also be computed; see, for instance, Ref.~\cite{Zuber}. We note that, by complex conjugation of $[E_\alpha,E_\beta]=N_{\alpha\beta}E_{\alpha+\beta}$ and, since $E^*_\alpha=-E_{-\alpha}$, we have $[E_{-\alpha},E_{-\beta}]=-N_{\alpha\beta}^*E_{-\alpha-\beta}$. It follows that $N_{\alpha\beta}^*=-N_{(-\alpha)(-\beta)}$. In particular, this implies that $f_{\kappa\lambda\tau}^*=-f_{(-\kappa)(-\lambda)(-\tau)}$ (recall that $[t_\lambda,t_\tau]=f_{\lambda\tau\kappa}t_{-\kappa}$).\\

Let us finally discuss the components of the form $(\,;\,)$ in the basis $\{iH_j,iE_\alpha\}$, that is, the various scalar products of the elements of the basis with respect to this form. We have
\beq
0=(E_\alpha;[H_j,H_k])=(H_j;[H_k,E_{\alpha}])=\alpha_k (H_j;E_{\alpha})\,.
\eeq
Because this is true for any $j$ and $k$ and because the roots are nonzero, it follows that $(H_j;E_\alpha)=\langle H_j;E_\alpha\rangle=0$. Similarly,
\bea
(H_j;[E_\alpha,E_\beta]) & = & (E_\beta;[H_j,E_\alpha])=\alpha_j (E_\alpha;E_\beta)\nonumber\\
& = & (E_\alpha,[E_\beta,H_j])=-\beta_j (E_{\alpha};E_\beta)\,.\nonumber\\
\eea
We conclude that, with our choice of normalization, $(E_\alpha,E_\beta)=\delta_{\alpha+\beta,0}$ or, equivalently, $\langle E_\alpha;E_\beta\rangle=-\delta_{\alpha,\beta}$. These results could have been anticipated since the $E_\alpha$ are eigenvectors of Hermitian operators and different $E_\alpha$ correspond to different eigenvalues. They are thus orthonormal with respect to the sesquilinear symmetric form $\langle\,\,;\,\rangle$.

We note finally that if $X=i(X_jH_j+X_{\alpha}E_\alpha)$ is an element of ${\cal G}$, $X_j=(iH_j;X)$ is real because $iH_j\in {\cal G}$, whereas our choice $E_{-\alpha}=-E_\alpha^*$ implies  $X_{-\alpha}=-(iE_{-\alpha};X)=(iE_\alpha^*;X)=-(iE_\alpha;X)^*=X_{\alpha}^*$.\\

\section{Casimirs}\label{appsec:casimir}

For a given irreducible unitary representation $t_\kappa\mapsto t^R_\kappa$, the Casimir operator is defined as
\beq
\sum_\kappa t^R_{-\kappa}t^R_\kappa=\sum_j t^R_{0^{(j)}} t^R_{0^{(j)}}+\sum_\alpha t^R_{-\alpha} t^R_{\alpha}=C_R\mathds{1}\,.
\eeq
Introducing the weights of the representation (see Sec.~\ref{sec:pol}), we can consider the matrix element $\langle \mu,a|...|\mu,a\rangle$ of the previous identity and we obtain
\beq
\label{appeq:cellela}
C_R = \mu^2+\sum_{\alpha,b} \left|\langle\mu+\alpha,b|t_\alpha^R|\mu,a\rangle\right|^2\quad\forall \,\mu,a\,,
\eeq
with $\mu^2=\sum_j \mu_j^2$. For the adjoint representation, the generators act through the commutator and the weights are either the roots or the zeros so that, choosing $\mu=0^{(j)}$, we obtain
\beq
C_{\rm ad}=\frac{1}{d_C}\sum_\alpha \alpha^2=\sum_\alpha \alpha_j^2\quad\forall j\,.
\eeq
From the relation $[t^R_\kappa,t^R_\lambda]=f_{\kappa\lambda\tau}t^R_{-\tau}$, we also have
\beq
C_{\rm ad} = \sum_{\kappa,\lambda} f_{(-\kappa)\tau(-\lambda)}f_{\kappa\lambda(-\tau)} = \sum_{\kappa,\lambda} |f_{\kappa\lambda\tau}|^2=\sum_{\kappa,\lambda} {\cal C}_{\kappa\lambda\tau}\,.
\eeq

\section{Background field potential at NLO}\label{appsec:final}
Starting from Eqs.~(\ref{eq:pot2loop})-(\ref{eq:pot2loop_fin}), we follow the steps of Ref.~\cite{Reinosa:2014zta} for computing Matsubara sums and angular momentum integrals in the SU($2$) case. We get

\begin{widetext}
\bea
V^{(2)} & = & m^2 C_G \sum_\kappa J_m^\kappa(1n)+{\sum_{\alpha}}^*\alpha^2\left\{\frac{15g^2}{8}\left(U^0 V^{\alpha}+U^{\alpha}V^0+U^{\alpha}V^{\alpha}\right)+\frac{21g^2}{4m^2}\tilde U^{\alpha} \tilde V^{\alpha}\right.\nonumber\\
&  & \hspace{1.0cm}+\frac{99g^2m^2}{256\pi^4}\int_0^\infty dq\,\frac{q}{\varepsilon_q}\int_0^\infty dk\,\frac{k}{\varepsilon_k} \Big[n_{\varepsilon_q+\varepsilon_k}+2(n_{\varepsilon_q+\varepsilon_k}+n_{\varepsilon_k}){\rm Re}\,n_{\varepsilon_q-i \hat r\cdot\alpha}\Big]\nonumber\\
&  & \hspace{6.0cm}\times\,{\rm Re}\,\ln\frac{(\varepsilon_q+\varepsilon_k+i0^+)^2-(\varepsilon_{k+q})^2}{(\varepsilon_q+\varepsilon_k+i0^+)^2-(\varepsilon_{k-q})^2}\nonumber\\
&  & \hspace{1.0cm}+\frac{99g^2m^2}{256\pi^4}\int_0^\infty dq\,\frac{q}{\varepsilon_q}\int_0^\infty dk\,\frac{k}{\varepsilon_k} \Big[2(n_{\varepsilon_k-\varepsilon_q}+n_{\varepsilon_k}){\rm Re}\,n_{\varepsilon_q-i \hat r\cdot\alpha}\Big]\nonumber\\
&  & \hspace{6.0cm}\times\,{\rm Re}\,\ln\frac{(\varepsilon_q-\varepsilon_k+i0^+)^2-(\varepsilon_{k+q})^2}{(\varepsilon_q-\varepsilon_k+i0^+)^2-(\varepsilon_{k-q})^2}\nonumber\\
&  & \hspace{1.0cm}+\frac{g^2m^2}{256\pi^4}\int_0^\infty dq\,\int_0^\infty dk\,\Big[n_{q+k}+2(n_{q+k}+n_{k}){\rm Re}\,n_{q-i \hat r\cdot\alpha}\Big]\nonumber\\
&  & \hspace{6.0cm}\times\,{\rm Re}\,\ln\frac{(q+k+i0^+)^2-(\varepsilon_{k+q})^2}{(q+k+i0^+)^2-(\varepsilon_{k-q})^2}\nonumber\\
&  & \hspace{1.0cm}+\frac{g^2m^2}{256\pi^4}\int_0^\infty dq\,\int_0^\infty dk\,\Big[2(n_{k-q}+n_{k}){\rm Re}\,n_{q-i \hat r\cdot\alpha}\Big]\nonumber\\
&  & \hspace{6.0cm}\times\,{\rm Re}\,\ln\frac{(q-k+i0^+)^2-(\varepsilon_{k+q})^2}{(q-k+i0^+)^2-(\varepsilon_{k-q})^2}\nonumber\\
&  & \hspace{1.0cm}+\frac{g^2m^2}{128\pi^4}\int_0^\infty dq\,\frac{q}{\varepsilon_q}\int_0^\infty dk\,\Big[n_{\varepsilon_q+k}+(n_{\varepsilon_q+k}+n_{k}){\rm Re}\,n_{\varepsilon_q-i \hat r\cdot\alpha}+(n_{\varepsilon_q+k}+n_{\varepsilon_q}){\rm Re}\,n_{k-i \hat r\cdot\alpha}\Big]\nonumber\\
& &\hspace{7.0cm}\times\,{\rm Re}\,\ln\frac{(\varepsilon_q+k+i0^+)^2-(k+q)^2}{(\varepsilon_q+k+i0^+)^2-(k-q)^2}\nonumber\\
&  & \hspace{1.0cm}+\frac{g^2m^2}{128\pi^4}\int_0^\infty dq\,\frac{q}{\varepsilon_q}\int_0^\infty dk\,\Big[(n_{k-\varepsilon_q}+n_{k}){\rm Re}\,n_{\varepsilon_q-i \hat r\cdot\alpha}+(n_{\varepsilon_q-k}+n_{\varepsilon_q}){\rm Re}\,n_{k-i \hat r\cdot\alpha}\Big]\nonumber\\
& & \left.\hspace{7.0cm}\times\,{\rm Re}\,\ln\frac{(\varepsilon_q-k+i0^+)^2-(k+q)^2}{(\varepsilon_q-k+i0^+)^2-(k-q)^2}\right\}\nonumber\\
& + & {\sum}^*_{\alpha,\beta,\gamma}2|N_{\alpha\beta}|^2\left\{\frac{15g^2}{16}(U^{\alpha} V^{\beta}+U^{\beta} V^{\gamma}+U^{\gamma} V^{\alpha}+U^{\beta} V^{\alpha}+U^{\alpha} V^{\gamma}+U^{\gamma} V^{\beta})\right.\nonumber\\
&  & \hspace{1.0cm}-\frac{21g^2}{8m^2}(\tilde U^{\alpha} \tilde V^{\beta}+\tilde U^{\beta} \tilde V^{\gamma}+\tilde U^{\gamma} \tilde V^{\alpha}+\tilde U^{\beta} \tilde V^{\alpha}+\tilde U^{\alpha} \tilde V^{\gamma}+\tilde U^{\gamma} \tilde V^{\beta})\nonumber\\
& & \hspace{1.0cm}+\frac{99g^2m^2}{256\pi^4}\sum_{s=\pm 1}\int_0^\infty dq\,\frac{q}{\varepsilon_q}\int_0^\infty dk\,\frac{k}{\varepsilon_k} {\rm Re}\,(n_{\varepsilon_q-i\hat r\cdot\alpha}n_{\varepsilon_k-is\hat r\cdot\beta}+n_{\varepsilon_q-i\hat r\cdot\beta}n_{\varepsilon_k-is\hat r\cdot\gamma}+n_{\varepsilon_q-i\hat r\cdot\gamma}n_{\varepsilon_k-is\hat r\cdot\alpha})\nonumber\\
& & \hspace{5.0cm}\times\,{\rm Re}\,\ln\frac{(\varepsilon_q+s\varepsilon_k+i\epsilon)^2-(\varepsilon_{k+q})^2}{(\varepsilon_q+s\varepsilon_k+i\epsilon)^2-(\varepsilon_{k-q})^2}\nonumber
\eea
\bea
&  & \hspace{1.0cm}+\frac{g^2m^2}{256\pi^4}\sum_{s=\pm 1}\int_0^\infty dq\,\int_0^\infty dk\,{\rm Re}\,(n_{q-i\hat r\cdot\alpha}n_{k-is\hat r\cdot\beta}+n_{q-i\hat r\cdot\beta}n_{k-is\hat r\cdot\gamma}+n_{q-i\hat r\cdot\gamma}n_{k-is\hat r\cdot\alpha})\nonumber\\
& & \hspace{5.0cm}\times\,{\rm Re}\,\ln\frac{(q+sk+i\epsilon)^2-(\varepsilon_{k+q})^2}{(q+sk+i\epsilon)^2-(\varepsilon_{k-q})^2}\nonumber\\
& & \hspace{1.0cm}+\frac{g^2m^2}{256\pi^4}\sum_{s=\pm 1}\int_0^\infty dq\,\frac{q}{\varepsilon_q}\int_0^\infty dk\,{\rm Re}\,(n_{\varepsilon_q-i\hat r\cdot\alpha}n_{k-is\hat r\cdot\beta}+n_{\varepsilon_q-i\hat r\cdot\beta}n_{k-is\hat r\cdot\gamma}+n_{\varepsilon_q-i\hat r\cdot\gamma}n_{k-is\hat r\cdot\alpha}\nonumber\\
& & \hspace{6.3cm}+\,n_{\varepsilon_q-i\hat r\cdot\beta}n_{k-is\hat r\cdot\alpha}+n_{\varepsilon_q-i\hat r\cdot\gamma}n_{k-is\hat r\cdot\beta}+n_{\varepsilon_q-i\hat r\cdot\alpha}n_{k-is\hat r\cdot\gamma})\nonumber\\
& & \hspace{5.0cm}\left.\times\,{\rm Re}\,\ln\frac{(\varepsilon_q+sk+i\epsilon)^2-(k+q)^2}{(\varepsilon_q+sk+i\epsilon)^2-(k-q)^2}\right\},
\eea
where $\sum^*_{\alpha}$ and ${\sum}^*_{\alpha,\beta,\gamma}$ denote, respectively, the sum over the pairs of roots $(\alpha,-\alpha)$ and the sum over the pairs of triplets of roots $((\alpha,\beta,\gamma),(-\alpha,-\beta,-\gamma))$ such that $\alpha+\beta+\gamma=0$. The first seven lines of the above formula are SU($2$)-like. The remaining five lines only appear when there is more than one pair of charged color states.\\

We can obtain the high-temperature limit by following the same approach as in Ref.~\cite{Reinosa:2014zta}. We arrive at
\bea
v_\infty(r) & \equiv & \lim_{T\to\infty}\frac{V(r,T)}{T^4}\nonumber\\
& = & -\frac{1}{3\pi^2}\int_0^\infty dx\,x^3\Big(d_C\,\tilde n_x+2\sum_{\alpha>0}{\rm Re}\,\tilde n_{x-ir\cdot\alpha}\Big)\nonumber\\
& + & \frac{g^2}{2\pi^4}{\sum_{\alpha}}^*\alpha^2\int_0^\infty \!\!\! dx\,x({\rm Re}\,\tilde n_{x-ir\cdot\alpha}+2\tilde n_x)\int_0^\infty \!\!\!dy\,y{\rm Re}\,\tilde n_{y-ir\cdot\alpha}\nonumber\\
& + & \frac{g^2}{\pi^4}{\sum_{\alpha}}^*\alpha^2\int_0^\infty\!\!\! dx\,x^2{\rm Im}\left.\frac{\partial\tilde n_{\sqrt{x^2+z}-ir\cdot\alpha}}{\partial z}\right|_{z=0}\int_0^\infty\!\!\!dy\,y^2 {\rm Im}\,\tilde n_{y-ir\cdot\alpha}\nonumber\\
& + & \frac{g^2}{\pi^4}{\sum}^*_{\alpha,\beta,\gamma}|N_{\alpha\beta}|^2\left(\int_0^\infty \!\!\! dx\,x{\rm Re}\,\tilde n_{x-ir\cdot\alpha}\int_0^\infty \!\!\!dy\,y{\rm Re}\,\tilde n_{y-ir\cdot\beta}+\mbox{cyclic perm.}\right)\nonumber\\
& - & \frac{g^2}{\pi^4}{\sum}^*_{\alpha,\beta,\gamma}|N_{\alpha\beta}|^2\left(\int_0^\infty\!\!\! dx\,x^2{\rm Im}\left.\frac{\partial\tilde n_{\sqrt{x^2+z}-ir\cdot\alpha}}{\partial z}\right|_{z=0}\int_0^\infty\!\!\!dy\,y^2 {\rm Im}\,\tilde n_{y-ir\cdot\beta}+\mbox{perm.}\right)\nonumber\\
\eea
and
\bea
v_\infty(r) & \equiv & -\frac{1}{3\pi^2}\Big(d_C\,P_4(0)+2\sum_{\alpha>0}P_4(\{r\cdot\alpha\})\Big)\nonumber\\
& + & \frac{g^2}{2\pi^4}{\sum_{\alpha}}^*\alpha^2\Big(\Big(P_2(\{r\cdot\alpha\})+2P_2(0)\Big)P_2(\{r\cdot\alpha\})+P'_2(\{r\cdot\alpha\})P_3(\{r\cdot\alpha\})\Big)\nonumber\\
& + & \frac{g^2}{\pi^4}{\sum}^*_{\alpha,\beta,\gamma}|N_{\alpha\beta}|^2\Big(P_2(\{r\cdot\alpha\})P_2(\{r\cdot\beta\})+P_2(\{r\cdot\beta\})P_2(\{r\cdot\gamma\})+P_2(\{r\cdot\gamma\})P_2(\{r\cdot\alpha\})\Big)\nonumber\\
& - & \frac{g^2}{2\pi^4}{\sum}^*_{\alpha,\beta,\gamma}|N_{\alpha\beta}|^2\Big(P'_2(\{r\cdot\alpha\})P_3(\{r\cdot\beta\})+P'_2(\{r\cdot\beta\})P_3(\{r\cdot\gamma\})+P'_2(\{r\cdot\gamma\})P_3(\{r\cdot\alpha\})\nonumber\\
& & \hspace{3.0cm}+\,P'_2(\{r\cdot\beta\})P_3(\{r\cdot\alpha\})+P'_2(\{r\cdot\gamma\})P_3(\{r\cdot\beta\})+P'_2(\{r\cdot\alpha\})P_3(\{r\cdot\gamma\})\Big),
\eea
with $\{r\cdot\alpha\}$ the remainder in the (Euclidean) division of $r\cdot\alpha$ by $2\pi$, and
\begin{align}
P_2(x) & =  \frac{(\pi-x)^2}{4}-\frac{\pi^2}{12}\,,\\
P_3(x) & =  -\frac{(\pi-x)^3}{6}+\frac{\pi^2(\pi-x)}{6}\,,\\
P_4(x) & =  -\frac{(\pi-x)^4}{8}+\frac{\pi^2(\pi-x)^2}{4}-\frac{7\pi^4}{120}\,.
\end{align}
\end{widetext}

\section{One-loop Polyakov loop}\label{appsec:PL}
We detail the evaluation of the one-loop expression (\ref{appeq:PL}) of the Polyakov loop. Using the weights of the representation, as introduced in Eq.~\eqn{eq:wei}, and writing $g\bar A=i\hat r_j t^R_{0^{(j)}}$, we have
\beq
g\bar A=\sum_{\mu,a}(i\hat r\cdot\mu) |\mu,a\rangle\langle\mu,a|
\eeq
and thus
\beq
e^{\tau g\bar A}=\sum_{\mu,a}e^{i\tau \hat r\cdot\mu} |\mu,a\rangle\langle\mu,a|\,.
\eeq
We then need to evaluate the color trace
\begin{align}
\label{eq:trace2}
 \tr \left\{t_\kappa^Re^{(\beta-\tau)g\bar A} t_\lambda^Re^{\tau g\bar A}\right\} =&\sum_{\mu,a,\mu',b}e^{i\tau \hat r\cdot\mu+i(\beta-\tau)\hat r\cdot\mu'}\nonumber\\
& \times\,\langle\mu,a|t^R_\kappa|\mu',b\rangle\langle\mu',b|t^R_\lambda|\mu,a\rangle\,.
\end{align}
Because this trace needs to be contracted with the propagator ${\cal G}^{\lambda\kappa}(\tau,{\bf 0})$, we only need to consider $\kappa=\lambda=0^{(j)}$ and $\lambda=-\kappa=\alpha$. In the first case, we have $\mu=\mu'$ and $a=b$ and the trace (\ref{eq:trace2}) becomes
\beq
 \sum_\mu {\rm mul}(\mu)\,\mu^2_j\,e^{i\beta \hat r\cdot\mu}\,,
\eeq
where ${\rm mul}(\mu)$ is the degeneracy of the weight $\mu$.  To deal with the second case, we remark that
\begin{align}
\label{eq:63}
t^R_{0^{(j)}}t^R_\alpha|\mu,a\rangle & = [t^R_{0^{(j)}},t^R_\alpha]|\mu,a\rangle+t^R_\alpha t^R_{0^{(j)}}|\mu,a\rangle\nonumber\\
& = \alpha_j t_\alpha^R|\mu,a\rangle+t^R_\alpha t^R_{0^{(j)}}|\mu,a\rangle\nonumber\\
& = (\alpha+\mu)_jt_\alpha^R|\mu,a\rangle\,.
\end{align}
It follows that $t^R_\alpha|\mu_a\rangle$ is either equal to zero or it has a definite weight equal to $\alpha+\mu$. The trace (\ref{eq:trace2}) is then such that $\mu'=\mu+\alpha$ and is rewritten as
\beq
\sum_{\mu,a,b}e^{i\tau \hat r\cdot\mu+i(\beta-\tau)\hat r \cdot(\mu+\alpha)}\,\left|\langle\mu+\alpha,b|t^R_\alpha|\mu,a\rangle\right|^2,
\eeq
where $|\mu+\alpha,b\rangle$ should be understood as $0$ if $\mu+\alpha$ is not a weight.
\begin{widetext}
We thus get for the one-loop contribution \eqn{appeq:PL} of the Polyakov loop
\beq
\label{appeq:bjuhb}
\ell^{(1)}(r)  = -\frac{g^2\beta}{2d_R}\sum_\mu{\rm mul}(\mu)\,\mu^2 e^{i r\cdot\mu}\!\int_0^\beta \!\!d\tau \,G_{00}^{00}(\tau,{\bf 0}) - \frac{g^2\beta}{2d_R}\sum_{\mu,\alpha,a,b}\left|\langle\mu+\alpha,b|t^R_\alpha|\mu,a\rangle\right|^2e^{ir\cdot\mu}\!\int_0^\beta \!\!d\tau \,G^{(-\alpha)\alpha}_{00}(\tau,{\bf 0})e^{i\tau\hat r \cdot\alpha}
\eeq
where we used $\beta \hat r=r$ as well as the $\beta$-periodicity of the propagator in imaginary time $G^{\kappa\lambda}_{\mu\nu}(\tau+\beta,{\bf x})=G^{\kappa\lambda}_{\mu\nu}(\tau,{\bf x})$ and the property $G^{\kappa\lambda}_{\mu\nu}(\tau,{\bf x})=G^{\lambda\kappa}_{\nu\mu}(-\tau,{\bf x})$ which follows from the definition of the two-point function and rotation invariance. In Fourier space, \Eqn{appeq:bjuhb} reads [we use $Q=(\omega_n,{\bf q})$ and we recall our convention (31) for the propagator in Fourier space]
\beq
\ell^{(1)} (r)= -\frac{g^2\beta}{2d_R}\sum_\mu{\rm mul}(\mu)\,\mu^2e^{ir\cdot\mu}\int_{\bf q} \frac{1}{q^2+m^2}-\frac{g^2\beta}{d_R}\sum_{\mu,\alpha,a,b}\left|\langle\mu+\alpha,b|t^R_\alpha|\mu,a\rangle\right|^2e^{ir\cdot\left(\mu+\frac{\alpha}{2}\right)}\sin\left(\frac{r\cdot\alpha}{2}\right)\int_Q\frac{G_{00}(Q^\alpha)}{\omega_n+\hat r\cdot\alpha}.
\eeq
Using \cite{Reinosa:2014zta}
\bea
\int_Q\frac{G_{00}(Q^\alpha)}{\omega_n+\hat r\cdot\alpha } & = & \frac{1}{2}{\rm cotan}\left(\frac{r\cdot\alpha}{2}\right)\int_{\bf q}\frac{1}{q^2+m^2}+\int_{\bf q} \frac{q^2}{m^2}\,{\rm Im}\left(\frac{n_{\varepsilon_q-i\hat r\cdot\alpha}}{\varepsilon^2_q}-\frac{n_{q-i\hat r\cdot\alpha}}{q^2}\right)\nonumber\\
& = & \frac{e^{i\frac{r\cdot\alpha}{2}}+e^{-i\frac{r\cdot\alpha}{2}}}{2}\left[\frac{1}{2}\frac{1}{\sin\left(\frac{r\cdot\alpha}{2}\right)}\int_{\bf q}\frac{1}{q^2+m^2}-\sin\left(\frac{r\cdot\alpha}{2}\right) \frac{m\,a(T,r\cdot\alpha)}{2\pi^2}\right],
\eea
with the function $a(T,x)$ defined in \Eqn{eq:ATX}, which is strictly positive if $m>0$, we get
\begin{align}
\ell^{(1)} (r)=& -\frac{g^2\beta}{2d_R}\sum_\mu{\rm mul}(\mu)\,\mu^2e^{ir\cdot\mu}\int_{\bf q} \frac{1}{q^2+m^2}\nonumber\\
& - \frac{g^2\beta}{2d_R}\sum_{\mu,\alpha,a,b}\left|\langle\mu+\alpha,b|t^R_\alpha|\mu,a\rangle\right|^2e^{ir\cdot\mu} \left[\frac{1}{2}\int_{\bf q} \frac{1}{q^2+m^2}-\sin^2\left(\frac{r\cdot\alpha}{2}\right)\frac{m\,a(T,r\cdot\alpha)}{2\pi^2}\right]\nonumber\\
& - \frac{g^2\beta}{2d_R}\sum_{\mu,\alpha,a,b} \left|\langle\mu+\alpha,b|t^R_\alpha|\mu,a\rangle\right|^2e^{ir\cdot(\mu+\alpha)}\left[\frac{1}{2}\int_{\bf q} \frac{1}{q^2+m^2}-\sin^2\left(\frac{r\cdot\alpha}{2}\right)\frac{m\,a(T,r\cdot\alpha)}{2\pi^2}\right].
\end{align}
We can change $\alpha$ to $-\alpha$, $\mu$ to $\mu+\alpha$ and $a\leftrightarrow b$ in the last line. We note that the set of roots is invariant under $\alpha\to-\alpha$. Moreover $\left|\langle\mu+\alpha,b|t^R_\alpha|\mu,a\rangle\right|^2$ selects those $\mu+\alpha$ which are weights and, for a given $\mu$, $\mu+\alpha$ explores all the weights of the representation as $\alpha$ is varied because the representation is irreducible. It follows that, upon the change of variables, the domains over which $\mu$ and $\alpha$ are summed over remain the same. Upon exploiting this fact, as well as the properties $a(T,x)=a(T,-x)$ and $\left|\langle\mu+\alpha,b|t^R_\alpha|\mu,a\rangle\right|^2=\left|\langle\mu,a|t^R_\alpha|\mu+\alpha,b\rangle\right|^2$, the last two lines are shown to be equal and we arrive at
\begin{align}
\ell^{(1)} (r)&= -\frac{g^2\beta}{2d_R}\sum_{\mu,a}\left\{\mu^2+\sum_{\alpha,b} \left|\langle\mu+\alpha,b|t^R_\alpha|\mu,a\rangle\right|^2\right\}e^{ir\cdot\mu}\int_{\bf q} \frac{1}{q^2+m^2}\nn&+\frac{g^2\beta}{d_R}\sum_{\mu,\alpha,a,b}e^{ir\cdot\mu} \left|\langle\mu+\alpha,b|t^R_\alpha|\mu,a\rangle\right|^2\sin^2\left(\frac{r\cdot\alpha}{2}\right)\frac{m\,a(T,r\cdot\alpha)}{2\pi^2}\,.
\end{align}
The contribution between curly brackets is independent of $\mu$ and is equal to the Casimir $C_R$ of the representation; see \Eqn{appeq:cellela}. We then end up with
\beq
\ell^{(1)} (r)= -\frac{g^2\beta C_R}{2d_R}\sum_\mu {\rm mul}(\mu)\,e^{ir\cdot\mu}\int_{\bf q} \frac{1}{q^2+m^2}+\frac{g^2\beta}{d_R}\sum_{\mu,\alpha,a,b}e^{ir\cdot\mu} \left|\langle\mu+\alpha,b|t^R_\alpha|\mu,a\rangle\right|^2\sin^2\left(\frac{r\cdot\alpha}{2}\right)\frac{m\,a(T,r\cdot\alpha)}{2\pi^2}.
\eeq
Using that $\int_{\bf q}1/(q^2+m^2)=-m/(4\pi)$ in dimensional regularization and using the expression \eqn{eq:l1l} of the LO Polyakov loop, we arrive at
\beq
\ell(r)=\frac{1}{d_R}\sum_\mu {\rm mul}(\mu)e^{ir\cdot\mu}\underbrace{\left\{1+g^2\frac{m}{T}\left(\frac{C_R}{8\pi}+\sum_{\alpha,a,b}\frac{\left|\langle\mu+\alpha,b|t^R_\alpha|\mu,a\rangle\right|^2}{{\rm mul}(\mu)}\sin^2\left(\frac{r\cdot\alpha}{2}\right)\frac{a(T,r\cdot\alpha)}{2\pi^2}\right)\right\}}_{>0}+{\cal O}(g^4).
\eeq
which is \Eqn{eq:PL_res}.
\end{widetext}

\section{Polyakov loops and the center symmetry in SU($N$) theories}
\label{appsec:ploopcounting}

As discussed in Sec.~\ref{sec:su4} for the SU($4$) case, demanding that the Polyakov loop functions $\ell_R(r)=0$ for only a subset of representations is usually not sufficient to uniquely select the (ensemble of) center-symmetric point(s) in the Weyl chamber. Clearly, a necessary requirement is to demand that the Polyakov loop functions vanish in at least one representative of each $N$-ality class, e.g., $\ell_{\bf 2}=0$ for SU($2$),  $\ell_{\bf 3}=\ell_{\bar{\bf 3}}=0$ for SU($3$), or  $\ell_{\bf 4}=\ell_{\bar{\bf 4}}=\ell_{\bf 6}=0$ for SU($4$). In the main text, we have proven that this is also a sufficient condition for center symmetry at LO in these groups. Here, we generalize this discussion to SU($N$).

The center group of SU($N$) is $Z_N$ and the Cartan subalgebra is of dimension $N-1$. The various irreducible representations of the group can be described in terms of Young tableaux, whose number of boxes gives the $N$-ality of the representation. Demanding that the Polyakov loop function in the first pair of (conjugate) fundamental representations ({\bf N}$,\bar{\bf N})$ vanishes gives two real constraints\footnote{In practice it is sufficient to demand that only one of these Polyakov loop vanishes since, for two conjugate representations $R$ and $\bar R$ one has $\ell_{\bar{R}}=\ell_R^*$.} and yields a hypersurface of dimension $N-3$ in the Cartan space. But these representations are of respective $N$-alities $1$ and $N-1$ and the associated Polyakov loops only probe a subset of $Z_N$. 

There are, in fact, $N-1$ independent fundamental representations, which span the whole set of nontrivial $N$-alities. These are given by all possible Young tableaux with one column of $k$ boxes, with $1\le k\le N-1$. Those representations have $N$-ality $k$ and their dimension is given by the binomial coefficient $C_N^k$. They come in pairs of conjugate representations, with $\mathbf{ C_N^{N-k}}=\overline{\mathbf {C_N^k}}$. For $N$ odd, there are $(N-1)/2$ pairs of conjugate fundamental representations and, demanding the vanishing of the corresponding Polyakov loop functions yields {\it a priori} $N-1$ real constraints. For $N$ even, there are $(N-2)/2$ pairs of conjugate representations and one real fundamental representation and the number of {\it a priori} independent constraints is, again, $N-1$. Assuming that these are indeed independent conditions, this reduces the ensemble of solutions to a countable set of points in the Weyl chamber. Further assuming that this only contains center-symmetric points would directly imply that all other Polyakov loops in representations of nonzero $N$-ality vanish. We believe these are reasonable assumptions although we have not found a rigorous proof.

In Sec.~\ref{sec:su4} we have also mentioned the possibility to probe the center symmetry through observables such as $\langle\tr (L_{\bf N})^k\rangle$, where $L_{\bf N}$ is the Polyakov loop operator \eqn{eq_pol_untrace} in the fundamental representation of lowest dimension \cite{Dumitru:2012fw,Pisarski:2002ji}. To probe all the elements of the center $Z_N$, one needs to require at least that $\langle\tr(L_{\bf N})^k\rangle=0$ for $1\le k\le N-1$. At LO, we have $\langle\tr (L_{\bf N})^k\rangle=\ell^{(0)}_{\bf N}(kr)$ and we have {\it a priori} $2(N-1)$ real constraints. 
However, using the Cayley-Hamilton theorem together with the fact that $L_{\bf N}$ is a unitary matrix, one reduces the number of {\em independent} constraints to at most $N-1$. As above, this reduces the ensemble of solutions to a countable set. The latter obviously includes all the center-symmetric points and it is reasonable to assume that there are no other solutions.


  \end{document}